%% file: after_revision_main_June2026.tex
\PassOptionsToPackage{table,xcdraw}{xcolor}
\documentclass[12pt]{article}
\input{final_preamble}

\begin{document}

\def\spacingset#1{\renewcommand{\baselinestretch}%
{#1}\small\normalsize} \spacingset{1}

\makeatletter
\g@addto@macro\normalsize{%
  \setlength\abovedisplayskip{10pt plus 2pt minus 5pt}%
  \setlength\belowdisplayskip{10pt plus 2pt minus 5pt}%
  \setlength\abovedisplayshortskip{6pt plus 2pt minus 3pt}%
  \setlength\belowdisplayshortskip{6pt plus 2pt minus 3pt}%
  \setlength\jot{3pt}
}
\makeatother

\if1\blind
{
  \title{Distributional Treatment Effect Transportability across Heterogeneous Sites}
\author{ Borna Bateni\\
Department of Statistics \& Data Science, University of California, \\ Los Angeles, Los Angeles, CA, USA\\
Yubai Yuan\\
Department of Statistics, The Pennsylvania State University, \\ University Park, PA, USA\\
Qi Xu\\
School of Statistics, University of Minnesota, Minneapolis, MN, USA\\
and \\
Annie Qu\thanks{
    Corresponding author. E-mail: \texttt{aqu2@ucsb.edu}. 
    The authors gratefully acknowledge support from the U.S. National Science Foundation (NSF) under Grants DMS-2515275 and DMS-2515698.
  }\\
Department of Statistics \& Applied Probability, University of California, Santa Barbara, Santa Barbara, CA, USA
}

  \date{}
  \maketitle
  \vspace{-10mm}
} \fi

\if0\blind
{
  \bigskip
  \bigskip
  \bigskip
  \title{Distributional Treatment Effect Transportability across Heterogeneous Sites}
  \author{}
  \date{}
  \maketitle
  \medskip
} \fi

\begin{abstract}
We study distributional transportability of treatment effects in a ``cross-site, one-armed target" design, where both treated and control units are observed in a source site, but only control units are observed in a target site. Our object of interest is not estimating the average treatment effect, but recovering the full treated distribution in the target site using transfer knowledge from the source site, while allowing cross-site heterogeneity in measurement systems, observed features, outcome reporting, population composition, and latent contextual factors. We model cross-site heterogeneity through a transformation between the sites that transports the joint feature--outcome distributions. This transformation is learned from comparing the observed control samples in the source and target sites, using an optimal transport criterion. The learned transformation is then applied to the source treated sample to construct a synthetic sample from the target treated distribution. We establish convergence of the resulting synthetic empirical distribution to the target treated distribution. Simulation studies across multiple data-generating scenarios and a real-world application to patient-derived xenograft (PDX) data demonstrate that our framework recovers the full distributional properties of the target treated population.
\end{abstract}

\noindent%
\vfill

\newpage
\spacingset{1.7} 
 \section{Introduction}

A central challenge in experimental research and policy evaluation is transporting treatment effects from a completed study at a site to a new site where the same intervention has not yet been fully observed. Even when an intervention is successful in one site, differences in site-specific characteristics can change the distribution of outcomes, so that treatment effects do not directly generalize to another site. 

A common version of this problem arises when a completed study at a \emph{source} site observes both treatment and control groups, whereas a prospective \emph{target} site only observes information from untreated units. This ``cross-site, one-armed target'' setting appears across several domains. In epidemiology and medicine, it appears when randomized trial results are extended to external patient populations  \citep{cole2010generalizing, Westreich2017AJE, Buchanan2018JRSSA, Hong2019AJE,  Dahabreh2020SIM}. In econometrics, it underlies program evaluation methods that predict impacts in new locations using prior experimental evidence \citep{hotz2005predicting, menzel2023transfer}. Analogous settings arise in political science \citep{EgamiHartman2021JRSSA} and education \citep{KernStuartHillGreen2016JREE}.

The literature above leaves two important limitations: first, cross-site transportability has largely been formulated in terms of transporting the \emph{average treatment effect} (ATE) from a source site to a target site. The ATE, however, is only a scalar summary of treatment effects and does not characterize the full distribution of counterfactuals. In many applications, the scientific question concerns how treatment reshapes the outcome distribution itself, including changes in quantiles, tail behavior, and heterogeneity in impacts across subpopulations \citep{heckman1997making, DJEBBARI200864, Bitler2017}. While \emph{distributional} causal estimands have been studied in other settings, such as panel designs through \emph{synthetic control} \citep{abadie2010synthetic, abadie2015comparative}, and causal distributional inference within a single site \citep{fan2010sharp, chernozhukov2013inference, firpo2016identification, lin2023causal}, the existing approaches do not directly address the cross-site, one-armed target setting, where the target treated distribution must be recovered using information from a distinct source site.

Second, cross-site transportability methods typically rely on a \emph{conditional invariance} assumption (called ``unconfounded location" in \citet{hotz2005predicting}), which requires potential outcome distributions to be invariant across sites, conditional on a common set of observed covariates. Under this assumption, cross-site heterogeneity is limited to differences in the distribution of observed covariates, so the target-site treatment effect can be recovered by reweighting or matching source units to the target covariate distribution. This assumption is restrictive when source and target sites differ not only in covariate distribution, but also in the variables recorded, measurement scales, outcome reporting conventions, and latent site-specific factors that change the feature--outcome geometry. In such cases, cross-site heterogeneity acts on the joint feature--outcome distribution itself, and treatment effects cannot be recovered by covariate reweighting alone. This motivates the development of methods that accommodate richer cross-site heterogeneity while using source-site data to recover treatment effects at the target site.

In this paper, we propose a framework for recovering the target treated distribution in the cross-site, one-armed target setting. The central idea is to view the source and target sites as distinct joint feature--outcome spaces with distinct covariate structures and similarity metrics, and represent cross-site heterogeneity through a transformation linking their distributions of treated and control units. Our method learns this transformation by comparing the observed control samples across sites and then applies it to the source treated sample to construct a synthetic treated sample in the target site.

To learn this transformation, we develop an \emph{optimal transport} (OT) criterion. OT-based methods have recently become popular in causal inference for constructing counterfactual distributions, with applications in difference-in-differences \citep{torous2024optimal}, synthetic control \citep{gunsilius2023distributional}, and balancing weights \citep{dunipace2021optimal}. The methodological novelty of our approach is to make the OT criterion transformation-aware: whereas standard OT learns a coupling between distributions, we modify the transport objective to also learn the cross-site transformation that maps the source distribution into the target space.

Accordingly, our contributions are as follows: \emph{(i) Conceptual.} We formulate the cross-site, one-armed target problem as a problem of \emph{distributional transportability}: the estimand is the full target treated distribution, rather than only the ATE. \emph{(ii) Identification. } We replace the classical conditional invariance (unconfounded location) assumption with a shared cross-site transformation of the joint feature--outcome distributions, allowing richer forms of cross-site heterogeneity than differences in covariate distributions alone. \emph{(iii) Methodology. }We develop a transformation-aware optimal transport criterion that learns a cross-site transformation from the source and target control samples and applies it to synthesize the target treated distribution from source treated data. \emph{(iv) Theory. }We establish  weak convergence of the resulting synthetic treated empirical distribution to the target treated distribution. \emph{(v) Empirical.} Through simulations and an application to patient-derived xenograft (PDX) data, we show that the proposed method recovers distributional features of the target treated population and improves on competing approaches across distributional metrics.

The remainder of the paper is organized as follows. Section~\ref{sec:1}
introduces the problem setup, notation, and identification assumptions. Section~\ref{sec:2} presents our estimation procedure and theoretical guarantees. Section~\ref{sec:3}
reports simulation results. Section~\ref{sec:4} presents an application
to patient-derived xenograft (PDX) data. Section~\ref{sec:5} concludes.

\section{Problem Setup and Identification}\label{sec:1}
\subsection{Cross-site, One-armed Target Setup}

We consider two sites, referred to as a \emph{source} and a \emph{target} site. Let $\mathcal Z:= \mathcal X\times \mathcal Y$ denote the joint space of features ($\mathcal X$) and outcomes ($\mathcal Y$) at the source site, and similarly $\mathcal Z':= \mathcal X'\times \mathcal Y'$ denote the joint feature--outcome space at the target site. We adopt a \emph{distributional causal inference} framework \citep{lin2023causal} in which the object of interest is the full distribution of potential outcomes. Accordingly, we represent treatment and control groups at each site by probability measures on the joint feature--outcome space. Let $\mu_0, \mu_1$ denote the control and treated distributions at the source site $\mathcal Z$, and $\mu'_0,\mu'_1$ denote the corresponding distributions at the target site $\mathcal Z'$. 

We assume that each site is equipped with a \emph{similarity metric}, $d_{\mathcal Z}$ on $\mathcal Z$ and $d_{\mathcal Z'}$ on $\mathcal Z'$, which quantify the similarity between the feature--outcome pairs within each site. Therefore, each site is a \emph{metric measure space} with a similarity metric and two probability measures, $\big(\mathcal Z, d_{\mathcal Z}, \{\mu_0,\mu_1\}\big)$ and $\big(\mathcal Z', d_{\mathcal Z'}, \{\mu_0',\mu_1'\}\big)$. Throughout, we assume that these spaces are \emph{Polish} (complete and separable under their respective metric), and all probability measures considered are Borel measures. Moreover, we assume that the source site corresponds to a well-studied environment, therefore, the source metric $d_\mathcal Z$ is known. By contrast, the target site represents a new environment in which the similarity metric $d_{\mathcal Z'}$ may differ, and is not assumed known a priori.

We study a cross-site design in which both treated and control arms are observed at the source site, but only the control arm is observed at the target site. Accordingly, we observe i.i.d. samples

\vspace{2.5mm}

\centerline{$\displaystyle Z_0:=\big\{z_{0i}\big\}_{i=1}^{n_0}\sim \mu_0 \,,\quad Z_1:=\big\{z_{1i}\big\}_{i=1}^{n_1}\sim \mu_1\,,\quad n_0,n_1<\infty\,,$}

\vspace{2.5mm}

\noindent drawn from the control and treated distributions in the source site, and a  single i.i.d. sample 
\vspace{2.5mm}

\centerline{$\displaystyle Z_0':=\big\{z_{0i}'\big\}_{i=1}^{n_0'}\sim \mu_0' \,,\quad n'_0<\infty\,,$}

\vspace{2.5mm}

\noindent drawn from the control distribution in the target site. The objective of the study is to generate a synthetic dataset, $Z_1^{'\,\text{(synth.)}}:=\big\{z'_{1i}\big\}_{i=1}^{n'_1}$, that resembles an i.i.d. sample from the target treated distribution $\mu'_1$.

\subsection{Identification Assumptions}\label{sec:identification}

The main identification challenge in the cross-site, one-armed target setting is that the target treated distribution $\mu'_1$ is unobserved. While the observed samples from the control distributions $\mu_0$ and $\mu'_0$ allow us to learn the cross-site shift among the control units, they do not, by themselves, determine how treated units would shift across sites. Identifying $\mu'_1$ therefore requires restrictions that link the observed control-side shift from $\mu_0$ to $\mu'_0$ to the unobserved treated-side shift from $\mu_1$ to $\mu'_1$. 

Classical cross-site transportability literature closes this gap by imposing conditional invariance across sites, referred to as ``unconfounded location" by \citet{hotz2005predicting}. In its standard form, unconfounded location requires

\vspace{2.5mm}

\centerline{$\displaystyle T_i\;\perp\; \Big(Y_i(0), Y_i(1)\Big)\; \Big|\; X_i\,,$} 

\vspace{2.5mm}

\noindent where $T_i$ denotes the site indicator. Under this assumption, cross-site differences arise only through the distribution of observed covariates, and site membership carries no residual information about the potential outcomes once those covariates are conditioned on. Therefore, the target-site treatment effect can be recovered by reweighting or matching source units to reproduce the target covariate distribution.

In many applications, however, source and target sites differ in covariate representation, measurement scales, outcome reporting conventions, and latent site-specific factors. Such differences induce structural shifts in the joint feature--outcome distribution that cannot be removed by covariate reweighting alone. In these settings, unconfounded location fails, and the target treated distribution is not identified without additional structure.

Our framework replaces unconfounded location with a transformation-based identification structure that allows cross-site differences to go beyond the observed covariates and act on the joint feature--outcome distribution itself. The assumptions below formalize this structure.

\begin{assumption}
\label{as:kernel}
The feature spaces in the source and target sites share a common ``anchor'' subspace, $\mathcal H\subseteq \mathcal X\cap \mathcal X'$, on which the features are directly comparable across sites. Accordingly, there exists a continuous alignment function

\vspace{2.5mm}

\centerline{$\displaystyle
\mathcal K:\mathcal Z\times\mathcal Z'\to\mathbb R_+\,,
$}

\vspace{2.5mm}

\noindent that depends on $z\in\mathcal Z$ and $z'\in \mathcal Z'$ only through their respective projections onto $\mathcal H$.
\end{assumption}

\begin{assumption}
\label{as:map}
There exists a map from the source to the target site, $\f:\mathcal Z \to \mathcal Z'$, such that $\f$ is bijective and measurable, and $\f^{-1}$ is also measurable, and $\f$ \emph{pushes forward} the source control distribution to the target control distribution

\vspace{2.5mm}

\centerline{$\displaystyle \f_{\#}\mu_0 = \mu_0'\,,$}

\vspace{2.5mm}

\noindent meaning for any measurable  subset of the target space $A\subseteq\mathcal Z'$, $\displaystyle \mu'_0(A)=\mu_0(\f^{-1}(A))\,$. Moreover, for the alignment function $\mathcal K$ defined above, $\f$ satisfies
\vspace{2.5mm}

\centerline{$\displaystyle
\mathcal K\!\left(z,\,
\f(z)\right)=0\,, \qquad \mu_0\text{-almost everywhere}\,.
$}
\end{assumption}

\begin{assumption}
    \label{as:trt} Any measurable bijection $f:\Z\to\Z'$ satisfying $f_\#\mu_0=\mu_0'$ and 
    
    \vspace{2.5mm}

\centerline{$\displaystyle \mathcal K(z,f(z))=0\,,\qquad \text{$\mu_0$-almost everywhere}\,,$}

\vspace{2.5mm}

\noindent is also a valid push-forward for the treated distributions across sites, satisfying

\vspace{2.5mm}

\centerline{$\displaystyle f_\#\mu_1=\mu'_1\,.$}

\vspace{2.5mm}

\noindent
\end{assumption}

\begin{assumption}\label{as:overlap}
The source treated distribution is absolutely continuous with respect to the source control distribution

\vspace{2.5mm}

\centerline{$\displaystyle
\mu_1 \ll \mu_0\, ,
$}

\vspace{2.5mm}

\noindent implying that the support of the treated distribution is contained in the support of the control distribution, $\operatorname{supp}(\mu_1)\subseteq \operatorname{supp}(\mu_0)$, up to $\mu_0$--null sets. By bijectivity of $\f$, absolute continuity is preserved under push-forward, implying $\mu'_1\ll\mu'_0$ as well.

\end{assumption}

The above assumptions impose a transformation-based structure on the cross-site heterogeneity. Assumption~\ref{as:kernel} requires that the source and target sites share a subset of ``anchor'' features, $\mathcal H$, on which the features are directly comparable across sites. Hence, although the two sites may differ in covariate representation, the anchor features provide a common reference frame in which the source and target observations can be compared directly.

Assumption~\ref{as:map} states that the observed shift from the source control distribution to the target control distribution is generated by a measurable transformation, $\f$, on the joint feature--outcome space. Thus, cross-site heterogeneity is allowed to act on the full feature--outcome distribution, rather than only through the covariates. However, while $\f$ may act flexibly in the full space, it preserves comparability on the anchor features through the alignment function $\mathcal K$.

Assumptions~\ref{as:kernel} and \ref{as:map} describe the observed control-side shift, but they do not, by themselves, determine how treated units should be transported across sites. In general, multiple transformations may agree on the source and target control distributions, while transporting the source treated distribution to different distributions in the target space. Assumption~\ref{as:trt} closes this identification gap by requiring that any transformation of the control distribution that remains stable on the anchor features also transports the source treated distribution to the target treated distribution.

Finally, Assumption ~\ref{as:overlap} is the support condition that ensures the treated distribution lies within the support of the control distribution, thus the transformation identified from the control-side distributions covers all the regions of the data space where treated observations may occur. Without this assumption, some treated units would lie in parts of the source space where the control data provide no information about how the transformation acts, making the transfer to the target space unidentifiable.

\subsubsection{Interpretation and Plausibility}

Together, Assumptions~\ref{as:map}--\ref{as:trt} assert that cross-site heterogeneity is governed by an \emph{anchor-compatible shared transformation}. The transformation $\f$ describes how the joint feature--outcome distribution of control units shifts from the source site to the target site. The anchor structure rules out transformations that violate comparability on anchor features. Thus, the identifying structure filters for control-side transport maps that remain stable on the anchor features. Assumption~\ref{as:trt} then requires this anchor-compatible control-side transformation to be \emph{shared} across treatment arms. Hence, once $\f$ is learned from comparing the observed control distributions $\mu_0$ and $\mu'_0$, subject to the anchor restrictions, the target treated distribution $\mu'_1$ is identified as the push-forward $\f_{\#}\mu_1$.

The shared-transformation assumption is most plausible when cross-site differences arise from broad site-level factors that affect treated and control units in similar ways. For example, if the source and target sites are two hospitals, such differences may include baseline care, monitoring quality, data collection protocols, calibration procedures, reporting conventions, or patient population composition. If one hospital provides stronger nurse support or more frequent follow-up for all patients, regardless of treatment arm, then a common transformation may reasonably describe how both the control and treated distributions shift across sites.

By contrast, the assumption becomes less credible when cross-site differences are \emph{treatment-specific}. This may occur when one site implements the treatment differently from the other, including through different dosing schedules, treatment adherence patterns, or treatment-specific follow-up protocols. In such cases, the observed control-side shift across sites may no longer reveal the treated-side shift, and a single shared transformation is unlikely to capture cross-site heterogeneity in both arms. Overall, because the target treated distribution is unobserved, the shared-transformation assumption cannot be verified from the available data alone; its plausibility depends on scientific knowledge of the application and whether the cross-site differences are expected to affect treated and control units similarly. A sensitivity result allowing the shared-transformation condition to hold only approximately is provided in Appendix~A.4 of the Supplementary Material.

The anchor subspace $\mathcal H$ consists of features such as age, gender, baseline disease stage, or pre-treatment biomarkers whose scientific definition is the same across sites. On these variables, matched source and target units are expected to remain directly comparable. By contrast, variables such as hospital-specific risk scores, scanner-dependent image features, or laboratory measurements produced under incompatible calibration standards may have different definitions and scales across sites, and therefore not be directly comparable. Such variables would not serve as anchors. Thus, the framework does not require the source and target sites to share the same full feature set, but only a subset of stable features that can support meaningful cross-site alignment.

To that end, the alignment function $\mathcal K$ encodes what ``comparable'' means on the anchor subspace. Different anchors may require different notions of comparability. For variables measured on a common scale across sites, such as age or gender, comparability means closeness in the \emph{original coordinate scale}: a patient in the source site should correspond to a patient in the target site with similar age or gender. For  anchor variables with the same scientific definition but different absolute scales across sites,
 such as blood pressure or biomarker concentrations, comparability may instead be more naturally expressed through closeness in \emph{quantiles}: a patient near the upper tail of the source distribution should still correspond to a patient near the upper tail of the target distribution. In our implementation, \(\mathcal K\) is constructed as a weighted sum of
coordinate-level differences on \(\mathcal H\), with each coordinate allowed to
use either a scale-based or quantile-based comparison. The precise forms used in the paper, including the
distance-based and quantile-based choices of $\mathcal K$, are given in Appendix~B.1 of Supplementary Material.

\begin{figure}[t]
\centering
\includegraphics[width=0.9\linewidth]{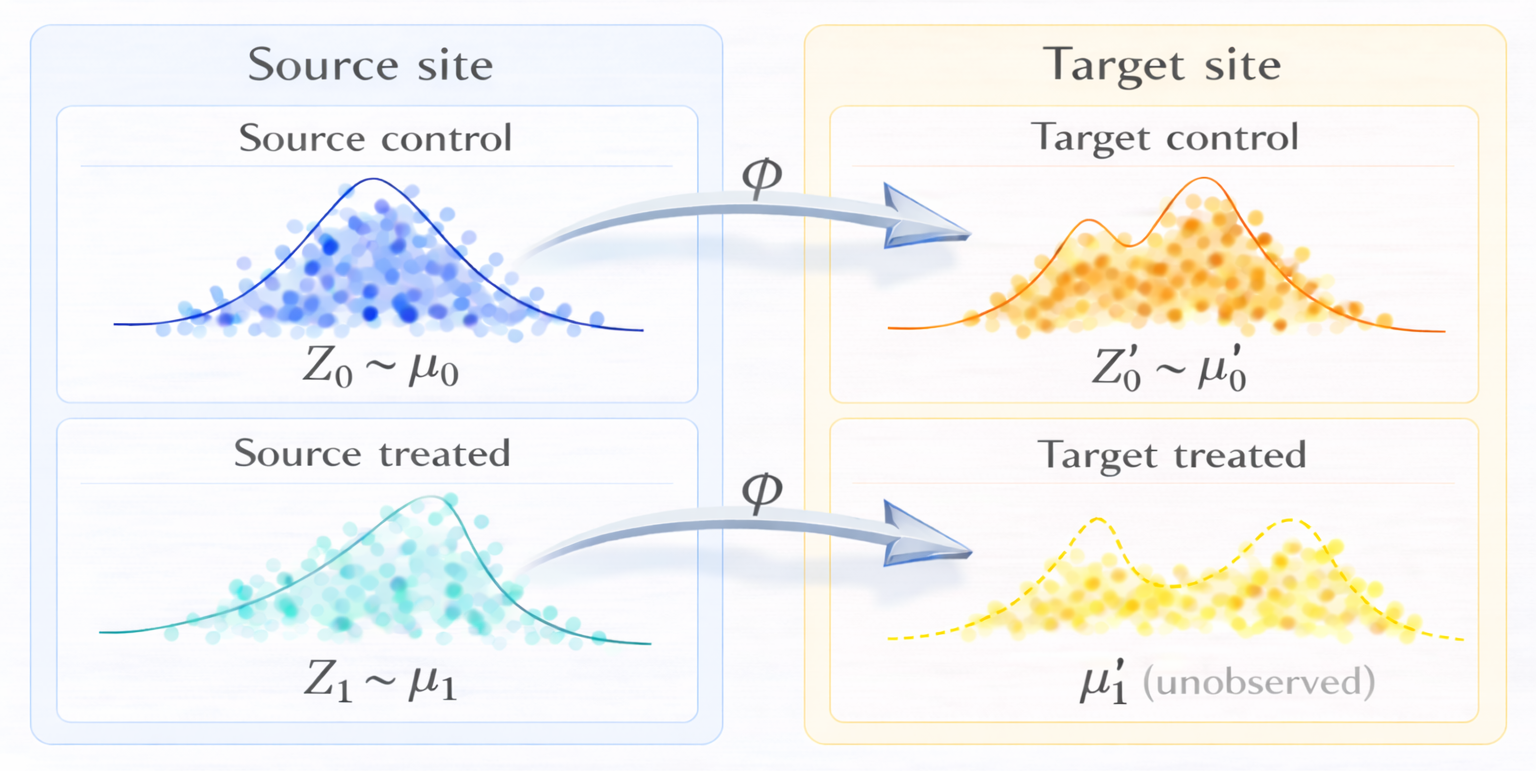}
\caption{Cross-site identification framework. In the source site, control and treated data are observed, while in the target site only the control data is observed. A shared $\f$ transfers the control and treated distributions across sites.}
\end{figure}

\subsubsection{Relation to Existing Literature}

The shared-transformation assumption has close analogues in causal inference literature. In population-adjusted indirect comparisons and network meta-analysis, \citet{phillippo2018methods,phillippo2020multilevel} introduce a ``shared effect modification'' assumption, requiring that the relevant effect modifiers alter treatment effects in the same way across treatments. Our assumptions play a similar role at the distributional level, requiring the entire joint feature--outcome distribution to be linked across sites by a common transformation shared across treatment arms. Related structure appears in distributional difference-in-differences, where \citet{torous2024optimal} formulate identification through a shared optimal transport map that transfers the treated and untreated outcome distributions across pre- and post-treatment periods. Our assumptions adopt the same transport map structure, but across sites rather than times. Our assumption framework is also comparable to the transportability framework of Pearl and Bareinboim \citeyearpar{bareinboim2016causal,pearl2022external}, where cross-environment causal effects are identified through invariance assumptions on structural mechanisms. Our assumptions instead operate directly on the joint feature--outcome distribution and its transformation across environments, without requiring specification of the underlying causal structure.

Our anchor features assumption is also natural within causal inference literature. In cross-environment causal analyses, valid comparisons across groups require that at least a subset of measurement parameters be equal across groups, a condition termed ``{partial measurement invariance}'' \citep{gregorich2006self,leitgob2023measurement}. The anchor subspace $\mathcal H$ plays an analogous role in our framework as it must remain sufficiently stable to permit meaningful cross-site comparison, while the remaining coordinates are allowed to differ across sites. A related idea also appears in the change-in-changes framework of \citet{athey2006identification}, where Assumption 3.2 requires the production function \(h(u,t)\) to be strictly increasing in the latent heterogeneity variable \(u\), so that the ordering induced by $u$ is preserved and the quantile transformation across periods is identified. Our quantile-based alignment uses the same rank-preservation intuition for the selected anchor coordinates, but across sites rather than time periods.

\section{Methodology}\label{sec:2}

In this section we describe the proposed methodology for generating a synthetic sample from the target treated distribution $\mu'_1$ using observations from the source control and treated distributions, $\mu_0$ and $\mu_1$, and the target control distribution, $\mu'_0$. The method proceeds in two steps. First, we estimate the cross-site distributional transformation between the source and target control distributions. Second, we apply the learned transformation to the source treated sample, to construct a synthetic target treated sample.

\subsection{Learning the Cross-site Transformation}

The first step is to use the control distributions, $\mu_0$ and $\mu'_0$, to learn a transformation, {$ f:\Z\to\Z'$}, that captures the distributional shift from $\mu_0$ to $\mu'_0$ through its push-forward,  

\vspace{2.5mm} 

\centerline{$\displaystyle f_\#\mu_0(A'):=\mu_0(f^{-1}(A'))=\mu'_0(A')\,,\qquad \text{for all measurable $A'\subseteq\mathcal Z'$}\,.$} 

\vspace{2.5mm}

\noindent We restrict our attention to a class of admissible maps

\vspace{2.5mm}

\centerline{$\displaystyle \Phi\subseteq\{f:\Z\to\Z'\}\,,$}

\vspace{2.5mm}

\noindent which represents the set of transformations considered plausible
for the application and computationally learnable. We assume that \(\Phi\) contains at
least one transformation satisfying $f_\#\mu_0=\mu'_0$, as well as other conditions in
Assumptions~\ref{as:map}--\ref{as:trt}. The regularity
conditions imposed on \(\Phi\) are stated in Appendix~A.2.2 of the
Supplementary Material. The specific parametrization of $\Phi$ and tuning choices used are in Appendix~B.2.

The transformation \(f\) enters the estimation problem through two induced objects: (i) its \emph{graph coupling}, which represents the source--target distributional alignment, and (ii) its \emph{pull-back metric}, which represents the target geometry under which alignment is evaluated. We first describe these two representations, then use them to define the optimization criterion.

\subsubsection{Coupling Representation}

The distributional alignment between $\mu_0$ and $\mu'_0$ can be represented through the notion of couplings. A \emph{coupling}, $\pi\in\Pi(\mu_0,\mu'_0)$ is a joint distribution on $\Z\times \Z'$ whose marginals are $\mu_0$ and $\mu'_0$:

\vspace{2.5mm} 

\centerline{$\displaystyle \pi(A\times \mathcal Z')=\mu_0(A)\,,\; \pi(\mathcal Z\times A')=\mu'_0(A')\,,\quad \text{for all measurable $A\subseteq\mathcal Z$ and $A'\subseteq\mathcal Z'$}\,,$} 

\vspace{2.5mm}

\noindent providing a flexible, probabilistic alignment between the distributions. Any measurable map $f:\Z\to\Z'$ that satisfies $f_\#\mu_0=\mu'_0$ induces a special ``graph'' coupling $\pi_f\in\Pi(\mu_0,\mu'_0)$, concentrated on the graph of $f$
\begin{equation} \label{eq:pi_f}\pi_f(A\times A')=\mu_0\big(A\cap f^{-1}(A')\big)\,,\qquad \text{for all measurable $A\subseteq\mathcal Z$ and $A'\subseteq\mathcal Z'$}\,.\end{equation}
\noindent Therefore, every admissible transformation determines a unique coupling, although not every coupling is induced by a transformation. We may therefore estimate the alignment between $\mu_0$ and $\mu'_0$ through a coupling, while requiring the coupling to remain consistent with the graph of an admissible transformation $f\in\Phi$.

\subsubsection{Metric Representation}

Recall that while the source metric $d_{\Z}$ is taken as known in our problem setup, the target metric $d_{\Z'}$ is not assumed known a priori. This matters because cross-site heterogeneity may change not only the location of the distributions $\mu_0$ and $\mu'_0$, but also the similarity structure of the target space. The target metric should thus be treated as part of the cross-site transformation mechanism to be learned, placing it within the broader \emph{metric-learning} perspective  \citep{kulis2013metric,hu2015deep}. 

Any bijective transformation map $f:\Z\to \Z'$ induces an associated ``pull-back'' metric on $\Z'$, defined as the distance, under the source metric $d_{\Z}$, of the $f$-pre-images of two points in $\Z'$
\begin{equation} \label{eq:pullback1}
 d_f'(z_1',z_2')
:=
d_{\mathcal Z}\big(f^{-1}(z_1'),f^{-1}(z_2')\big),
\qquad z_1',z_2'\in\mathcal Z' \,.
\end{equation}
\noindent Therefore, we may estimate the target metric by optimizing over the set of admissible pull-back metrics, $\{d'_f: f\in \Phi\}$.

\subsubsection{Optimization Criterion}

We now define the criterion used to learn the cross-site transformation. Because the source and target control distributions are supported on different spaces, \(\mathcal Z\) and \(\mathcal Z'\), the alignment must compare their within-space geometries rather than their coordinates directly. At the same time, the anchor features in Assumption~\ref{as:kernel} must remain directly comparable across sites. These two requirements are naturally combined by the \emph{Fused Gromov--Wasserstein} (FGW) distance \citep{vayer2020fused}, which aligns \(\mu_0\) and \(\mu'_0\) by minimizing over couplings

\vspace{2.5mm}

\centerline{$\displaystyle \mathscr L_{\FGW}\left(\mu_0,\mu'_0\;;\; d_\Z, d_{\Z'},\mathcal K\right)
:=
\inf_{\pi\in\Pi(\mu_0,\mu'_0)}
\Big[\mathcal D_{d_{\mathcal Z},d_{\mathcal Z'}}(\pi) + \alpha\,\mathcal A_{\mathcal K}(\pi)\Big]\,,\qquad \alpha\in(0,1)\,.$}

\vspace{2.5mm}

\noindent The distortion functional   

\vspace{2.5mm}

\centerline{$\displaystyle \mathcal D_{d_{\mathcal Z},d_{\mathcal Z'}}(\pi)
:=
\int_{\mathcal Z\times\mathcal Z'}
\int_{\mathcal Z\times\mathcal Z'}
\big|d_{\mathcal Z}(z_1,z_2)-d_{\mathcal Z'}(z'_1,z'_2)\big|
\,d\pi(z_1,z'_1)\,d\pi(z_2,z'_2)\,,$}

\vspace{2.5mm}

\noindent measures the extent to which the coupling \(\pi\) preserves relational geometry, meaning pairs of source units that are close or far under \(d_{\mathcal Z}\) should be matched to pairs of target units that are similarly close or far under \(d_{\mathcal Z'}\). The anchor-alignment functional

\vspace{2.5mm}

\centerline{$\displaystyle \mathcal A_{\mathcal K}(\pi)
:=
\int_{\mathcal Z\times\mathcal Z'} 
\mathcal K\!\left(z,
z'\right)
\,d\pi(z,z')\,,$}

\vspace{2.5mm}

\noindent penalizes couplings that match source and target units with incompatible anchor features, where \(\mathcal K\) is the anchor discrepancy function defined in Assumption~\ref{as:kernel}. Thus, the FGW criterion combines geometric alignment of the full control distributions with direct comparability on the anchor subspace.

The FGW criterion provides the appropriate starting point, but it ultimately learns an optimal coupling, not the cross-site transformation \(f\) itself. To learn \(f\), we use the two induced representations described above: the graph coupling links \(f\) to the coupling alignment, and the pull-back metric links \(f\) to the target geometry. 

First, by the coupling representation, the learned coupling should be compatible with the graph of an admissible transformation. We therefore introduce a graph-consistency functional

\vspace{2.5mm}

\centerline{$\displaystyle \mathcal G_{f,d_{\mathcal Z'}}(\pi):=\int_{\mathcal Z\times \mathcal Z'} d_{\mathcal Z'}\left(f(z),z'\right)\,d\pi(z,z')\,,\qquad f\in\Phi\,,$}

\vspace{2.5mm}

\noindent that encourages the coupling $\pi$ to concentrate near the graph coupling $\pi_f$ defined in \eqref{eq:pi_f}. 

Second, by the metric representation, the target metric \(d_{\mathcal Z'}\) is learned by searching over the set of admissible pull-back metrics $d'_f$ defined in \eqref{eq:pullback1}. Hence, for each \(f\in\Phi\), the target metric is evaluated using \(d'_f\), replacing 

\vspace{2.5mm}

\centerline{$\displaystyle \mathcal D_{d_{\mathcal Z},d_{\mathcal Z'}}(\pi)
\quad\text{and}\quad
\mathcal G_{f,d_{\mathcal Z'}}(\pi)\,,$}

\noindent by

\centerline{$\displaystyle \mathcal D_{d_{\mathcal Z},d'_f\,}(\pi)
\quad\text{and}\quad
\mathcal G_{f,d'_f\,}(\pi)\,.$}

\vspace{2.5mm}

Combining the FGW criterion with the graph-consistency functional and the pull-back metric, we define a \emph{regularized fused Gromov--Wasserstein} (RFGW) criterion
\begin{equation}\label{eq:RFGW_definition} \RFGW\left(\pi,d_{\Z},\mathcal K,f\right)
\;:=\;
\Big[
(1\!-\!\lambda\!-\!\alpha)\,\mathcal D_{d_{\mathcal Z},d'_{f}}(\pi)
+
\alpha\,\mathcal A_{\mathcal K}(\pi) + \lambda\,\mathcal G_{f,d'_{f}}(\pi)
\Big]\,,\end{equation}

\noindent where $f\in\Phi$, $\pi\in\Pi(\mu_0,\mu'_0)$, and  $\alpha,\lambda\in[0,1]$ are tuning parameters with $\alpha+\lambda\leq 1$.  The population optimization problem then becomes

\vspace{-7.5mm}

\begin{equation}\label{eq: (pi,f)=min RFGW}
(\pi^\ast,f^\ast)\;\in\; \operatorname*{arg\,inf}_{\substack{
\pi \in \Pi(\mu_0,\mu_0')\\
f \in \Phi
}}  
    \RFGW(\pi, d_{\Z}, \mathcal K, f)\,,
\end{equation}
\noindent which learns the cross-site transformation \(f\) through the two structures it induces: its graph coupling and its pull-back metric.

\subsubsection{Empirical Estimator}

The population criterion above defines the cross-site transformation in terms of the control distributions $\mu_0$ and $\mu'_0$. In practice, these distributions are not observed directly. Rather, we observe control samples
$Z_0=\{z_{0i}\}_{i=1}^{n_0}\sim\mu_0$ and
$Z_0'=\{z_{0j}'\}_{j=1}^{n_0'}\sim\mu_0'$. We therefore replace the population measures by the empirical measures

\vspace{2.5mm}

\centerline{$\displaystyle {\mh}_0 = \frac{1}{n_0}\sum_{i=1}^{n_0}\delta_{z_{0i}}\,,
\qquad
{\mh}'_0 = \frac{1}{n'_0}\sum_{j=1}^{n'_0}\delta_{z'_{0j}}\,,$}

\vspace{2.5mm}

\noindent where $\delta_z$ denotes the Dirac measure at $z$. For these empirical measures, any coupling $\pi\in\Pi(\widehat{\mu}_0,\widehat{\mu}_0')$ can be represented by a nonnegative matrix of weights, 
$\pi=\{p_{ij}\}_{i\leq n_0}^{j\leq n'_0}\in\mathbb{R}_+^{n_0\times n_0'}$, satisfying the marginal constraints

\vspace{2.5mm}

\centerline{$\displaystyle 
\sum_{j=1}^{n_0'}p_{ij}=\frac{1}{n_0}\quad\text{for each }i=1,\dots,n_0,
\qquad
\sum_{i=1}^{n_0}p_{ij}=\frac{1}{n_0'}\quad\text{for each }j=1,\dots,n_0'\,.$}

\vspace{2.5mm} 

\noindent Thus, for any $\pi\in\Pi(\mh_0,\mh'_0)$, the RFGW criterion in \eqref{eq:RFGW_definition} admits the discrete representation
\begin{equation}\label{eq: discrete RFGW}
\begin{aligned}
    \RFGW\left(\pi, d_\Z,\mathcal K,f\right)
\;=\;
(1\!-\!\lambda\!-\!\alpha&)\;\Big[\;\;\sum_{i,k=1}^{n_0}\sum_{j,m=1}^{n'_0} \Big|d_{\mathcal Z}(z_{0i},z_{0k})-d'_{f}(z'_{0j},z'_{0m})\Big|\,p_{ij}\,p_{km}\;\;\Big]
\\+\;
\alpha&\;\;\Big[\;\;\sum_{i=1}^{n_0}\sum_{j=1}^{n'_0}\mathcal K\!\left(z_{0i},
z'_{0j}\right)\,p_{ij}\;\;\Big]\\+\;\lambda&\;\; \Big[\;\;\sum_{i=1}^{n_0}\sum_{j=1}^{n'_0} d'_{f}\left(f(z_{0i}),z'_{0j}\right)\,p_{ij}\;\;\Big]\,,\qquad p_{ij},p_{km}\in \pi\,.
\end{aligned}
\end{equation}

 \noindent Solving the discrete optimization problem in \eqref{eq: discrete RFGW} yields the empirical estimator \begin{equation}\label{eq: (pi,f)=min RFGW for mh}
(\widehat\pi,\widehat f)\;\in\; \operatorname*{arg\,inf}_{\substack{
\pi \in \Pi(\widehat\mu_0,\widehat\mu_0')\\
f \in \Phi
}}
    \RFGW(\pi, d_{\Z},  \mathcal K, f)\,.
\end{equation}
\noindent The pair $(\widehat{\pi},\widehat{f})$ forms the basis of the second step of the method, in which we construct a synthetic sample from the target treated distribution.

\subsection{Constructing the Synthetic Target Treated Distribution}

The second step uses the estimated cross-site transformation to construct a synthetic sample of the unobserved target treated distribution \(\mu'_1\). Under the identification conditions in Section~\ref{sec:identification}, the target treated distribution is the push-forward of the source treated distribution $\mu_1$, under the same cross-site transformation learned from the control distributions. We therefore use the empirical estimator \((\widehat \pi,\widehat f)\) from the control samples to transport the observed source treated sample, $ 
Z_1=\{z_{1k}\}_{k=1}^{n_1}\sim\mu_1\,,$ into the target space.

For each treated source observation \(z_{1k}\in Z_1\), we define its synthetic target analogue, $z'_{1k}{}^{\mathrm{(synth.)}}\in \mathcal Z'$, as any solution to
\begin{equation}\label{eq: z1k(synth)}
\begin{aligned}
{z'_{1k}}^{\mathrm{(synth.)}}
\in
\operatorname*{arg\,inf}_{z'\in\mathcal Z'}
\Big\{\;\;
(1-\kappa-\gamma&)\;\Big[\;\;
\sum_{i=1}^{n_0}\sum_{j=1}^{n'_0}\;\;
\Big|
d_{\mathcal Z}(z_{0i},z_{1k})
-
d'_{\widehat f}(z'_{0j},z')
\Big|\;\;
\,\widehat p_{ij}\;\;\Big]\\
\;+\; \kappa&\;\;\Big[\;\;\mathcal K \big(z_{1k},z'\big)\;\;\Big]
\\
\;+\;
\gamma&\;\;\Big[\;\;d'_{\widehat f}\big(\widehat f(z_{1k}),z'\big)\;\;\Big]
\;\;\Big\}\,,\quad z_{0i}\in Z_0,z'_{0j}\in Z'_0,\;\widehat p_{ij}\in \widehat\pi,
\end{aligned}
\end{equation}
where \(\kappa, \gamma\in[0,1]\) are tuning parameters with $\kappa + \gamma \leq 1$. 

The three terms in \eqref{eq: z1k(synth)} mirror the three components of the RFGW criterion used to learn \((\widehat\pi,\widehat f)\). The first term preserves relational geometry: it compares the distance profile of the treated source unit \(z_{1k}\) relative to the source controls \(\{z_{0i}\}_{i=1}^{n_0}\) with the distance profile of a candidate target point \(z'\) relative to the target controls \(\{z'_{0j}\}_{j=1}^{n'_0}\), weighted by the learned control coupling \(\widehat\pi\). The second term enforces anchor compatibility through the discrepancy \(\mathcal K(z_{1k},z')\). The third term enforces consistency with the estimated transformation by encouraging \(z'\) to remain close to the direct image \(\widehat f(z_{1k})\) under the learned pull-back metric \(d'_{\widehat f}\). Thus, \({z'_{1k}}^{\mathrm{(synth.)}}\) is chosen to preserve the treated unit's control-relative geometry, maintain anchor-feature comparability, and remain consistent with the estimated cross-site transformation.

Applying \eqref{eq: z1k(synth)} to each treated observation yields the synthetic target treated sample $ 
{Z'_1}^{\mathrm{(synth.)}}
:=
\{{z'_{1k}}^{\mathrm{(synth.)}}\}_{k=1}^{n_1}$, and the resulting empirical distribution

\vspace{2.5mm}

\centerline{$\displaystyle 
\widehat\mu_1'{}^{\mathrm{(synth.)}}
:=
\frac{1}{n_1}
\sum_{k=1}^{n_1}
\delta_{{z'_{1k}}^{\mathrm{(synth.)}}}\,,$}

\vspace{2.5mm} 

\noindent which serves as the estimator of the unobserved target treated distribution \(\mu'_1\). Further implementation details, including the optimization procedure and tuning of \((\alpha,\lambda,\kappa,\gamma)\), are provided in Appendix~B.3 of Supplementary Material.

\subsection{Theoretical Results}

We now establish the asymptotic properties of the proposed procedure. We first analyze the empirical minimizers of the control-alignment problem, $(\widehat \pi, \widehat f)$, and show that their asymptotic limits correspond to a cross-site transformation that transports the data distributions across sites, and the graph coupling induced by it. We then show that the resulting synthetic treated sample ${Z'_1}^{\mathrm{(synth.)}}$ converges to the treated distribution that would be observed at the target site, $\mu'_1$.

The following lemma establishes the asymptotic behavior of the empirical minimizers $(\widehat\pi, \widehat f)$. 

\begin{lemma}\label{lem:control_alignment_consistency}
For each pair \((n_0,n_0')\), let \((\widehat{\pi}_{n_0,n_0'},\widehat{f}_{n_0,n_0'})\) denote an empirical minimizer of \eqref{eq: (pi,f)=min RFGW for mh} based on control samples of sizes \((n_0,n_0')\). Under Assumptions~\textnormal{(\ref{as:map}--\ref{as:trt})} and the regularity conditions on \(\Phi\) stated in Appendix~A.2.2, with probability one, every convergent subsequence of
\((\widehat{\pi}_{n_0,n_0'},\widehat f_{n_0,n_0'})\) converges to an admissible bijection and its induced coupling, 

\vspace{2.5mm}

\centerline{$\displaystyle
\widehat{\pi}_{n_0,n_0'}
{\xrightarrow[n_0,n'_0\to \infty]{}}\,
\pi_g
\;\text{ weakly on }\mathcal Z\times\mathcal Z'\,,\;\,\text{ and}
\quad
\widehat f_{n_0,n_0'}(z) {\xrightarrow[n_0,n'_0\to \infty]{}}\, g(z)
\;\text{ for every }z\in\Z\, ,
$}

\vspace{2.5mm}

\noindent where $g\in\Phi$ satisfies

\vspace{2.5mm}

\centerline{$\displaystyle
g_{\#}\mu_0=\mu_0'\,,\quad g_\#\mu_1=\mu'_1\,,
\qquad\text{and}\qquad
\mathcal K\!\left(z,g(z)\right)=0
\quad \mu_0\text{-almost everywhere}\,,
$}

\vspace{2.5mm}

\noindent and $\pi_g\in\Pi(\mu_0,\mu'_0)$ is the graph coupling induced by $g$.

\end{lemma}

Building on Lemma~\ref{lem:control_alignment_consistency}, we now show that the synthetic treated construction is also consistent.

\begin{theorem}\label{thm:1}
Let $(\widehat{\pi}_{n_0,n_0'},\widehat{f}_{n_0,n_0'})$ denote an empirical minimizer of \eqref{eq: (pi,f)=min RFGW for mh} based on control samples of sizes $(n_0,n_0')$. For each $z\in\mathcal Z$, let $\widehat z'_{n_0,n_0'}(z)$ denote a synthetic target point constructed from $z$ using  $(\widehat\pi_{n_0,n_0'},\widehat f_{n_0,n_0'})$ through the synthesis problem in \eqref{eq: z1k(synth)}. Then, under Assumptions~\textnormal{(\ref{as:map}--\ref{as:overlap})}, and the regularity conditions on $\Phi$ in Appendix~A.2.2, with probability one, along every
convergent subsequence of
\((\widehat\pi_{n_0,n_0'},\widehat f_{n_0,n_0'})\) with limit \((\pi_g,g)\) as
characterized in Lemma~\ref{lem:control_alignment_consistency}, we have

\vspace{2.5mm}

\centerline{$\displaystyle
d'_{g}\!\big(\widehat z'_{n_0,n_0'}(z),\,g(z)\big)
\xrightarrow[n_0,n'_0\to\infty]{} 0
\qquad
\text{for }\mu_1\text{-almost every } z\in\mathcal Z\,.
$}

\vspace{2.5mm}

\noindent Consequently, if $\{z_{1k}\}_{k=1}^{n_1}$ are i.i.d.\ samples from $\mu_1$ and
\(
z_{1k}'{}^{\mathrm{(synth.)}}
:=
\widehat z'_{n_0,n_0'}(z_{1k}),
\)
then the empirical measure $\displaystyle \widehat\mu_1'{}^{\mathrm{(synth.)}}
:=
\frac{1}{n_1}\sum_{k=1}^{n_1}\delta_{z_{1k}'{}^{\mathrm{(synth.)}}}$ satisfies, with probability one,

\vspace{2.5mm}

\centerline{$\displaystyle \widehat\mu_1'{}^{\mathrm{(synth.)}}
\;\xrightarrow[n_0,n'_0,n_1\to\infty]{}\;
\mu_1'
\quad \text{weakly on } \Z'\,.$}

\end{theorem}

Theorem~\ref{thm:1} shows that the proposed synthetic construction is consistent at both the pointwise and distributional levels. In particular, every convergent subsequential limit of the empirical control-side alignment is a transport map $g$ satisfying $g_\#\mu_1=\mu'_1$. Along these subsequences, the synthetic points constructed from the treated source observations converge to their image under the corresponding limit map $g$, making the empirical distribution of the full synthetic treated sample converge weakly to the target-site treated distribution \(\mu_1'\). The proofs of Lemma~\ref{lem:control_alignment_consistency} and Theorem~\ref{thm:1} are given in the Supplementary Material. Also, a sensitivity extension of Theorem~\ref{thm:1} is given in Appendix~A.4, showing that approximate violations of the shared-transformation condition yield a correspondingly bounded Wasserstein error.

\section{Simulations}\label{sec:3}

We evaluate the performance of the proposed method through a series of numerical experiments designed to reflect the cross-site, one-armed target setting in Section~\ref{sec:1}. In each experiment, the competing methods observe samples from the source control, source treated, and target control distributions, $(Z_0,Z_1,Z'_0)$, and use these samples to construct a synthetic target treated sample, ${Z'_1}^{\mathrm{(synth.)}}$. The target treated sample $Z'_1$ is withheld during estimation and used only as an oracle benchmark for evaluation. We assess how well each method recovers the target treated distribution along three dimensions: the marginal distribution of the outcome $Y'_1$, the distribution of the features $X'_1$, and the dependence between the features and outcome within the joint distribution $Z'_1=(X'_1,Y'_1)$. 

\subsection{Design overview}

We consider the source and target sites to be two hospitals and construct a patient-level data-generating process reflecting their differences. The two hospitals record eight \emph{quantile-comparable anchors: }$X_Q=($age, systolic blood pressure, heart rate, lactate, creatinine, GCS, BUN, PaO$_2$/FiO$_2$ ratio$)$, as well as four hospital-specific scores: $X_U=($disease severity, comorbidity, acuity, respiratory status$)$. These scores are constructed from different combinations of patient measurements and, while correlated across hospitals, are not directly or one-to-one comparable. The patient features $(X_Q,X_U)$ jointly determine the outcome $Y$, with treatment effects allowed to vary across patients.

Cross-site heterogeneity is introduced through a  transformation $\f:\mathcal Z\rightarrow \mathcal Z'$. For the anchor coordinates, this transformation changes their distributions while preserving their coordinate-level comparability. The same transformation is applied to the control and treated outcomes in the baseline designs, consistent with the shared transformation structure in Assumption~\ref{as:trt}. For each site, we generate control and treated samples of size $n=1000$. Exact specifications of the features and outcome model and cross-site transformations are provided in Appendix~C.1 of the Supplementary Material.

\subsection{Methods compared}

We evaluate the performance of our proposed method, denoted \model{OTSynth}. As described in Section~\ref{sec:2}, \model{OTSynth} first estimates the cross-site transformation from the observed control distributions by solving the regularized fused Gromov--Wasserstein problem in~\eqref{eq: (pi,f)=min RFGW}. The resulting estimator is then used in~\eqref{eq: z1k(synth)} to transport each source treated observation into the target space and construct ${Z'_1}^{\mathrm{(synth.)}}$. In the simulations, the admissible transformation $\widehat f$ is parameterized by a neural network. Architecture, optimization, and tuning details are provided in Appendix~B of the Supplementary Material.

We compare \model{OTSynth} with several alternative methods. The first method is the two-way fixed-effects regression (\model{TWFE}) that models the outcome as a function of treatment, site, and observed patient characteristics. We include three other methods adapted from related literature: a version of the nearest-neighbor \model{Matching} procedure adapted from \citet{hotz2005predicting}; an extension of Matching that incorporates the Generalized Synthetic Control (\model{GenSynth}) structure following \citet{xu2017generalized}; and finally, the cyclic generative adversarial network (\model{CycleGAN}) of \citet{zhu2017unpaired}, adapted to learn an unpaired transformation between the source and target control distributions. These latter methods are adaptations of techniques originally developed for related but distinct tasks: \cite{hotz2005predicting} assume having source and target sites, but only target-site features are observed; \citep{xu2017generalized} is designed for panel data with observed treated and control units pre- and post-intervention; and \cite{zhu2017unpaired} propose CycleGANs for unpaired image-to-image translation. Because no established benchmark exists for the present problem, we utilize these approaches to construct reasonable benchmarks specifically designed to generate synthetic target treated datasets. Complete specifications and implementation details for all competing methods are provided in Appendix~C.2 of the Supplementary Material.

\subsection{Performance metrics}

We evaluate the synthetic target treated sample along the three dimensions: recovery of the outcome distribution, recovery of the feature distribution, and preservation of the dependence between features and outcome. To make the reported discrepancies comparable across simulation designs, scale-dependent quantities are standardized relative to the oracle target treated distribution.

For the marginal outcome distribution, we report the standardized absolute errors in the mean and standard deviation, 

\vspace{2.5mm}

\centerline{$\displaystyle E_{\mathrm{mean}}
=
\tfrac{\left|\overline Y^{\prime\mathrm{(synth.)}}_1-\overline Y^{\prime\mathrm{(oracle)}}_1\right|}
{\mathrm{sd}(Y^{\prime\mathrm{(oracle)}}_1)}\,,
\text{ and }
E_{\mathrm{sd}}
=
\tfrac{\left|\mathrm{sd}(Y_1^{\prime\mathrm{(synth.)}})-\mathrm{sd}(Y^{\prime\mathrm{(oracle)}}_1)\right|}
{\mathrm{sd}(Y^{\prime\mathrm{(oracle)}}_1)}\,.$}

\vspace{2.5mm}

\noindent We measure recovery of the two tails through

\vspace{2.5mm}

\centerline{$\displaystyle E_{\mathrm{tail}}
=
\frac{1}{2}\left[
\left|\Pr\!\left({Y'_1}^{\mathrm{(synth.)}}\leq q_{.05}\right)-.05\right|
+
\left|\Pr\!\left({Y'_1}^{\mathrm{(synth.)}}\geq q_{.95}\right)-.05\right|
\right]\,,$}

\vspace{2.5mm}

\noindent where $q_{.05}$ and $q_{.95}$ are the corresponding quantiles of $Y^{\prime\mathrm{(oracle)}}_1\!\!$. Also, we report the 1-Wasserstein and Energy distances between ${Y'_1}^{\mathrm{(synth.)}}\!$ and $Y^{\prime\mathrm{(oracle)}}_1\!\!$, normalized by $\mathrm{sd}(Y^{\prime\mathrm{(oracle)}}_1)$.

To evaluate recovery of the multivariate feature distribution, each coordinate of $X$ is standardized using its mean and standard deviation in the oracle target treated sample, $X^{\prime\mathrm{(oracle)}}_1$. We then compute the average coordinate-wise 1-Wasserstein distance and the Energy distance between $X^{\prime\mathrm{(synth.)}}_1$ and $X^{\prime\mathrm{(oracle)}}_1$.  Finally, we assess recovery of the outcome--feature dependence through

\vspace{2.5mm}

\centerline{$\displaystyle E_{\mathrm{corr}}
=
\frac{1}{\sqrt{p}}
\left\|
\operatorname{Corr}\!\left({Y_1}^{\prime\mathrm{(synth.)}},{X_1}^{\prime\mathrm{(synth.)}}\right)
-
\operatorname{Corr}\!\left(Y^{\prime\mathrm{(oracle)}}_1,X^{\prime\mathrm{(oracle)}}_1\right)
\right\|_2\,,$}

\vspace{2.5mm}

\noindent where $p$ is the number of observed features. Thus, smaller values indicate better recovery for all reported metrics. Additional distributional, quantile, tail, feature-specific, and dependence diagnostics are reported in Appendix~C.3 of the Supplementary Material.

\subsection{Baseline Results}

We first consider a baseline design in which patients are assigned randomly across treated and control, and randomly across the source and target sites, so that neither treatment assignment nor site membership depends on patient characteristics. Within this setting, we compare performance under linear and nonlinear forms of cross-site heterogeneity. In the \emph{linear} design, the transformation $\f$ induces affine shifts in the patient features and outcome. The \emph{nonlinear} design introduces nonlinear transformations and interactions in the patient features and outcome. In both cases, the same transformation $\f$ applies to the control and treated populations, so that the shared-transformation assumption remains satisfied. Exact specifications are given in Appendix~C.1 of the Supplementary Material.

\definecolor{Yshade}{RGB}{244,248,244}
\definecolor{Xshade}{RGB}{240,242,248}
\definecolor{Dshade}{RGB}{250,248,238}

\newcommand{\estold}[2]{%
\begingroup
\renewcommand{\arraystretch}{0.75}%
\begin{tabular}[c]{@{}c@{}}
#1\\[-2.5pt]
{\scriptsize\color{black!55}#2}
\end{tabular}%
\endgroup
}

\begin{table}[H]
\centering
\captionsetup{justification=centering}

\caption{Performance in the linear and nonlinear baseline designs.}
\label{tab:dgp_baseline_results}

\setlength{\tabcolsep}{4.5pt}
\renewcommand{\arraystretch}{1.55}

\begin{adjustbox}{max width=\textwidth}
\begin{tabular}{
ll
*{5}{>{\columncolor{Yshade}}c}
*{2}{>{\columncolor{Xshade}}c}
>{\columncolor{Dshade}}c
}
\toprule
&
&
\multicolumn{5}{c}{\cellcolor{Yshade}Outcome \(Y\)}
&
\multicolumn{2}{c}{\cellcolor{Xshade}Features \(X\)}
&
\multicolumn{1}{c}{\cellcolor{Dshade}Dependence}
\\
\cmidrule(lr){3-7}
\cmidrule(lr){8-9}
\cmidrule(lr){10-10}

DGP type
& Model
& \(E_{\mathrm{mean}}\)
& \(E_{\mathrm{sd}}\)
& \(E_{\mathrm{tail}}\)
& Std.\ \(W_1\)
& Std.\ Energy
& Avg.\ \(W_1\)
& Energy
& \(E_{\mathrm{corr}}\)
\\
\midrule


\multirow{5}{*}{Linear}
& TWFE
& \estold{\textbf{0.0460}}{(0.0040)}
& \estold{0.3389}{(0.0092)}
& \estold{0.0319}{(0.0007)}
& \estold{0.2398}{(0.0040)}
& \estold{0.0354}{(0.0015)}
& \estold{\textbf{0.0556}}{(0.0013)}
& \estold{\textbf{0.0096}}{(0.0004)}
& \estold{0.1576}{(0.0029)}
\\

& Matching
& \estold{0.4085}{(0.0121)}
& \estold{0.2066}{(0.0176)}
& \estold{0.0330}{(0.0012)}
& \estold{0.4148}{(0.0117)}
& \estold{0.1013}{(0.0045)}
& \estold{\textbf{0.0556}}{(0.0013)}
& \estold{\textbf{0.0096}}{(0.0004)}
& \estold{\textbf{0.0638}}{(0.0021)}
\\

& GenSynth
& \estold{0.1759}{(0.0081)}
& \estold{0.6729}{(0.0192)}
& \estold{0.1545}{(0.0018)}
& \estold{0.5177}{(0.0106)}
& \estold{0.1366}{(0.0040)}
& \estold{0.2840}{(0.0019)}
& \estold{0.2487}{(0.0042)}
& \estold{0.1577}{(0.0023)}
\\

& CycleGAN
& \estold{0.0999}{(0.0102)}
& \estold{0.1291}{(0.0134)}
& \estold{0.0160}{(0.0015)}
& \estold{0.1218}{(0.0091)}
& \estold{\textbf{0.0090}}{(0.0012)}
& \estold{0.0900}{(0.0016)}
& \estold{0.0176}{(0.0009)}
& \estold{0.0674}{(0.0019)}
\\

& \textbf{OTSynth}
& \estold{0.0708}{(0.0072)}
& \estold{\textbf{0.0807}}{(0.0096)}
& \estold{\textbf{0.0128}}{(0.0010)}
& \estold{\textbf{0.1136}}{(0.0056)}
& \estold{0.0091}{(0.0010)}
& \estold{0.0831}{(0.0017)}
& \estold{0.0173}{(0.0009)}
& \estold{0.0680}{(0.0020)}
\\

\addlinespace[2pt]
\midrule
\addlinespace[2pt]


\multirow{5}{*}{Nonlinear}
& TWFE
& \estold{\textbf{0.0508}}{(0.0062)}
& \estold{0.5051}{(0.0197)}
& \estold{0.0339}{(0.0009)}
& \estold{0.1851}{(0.0048)}
& \estold{0.0215}{(0.0013)}
& \estold{\textbf{0.0554}}{(0.0012)}
& \estold{\textbf{0.0093}}{(0.0004)}
& \estold{0.2270}{(0.0045)}
\\

& Matching
& \estold{0.5570}{(0.0223)}
& \estold{0.6445}{(0.0132)}
& \estold{0.0654}{(0.0028)}
& \estold{0.6720}{(0.0189)}
& \estold{0.5117}{(0.0171)}
& \estold{\textbf{0.0554}}{(0.0012)}
& \estold{\textbf{0.0093}}{(0.0004)}
& \estold{0.1048}{(0.0041)}
\\

& GenSynth
& \estold{0.0779}{(0.0049)}
& \estold{0.4399}{(0.0181)}
& \estold{0.0222}{(0.0008)}
& \estold{0.1431}{(0.0034)}
& \estold{0.0105}{(0.0006)}
& \estold{0.3365}{(0.0014)}
& \estold{0.3859}{(0.0037)}
& \estold{0.1647}{(0.0031)}
\\

& CycleGAN
& \estold{0.0857}{(0.0102)}
& \estold{\textbf{0.2071}}{(0.0210)}
& \estold{0.0547}{(0.0017)}
& \estold{0.2145}{(0.0084)}
& \estold{0.0259}{(0.0017)}
& \estold{0.1136}{(0.0016)}
& \estold{0.0216}{(0.0010)}
& \estold{0.1242}{(0.0038)}
\\

& \textbf{OTSynth}
& \estold{0.0606}{(0.0072)}
& \estold{0.2524}{(0.0228)}
& \estold{\textbf{0.0212}}{(0.0016)}
& \estold{\textbf{0.1151}}{(0.0057)}
& \estold{\textbf{0.0079}}{(0.0009)}
& \estold{0.0923}{(0.0019)}
& \estold{0.0166}{(0.0009)}
& \estold{\textbf{0.0825}}{(0.0042)}
\\

\bottomrule
\end{tabular}
\end{adjustbox}

\vspace{2mm}

\begin{minipage}{0.98\textwidth}
\centering
\footnotesize
Entries report means over $R=50$ Monte Carlo runs, with standard errors in parentheses.
\end{minipage}

\end{table}

\begin{figure}[!t]
\centering
\includegraphics[width=\textwidth]{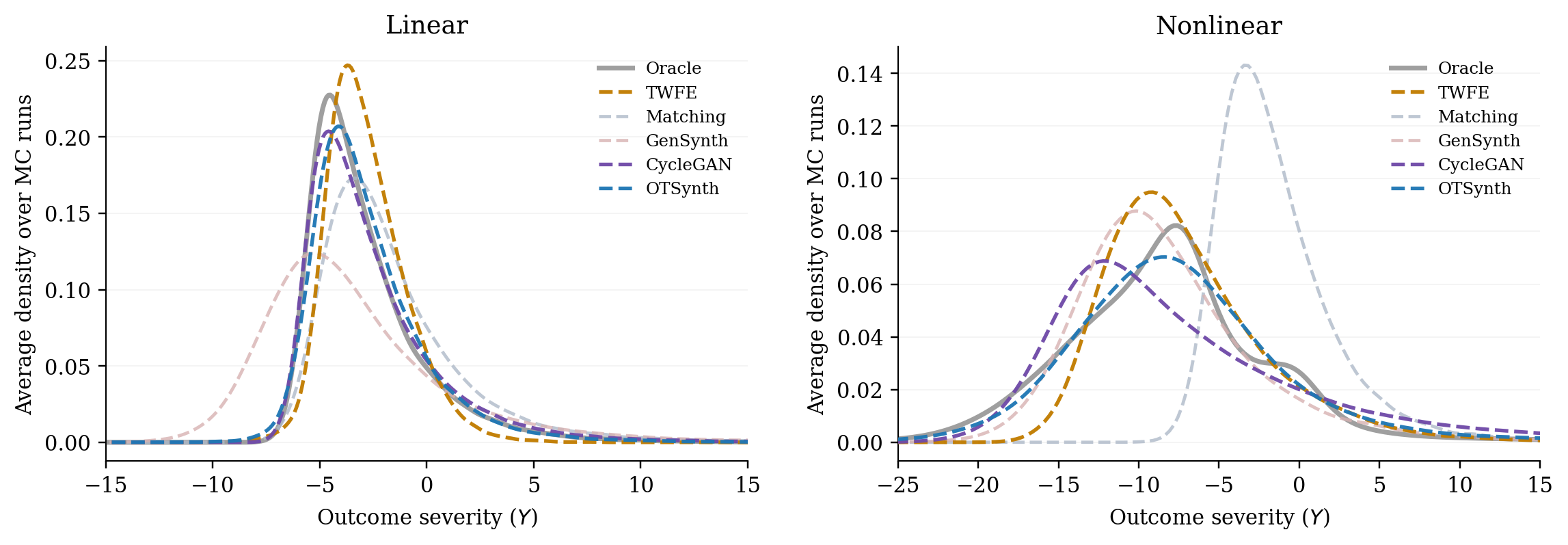}
\caption{Estimated outcome densities under the linear and nonlinear baseline designs.}
\label{fig:y_density}
\end{figure}

Table~\ref{tab:dgp_baseline_results} and Figure~\ref{fig:y_density} summarize performance under the linear and nonlinear baseline designs. Under the linear setting, \model{CycleGAN}, and \model{OTSynth} both recover the target treated distribution reasonably well. \model{OTSynth} performs best on the outcome standard deviation, tail probabilities, and 1-Wasserstein distance and \model{CycleGAN} leads marginally on Energy distance; overall, the differences among the strongest methods are modest. This suggests that when the cross-site transformation is sufficiently simple, the additional geometric structure of \model{OTSynth} offers limited improvement. The particularly low feature-distribution errors of \model{TWFE} and \model{Matching} in both the linear and nonlinear designs are due to the fact that both methods use the target control observations directly, leaving the features unchanged and constructing only counterfactual outcomes. Under randomized treatment assignment, those target control features already follow the same distribution as the oracle target treated features. Nevertheless, \model{OTSynth} remains competitive on the feature metrics.

The advantage of \model{OTSynth} becomes more pronounced under nonlinear cross-site heterogeneity. It achieves substantially smaller 1-Wasserstein and Energy distances for the outcome distribution and feature--outcome correlation error. Notably, \model{CycleGAN}, which is competitive in the linear design and likewise attempts to learn a cross-site transformation, deteriorates on both the outcome distribution and the feature--outcome dependence under nonlinearity, suggesting that the anchor information and the RFGW-based objective of \model{OTSynth} are important for identifying more complex cross-site transformations. Consistent with the results, Figure~\ref{fig:y_density} shows that \model{OTSynth} is the only method that closely reconstructs the overall shape of the oracle outcome distribution. Additional evaluation metrics and simulation results are reported in Appendix~C.3 of the Supplementary Material.

\subsection{Results under Non-Random Treatment and Site Selection}
\label{sec:robustness}

We next examine performance under non-random treatment assignment and site membership. These settings introduce feature differences across treatment groups and sites, while preserving the identifying assumptions of \model{OTSynth}. Results are reported in the first two blocks of Table~\ref{tab:dgp_sensitivity_results}.

\newcommand{\est}[2]{%
\begingroup
\renewcommand{\arraystretch}{0.68}%
\begin{tabular}[c]{@{}c@{}}
{\fontsize{7.5}{7.8}\selectfont #1}\\[0.55pt]
{\fontsize{5.0}{5.4}\selectfont\color{black!55}#2}\\[1.8pt]
\end{tabular}%
\endgroup
}

\newcommand{\designrule}{%
\specialrule{\heavyrulewidth}{2pt}{2pt} 
}

\begin{table}[!p]
\centering
\captionsetup{justification=centering}

\caption{Performance under departures from the baseline design.}
\label{tab:dgp_sensitivity_results}

\fontsize{7.7}{8.2}\selectfont
\setlength{\tabcolsep}{2.7pt}
\renewcommand{\arraystretch}{1.24}

\begin{adjustbox}{
    max width=\textwidth,
    max height=0.88\textheight,
    keepaspectratio
}
\begin{tabular}{
lll
*{5}{>{\columncolor{Yshade}}c}
*{2}{>{\columncolor{Xshade}}c}
>{\columncolor{Dshade}}c
}
\toprule
&
&
&
\multicolumn{5}{c}{\cellcolor{Yshade}Outcome \(Y\)}
&
\multicolumn{2}{c}{\cellcolor{Xshade}Features \(X\)}
&
\multicolumn{1}{c}{\cellcolor{Dshade}Dependence}
\\

\cmidrule(lr){4-8}
\cmidrule(lr){9-10}
\cmidrule(lr){11-11}

Departure
& DGP
& Model
& \(E_{\mathrm{mean}}\)
& \(E_{\mathrm{sd}}\)
& \(E_{\mathrm{tail}}\)
& Std.\ \(W_1\)
& Std.\ Energy
& Avg.\ \(W_1\)
& Energy
& \(E_{\mathrm{corr}}\)
\\
\midrule


\multirow{10}{*}{\shortstack[l]{Observational\\treatment\\assignment}}
& \multirow{5}{*}{Linear}
& TWFE
& \est{0.2272}{(0.0067)}
& \est{0.4509}{(0.0058)}
& \est{0.0314}{(0.0006)}
& \est{0.2820}{(0.0039)}
& \est{0.0445}{(0.0016)}
& \est{0.1731}{(0.0024)}
& \est{0.1228}{(0.0027)}
& \est{0.1725}{(0.0031)}
\\

&& Matching
& \est{0.1563}{(0.0113)}
& \est{\textbf{0.0787}}{(0.0115)}
& \est{\textbf{0.0110}}{(0.0009)}
& \est{0.1750}{(0.0096)}
& \est{0.0238}{(0.0022)}
& \est{0.1731}{(0.0024)}
& \est{0.1228}{(0.0027)}
& \est{\textbf{0.0608}}{(0.0024)}
\\

&& GenSynth
& \est{0.1703}{(0.0082)}
& \est{0.6538}{(0.0151)}
& \est{0.1557}{(0.0016)}
& \est{0.5049}{(0.0072)}
& \est{0.1331}{(0.0029)}
& \est{0.3049}{(0.0022)}
& \est{0.2984}{(0.0048)}
& \est{0.1480}{(0.0021)}
\\

&& CycleGAN
& \est{0.1011}{(0.0093)}
& \est{0.0994}{(0.0117)}
& \est{0.0144}{(0.0011)}
& \est{\textbf{0.1147}}{(0.0085)}
& \est{\textbf{0.0088}}{(0.0013)}
& \est{0.0952}{(0.0019)}
& \est{\textbf{0.0223}}{(0.0011)}
& \est{0.0684}{(0.0022)}
\\

&& \textbf{OTSynth}
& \est{\textbf{0.0837}}{(0.0110)}
& \est{0.1128}{(0.0176)}
& \est{0.0185}{(0.0013)}
& \est{0.1287}{(0.0093)}
& \est{0.0105}{(0.0015)}
& \est{\textbf{0.0940}}{(0.0018)}
& \est{0.0232}{(0.0009)}
& \est{0.0843}{(0.0032)}
\\

\cmidrule(lr){2-11}

& \multirow{5}{*}{Nonlinear}
& TWFE
& \est{0.1456}{(0.0058)}
& \est{0.6262}{(0.0128)}
& \est{0.0401}{(0.0009)}
& \est{0.2196}{(0.0051)}
& \est{0.0252}{(0.0012)}
& \est{0.1745}{(0.0023)}
& \est{0.1188}{(0.0026)}
& \est{0.2435}{(0.0035)}
\\

&& Matching
& \est{0.3517}{(0.0136)}
& \est{0.7137}{(0.0085)}
& \est{0.0347}{(0.0010)}
& \est{0.5154}{(0.0119)}
& \est{0.3343}{(0.0101)}
& \est{0.1745}{(0.0023)}
& \est{0.1188}{(0.0026)}
& \est{0.1077}{(0.0035)}
\\

&& GenSynth
& \est{0.1634}{(0.0065)}
& \est{0.4666}{(0.0140)}
& \est{\textbf{0.0143}}{(0.0012)}
& \est{0.1813}{(0.0045)}
& \est{0.0235}{(0.0014)}
& \est{0.3686}{(0.0014)}
& \est{0.4742}{(0.0045)}
& \est{0.1743}{(0.0034)}
\\

&& CycleGAN
& \est{0.0740}{(0.0100)}
& \est{\textbf{0.2476}}{(0.0206)}
& \est{0.0473}{(0.0024)}
& \est{0.1967}{(0.0094)}
& \est{0.0234}{(0.0023)}
& \est{0.1235}{(0.0017)}
& \est{0.0273}{(0.0011)}
& \est{0.1465}{(0.0038)}
\\

&& \textbf{OTSynth}
& \est{\textbf{0.0638}}{(0.0063)}
& \est{0.2920}{(0.0239)}
& \est{0.0206}{(0.0016)}
& \est{\textbf{0.1232}}{(0.0049)}
& \est{\textbf{0.0076}}{(0.0007)}
& \est{\textbf{0.1024}}{(0.0016)}
& \est{\textbf{0.0210}}{(0.0009)}
& \est{\textbf{0.0956}}{(0.0038)}
\\

\designrule


\multirow{10}{*}{\shortstack[l]{Limited\\cross-site\\overlap}}
& \multirow{5}{*}{Linear}
& TWFE
& \est{\textbf{0.0473}}{(0.0056)}
& \est{0.3804}{(0.0081)}
& \est{0.0365}{(0.0005)}
& \est{0.2870}{(0.0039)}
& \est{0.0546}{(0.0019)}
& \est{\textbf{0.0547}}{(0.0012)}
& \est{\textbf{0.0092}}{(0.0003)}
& \est{0.1804}{(0.0034)}
\\

&& Matching
& \est{0.1941}{(0.0186)}
& \est{0.2151}{(0.0138)}
& \est{0.0359}{(0.0032)}
& \est{0.2433}{(0.0143)}
& \est{0.0435}{(0.0063)}
& \est{\textbf{0.0547}}{(0.0012)}
& \est{\textbf{0.0092}}{(0.0003)}
& \est{0.1124}{(0.0044)}
\\

&& GenSynth
& \est{0.2847}{(0.0184)}
& \est{0.1426}{(0.0136)}
& \est{0.1034}{(0.0028)}
& \est{0.3295}{(0.0113)}
& \est{0.0805}{(0.0053)}
& \est{0.4588}{(0.0025)}
& \est{0.7498}{(0.0073)}
& \est{0.1804}{(0.0031)}
\\

&& CycleGAN
& \est{0.0709}{(0.0088)}
& \est{0.1037}{(0.0105)}
& \est{0.0162}{(0.0012)}
& \est{\textbf{0.1029}}{(0.0074)}
& \est{\textbf{0.0064}}{(0.0009)}
& \est{0.1111}{(0.0016)}
& \est{0.0207}{(0.0008)}
& \est{0.0745}{(0.0024)}
\\

&& \textbf{OTSynth}
& \est{0.0719}{(0.0077)}
& \est{\textbf{0.0985}}{(0.0101)}
& \est{\textbf{0.0153}}{(0.0010)}
& \est{0.1264}{(0.0056)}
& \est{0.0100}{(0.0009)}
& \est{0.1092}{(0.0017)}
& \est{0.0249}{(0.0010)}
& \est{\textbf{0.0701}}{(0.0026)}
\\

\cmidrule(lr){2-11}

& \multirow{5}{*}{Nonlinear}
& TWFE
& \est{\textbf{0.0540}}{(0.0069)}
& \est{0.5764}{(0.0165)}
& \est{0.0373}{(0.0011)}
& \est{0.2192}{(0.0062)}
& \est{0.0297}{(0.0021)}
& \est{\textbf{0.0550}}{(0.0011)}
& \est{\textbf{0.0092}}{(0.0003)}
& \est{0.2609}{(0.0036)}
\\

&& Matching
& \est{0.1974}{(0.0102)}
& \est{0.7956}{(0.0061)}
& \est{0.0470}{(0.0007)}
& \est{0.4327}{(0.0086)}
& \est{0.2334}{(0.0076)}
& \est{\textbf{0.0550}}{(0.0011)}
& \est{\textbf{0.0092}}{(0.0003)}
& \est{0.1707}{(0.0036)}
\\

&& GenSynth
& \est{0.1647}{(0.0106)}
& \est{0.5900}{(0.0229)}
& \est{0.0417}{(0.0042)}
& \est{0.2431}{(0.0145)}
& \est{0.0384}{(0.0059)}
& \est{0.4419}{(0.0021)}
& \est{0.7243}{(0.0064)}
& \est{0.1956}{(0.0061)}
\\

&& CycleGAN
& \est{0.0856}{(0.0111)}
& \est{\textbf{0.2199}}{(0.0209)}
& \est{0.0529}{(0.0035)}
& \est{0.2342}{(0.0117)}
& \est{0.0344}{(0.0032)}
& \est{0.1401}{(0.0016)}
& \est{0.0239}{(0.0009)}
& \est{0.1327}{(0.0030)}
\\

&& \textbf{OTSynth}
& \est{0.0733}{(0.0073)}
& \est{0.2367}{(0.0210)}
& \est{\textbf{0.0157}}{(0.0014)}
& \est{\textbf{0.1186}}{(0.0058)}
& \est{\textbf{0.0095}}{(0.0010)}
& \est{0.1185}{(0.0017)}
& \est{0.0243}{(0.0009)}
& \est{\textbf{0.0828}}{(0.0030)}
\\

\designrule


\multirow{10}{*}{\shortstack[l]{Arm-specific\\cross-site\\transformation}}
& \multirow{5}{*}{Linear}
& TWFE
& \est{0.3976}{(0.0101)}
& \est{0.2305}{(0.0100)}
& \est{\textbf{0.0253}}{(0.0007)}
& \est{0.4771}{(0.0071)}
& \est{0.1997}{(0.0054)}
& \est{\textbf{0.0556}}{(0.0013)}
& \est{\textbf{0.0096}}{(0.0004)}
& \est{0.1784}{(0.0029)}
\\

&& Matching
& \est{0.4085}{(0.0121)}
& \est{0.2066}{(0.0176)}
& \est{0.0330}{(0.0012)}
& \est{\textbf{0.4148}}{(0.0117)}
& \est{0.1013}{(0.0045)}
& \est{\textbf{0.0556}}{(0.0013)}
& \est{\textbf{0.0096}}{(0.0004)}
& \est{\textbf{0.0638}}{(0.0021)}
\\

&& GenSynth
& \est{\textbf{0.0803}}{(0.0080)}
& \est{0.6698}{(0.0189)}
& \est{0.1190}{(0.0020)}
& \est{0.4620}{(0.0106)}
& \est{\textbf{0.0760}}{(0.0028)}
& \est{0.2840}{(0.0019)}
& \est{0.2487}{(0.0042)}
& \est{0.1594}{(0.0024)}
\\

&& CycleGAN
& \est{0.4881}{(0.0142)}
& \est{0.2063}{(0.0164)}
& \est{0.0505}{(0.0015)}
& \est{0.4902}{(0.0140)}
& \est{0.1374}{(0.0061)}
& \est{0.0903}{(0.0016)}
& \est{0.0177}{(0.0010)}
& \est{0.0767}{(0.0018)}
\\

&& \textbf{OTSynth}
& \est{0.4421}{(0.0139)}
& \est{\textbf{0.1184}}{(0.0128)}
& \est{0.0359}{(0.0016)}
& \est{0.4523}{(0.0130)}
& \est{0.1424}{(0.0065)}
& \est{0.0830}{(0.0015)}
& \est{0.0169}{(0.0008)}
& \est{0.0919}{(0.0025)}
\\

\cmidrule(lr){2-11}

& \multirow{5}{*}{Nonlinear}
& TWFE
& \est{0.2919}{(0.0168)}
& \est{\textbf{0.2147}}{(0.0218)}
& \est{0.0884}{(0.0024)}
& \est{0.3577}{(0.0133)}
& \est{0.0884}{(0.0053)}
& \est{\textbf{0.0554}}{(0.0012)}
& \est{\textbf{0.0093}}{(0.0004)}
& \est{0.2530}{(0.0040)}
\\

&& Matching
& \est{0.5570}{(0.0223)}
& \est{0.6445}{(0.0132)}
& \est{0.0654}{(0.0028)}
& \est{0.6720}{(0.0189)}
& \est{0.5117}{(0.0171)}
& \est{\textbf{0.0554}}{(0.0012)}
& \est{\textbf{0.0093}}{(0.0004)}
& \est{\textbf{0.1048}}{(0.0041)}
\\

&& GenSynth
& \est{\textbf{0.1092}}{(0.0069)}
& \est{0.4141}{(0.0190)}
& \est{\textbf{0.0146}}{(0.0010)}
& \est{\textbf{0.1437}}{(0.0042)}
& \est{\textbf{0.0121}}{(0.0011)}
& \est{0.3365}{(0.0014)}
& \est{0.3859}{(0.0037)}
& \est{0.1817}{(0.0033)}
\\

&& CycleGAN
& \est{0.3156}{(0.0217)}
& \est{0.5408}{(0.0496)}
& \est{0.1262}{(0.0031)}
& \est{0.5299}{(0.0232)}
& \est{0.1216}{(0.0070)}
& \est{0.1131}{(0.0016)}
& \est{0.0217}{(0.0010)}
& \est{0.1404}{(0.0038)}
\\

&& \textbf{OTSynth}
& \est{0.3037}{(0.0196)}
& \est{0.4857}{(0.0549)}
& \est{0.1098}{(0.0027)}
& \est{0.4401}{(0.0199)}
& \est{0.1007}{(0.0056)}
& \est{0.0917}{(0.0018)}
& \est{0.0162}{(0.0008)}
& \est{0.1210}{(0.0042)}
\\

\designrule


\multirow{10}{*}{\shortstack[l]{Heterogeneous\\care\\regimes}}
& \multirow{5}{*}{Linear}
& TWFE
& \est{\textbf{0.0462}}{(0.0041)}
& \est{0.3889}{(0.0068)}
& \est{0.0419}{(0.0005)}
& \est{0.3347}{(0.0040)}
& \est{0.0690}{(0.0014)}
& \est{\textbf{0.0556}}{(0.0013)}
& \est{\textbf{0.0096}}{(0.0004)}
& \est{0.1269}{(0.0030)}
\\

&& Matching
& \est{1.1087}{(0.0108)}
& \est{0.2158}{(0.0155)}
& \est{0.0734}{(0.0019)}
& \est{1.1698}{(0.0100)}
& \est{0.9873}{(0.0127)}
& \est{\textbf{0.0556}}{(0.0013)}
& \est{\textbf{0.0096}}{(0.0004)}
& \est{0.2602}{(0.0037)}
\\

&& GenSynth
& \est{0.1892}{(0.0096)}
& \est{0.9620}{(0.0161)}
& \est{0.0949}{(0.0015)}
& \est{0.8718}{(0.0110)}
& \est{0.2821}{(0.0059)}
& \est{0.2840}{(0.0019)}
& \est{0.2487}{(0.0042)}
& \est{0.2951}{(0.0026)}
\\

&& CycleGAN
& \est{0.1655}{(0.0104)}
& \est{0.1443}{(0.0114)}
& \est{0.0643}{(0.0013)}
& \est{0.3939}{(0.0064)}
& \est{0.0724}{(0.0016)}
& \est{0.0928}{(0.0016)}
& \est{0.0185}{(0.0009)}
& \est{0.1103}{(0.0030)}
\\

&& \textbf{OTSynth}
& \est{0.0633}{(0.0059)}
& \est{\textbf{0.0658}}{(0.0064)}
& \est{\textbf{0.0126}}{(0.0009)}
& \est{\textbf{0.1485}}{(0.0031)}
& \est{\textbf{0.0126}}{(0.0005)}
& \est{0.0842}{(0.0018)}
& \est{0.0175}{(0.0009)}
& \est{\textbf{0.0719}}{(0.0027)}
\\

\cmidrule(lr){2-11}

& \multirow{5}{*}{Nonlinear}
& TWFE
& \est{\textbf{0.0422}}{(0.0047)}
& \est{0.5039}{(0.0141)}
& \est{0.0432}{(0.0006)}
& \est{0.2586}{(0.0045)}
& \est{0.0372}{(0.0014)}
& \est{\textbf{0.0554}}{(0.0012)}
& \est{\textbf{0.0093}}{(0.0004)}
& \est{0.1906}{(0.0036)}
\\

&& Matching
& \est{0.4399}{(0.0085)}
& \est{0.6761}{(0.0090)}
& \est{0.0490}{(0.0002)}
& \est{0.5506}{(0.0102)}
& \est{0.2587}{(0.0065)}
& \est{\textbf{0.0554}}{(0.0012)}
& \est{\textbf{0.0093}}{(0.0004)}
& \est{0.2236}{(0.0036)}
\\

&& GenSynth
& \est{0.1666}{(0.0052)}
& \est{0.4585}{(0.0132)}
& \est{0.0383}{(0.0005)}
& \est{0.2569}{(0.0050)}
& \est{0.0368}{(0.0015)}
& \est{0.3365}{(0.0014)}
& \est{0.3859}{(0.0037)}
& \est{0.2838}{(0.0032)}
\\

&& CycleGAN
& \est{0.0992}{(0.0100)}
& \est{\textbf{0.1463}}{(0.0175)}
& \est{0.0304}{(0.0015)}
& \est{0.1965}{(0.0085)}
& \est{0.0241}{(0.0013)}
& \est{0.1141}{(0.0016)}
& \est{0.0213}{(0.0010)}
& \est{0.1336}{(0.0035)}
\\

&& \textbf{OTSynth}
& \est{0.0629}{(0.0069)}
& \est{0.1708}{(0.0204)}
& \est{\textbf{0.0100}}{(0.0008)}
& \est{\textbf{0.1361}}{(0.0048)}
& \est{\textbf{0.0130}}{(0.0009)}
& \est{0.0928}{(0.0017)}
& \est{0.0165}{(0.0008)}
& \est{\textbf{0.0833}}{(0.0030)}
\\

\bottomrule
\end{tabular}
\end{adjustbox}

\vspace{1.5mm}

\begin{minipage}{0.99\textwidth}
\centering
\footnotesize
Entries report means over $R=50$ Monte Carlo runs, with standard errors in parentheses.
\end{minipage}

\end{table}

First, we replace randomized treatment assignment with an observational mechanism in which treatment probabilities are governed by a propensity score based on patient characteristics and disease severity, so that sicker patients are more likely to receive treatment. Consequently, the control and treated populations no longer share the same feature distribution. This removes the feature-distribution advantage of \model{TWFE} and it deteriorates substantially in reconstructing the target treated distribution. \model{CycleGAN} and \model{OTSynth} remain the strongest performers in the linear setting, while \model{OTSynth} is the only consistently accurate performer under nonlinearity.

We next consider limited cross-site overlap by allowing site membership to depend on a propensity score, so that sicker patients are more likely to be observed at the target hospital. Treatment remains randomized within each site, and therefore \model{TWFE} and \model{Matching} retain their advantage on the feature metrics by using the target control features directly. However, the reduced overlap between source and target populations makes outcome transport more difficult, and \model{TWFE} deteriorates on the outcome and feature--outcome dependence metrics. As in the observational design, \model{CycleGAN} and \model{OTSynth} perform well under the linear transformation, whereas only \model{OTSynth} remains consistently accurate under nonlinearity.

\subsection{Sensitivity to Assumption Violations}
\label{sec:sensitivity}

We next examine sensitivity to violations of assumptions underlying \model{OTSynth}. We first consider violation of the shared-transformation assumption, followed by a departure from the regularity condition requiring the cross-site transformation to be continuous. Results are reported in the third and fourth blocks of Table~\ref{tab:dgp_sensitivity_results}.

First, we define two distinct cross-site outcome transformations and allow the control and treated populations to follow different transformations, with treated patients receiving the more favorable target-site care regime. This directly violates the shared-transformation assumption, since the transformation learned from the source and target control samples no longer identifies the transformation governing the treated samples. Correspondingly, this is the setting in which \model{OTSynth} deteriorates substantially. However, none of the competing methods reliably recover the target treated distribution either, indicating that the violation creates a common identification failure.

We next use the same two cross-site outcome transformations and assign patients between the two maps according to a propensity score, thereby constructing heterogeneous target-site care regimes. This represents settings in which the target hospital may provide enhanced care selectively to sicker patients, or conversely reserve limited resources for less severe cases so that sicker patients receive less favorable care. The resulting cross-site transformation is therefore discontinuous at the regime boundary and violates the continuity condition imposed on the transformation class. Despite this, \model{OTSynth} remains the only method that is consistently accurate across both the linear and nonlinear settings, with substantially better reconstruction of the outcome distribution and feature--outcome dependence than the competing approaches.

Overall, the results suggest that \model{OTSynth} deteriorates when the shared-transformation assumption is violated. However, as long as that assumption remains valid, \model{OTSynth} is robust to discontinuous transformations and non-random treatment and site selection, maintaining accuracy where competing methods often degrade. Additional simulation results are reported in Appendix~C.3 of the Supplementary Material.

\section{Real Data Application} \label{sec:4}

To evaluate the practical applicability of our proposed method, we apply it to patient-derived xenograft (PDX) data \citep{gao2015high}, a rich and highly controlled resource for understanding cancer treatment responses. PDX provides a framework where patient tumor samples are implanted into mice, enabling comprehensive screening of multiple treatment options while controlling for genetic and environmental variability. 

The dataset includes five cancer types—breast cancer (BRCA), melanoma (CM), colorectal cancer (CRC), non-small cell lung cancer (NSCLC), and pancreatic cancer (PDAC)—and three genomic feature modalities: gene expression (RNA-seq), copy number variants, and mutation indicators. Given the high dimensionality of the genomic data, we applied sparse principal component analysis \citep{zou2006sparse} separately by modality, retaining 10 RNA, 20 copy-number, and 10 mutation components. For \model{OTSynth}, all 40 components were used as quantile-comparable anchors.

Our primary outcome variable is the log-transformed time to tumor doubling (\emph{TTD}, in days), defined as the time from baseline until tumor volume doubled. \emph{TTD} reflects the treatment efficacy in delaying tumor progression and is recognized as one of the important predictors of survival in oncology \citep{tanaka2004metastatic, miura2019solid}. The scaled tumor shrinkage metric (\emph{BAR}) was also available in the dataset, but was not used. Among the treated groups, we defined observations receiving \emph{BKM120} (Buparlisib), a PI3k inhibitor and well-studied anticancer agent, as the treatment group and untreated observations as the control group. 

We used observations with \emph{BRCA} as the source site and observations with \emph{NSCLC, CM, CRC}, and \emph{PDAC} as separate target sites. For each source--target pair, we fit \model{OTSynth} and the competing methods using the source control, source treated, and target control samples to construct the synthetic target-treated distribution. The observed target-treated sample was held out from model fitting and used only for final oracle evaluation. The untreated/BKM120 sample sizes were BRCA 38/38, CM 32/32, CRC 43/40, NSCLC 25/25, and PDAC 35/34, and there were no missing values.

\begin{table}[!t]
\centering
\captionsetup{justification=centering}

\caption{Comparison of model performance across cancer types.}
\label{tab:realdata_results}

\setlength{\tabcolsep}{5pt}
\renewcommand{\arraystretch}{1.15}

\begin{adjustbox}{max width=0.84\textwidth}
\begin{tabular}{
ll
*{5}{>{\columncolor{Yshade}}c}
}
\toprule
&
&
\multicolumn{5}{c}{\cellcolor{Yshade}Outcome \(Y\)}
\\
\cmidrule(lr){3-7}

Source:Target
& Model
& Mean
& Std.\ dev.
& $(q_{.25},q_{.5},q_{.75})$
& \(W_1\)
& KL div.
\\
\midrule


\multirow{6}{*}{BRCA:NSCLC}
& TWFE
& 3.552
& 0.597
& (3.065, 3.620, 4.065)
& 0.564
& 2.071
\\

& Matching
& 4.350
& 0.441
& (4.038, 4.409, \textbf{4.591})
& 1.077
& 3.310
\\

& GenSynth
& 3.484
& 0.674
& (2.999, 3.255, 3.968)
& 0.491
& 2.434
\\

& CycleGAN
& 2.398
& \textbf{1.239}
& (1.165, 2.215, 3.366)
& 0.935
& 12.075
\\

& \textbf{OTSynth}
& \textbf{3.237}
& 0.629
& (\textbf{2.848}, \textbf{3.122}, 3.620)
& \textbf{0.470}
& \textbf{1.684}
\\

\cmidrule(lr){2-7}

& \textbf{Oracle}
& 3.273
& 1.130
& (2.451, 3.138, 4.465)
& ---
& ---
\\

\designrule


\multirow{6}{*}{BRCA:CM}
& TWFE
& 3.367
& 0.484
& (3.106, 3.297, \textbf{3.626})
& 0.325
& 5.411
\\

& Matching
& 3.951
& 0.494
& (3.604, 3.947, 4.174)
& 0.902
& 10.838
\\

& GenSynth
& 3.431
& 0.544
& (3.117, 3.200, 3.735)
& 0.381
& 4.599
\\

& CycleGAN
& 1.556
& 1.186
& (0.671, 0.882, 2.713)
& 1.494
& 33.559
\\

& \textbf{OTSynth}
& \textbf{2.900}
& \textbf{0.679}
& (\textbf{2.417}, \textbf{2.859}, 3.418)
& \textbf{0.185}
& \textbf{1.997}
\\

\cmidrule(lr){2-7}

& \textbf{Oracle}
& 3.049
& 0.640
& (2.589, 3.042, 3.602)
& ---
& ---
\\

\designrule


\multirow{6}{*}{BRCA:CRC}
& TWFE
& 3.962
& 0.413
& (3.711, 4.002, \textbf{4.201})
& 0.511
& 1.424
\\

& Matching
& 4.137
& 1.427
& (\textbf{3.010}, 4.124, 5.008)
& 0.683
& 7.419
\\

& GenSynth
& 4.163
& 0.719
& (3.717, 3.920, 4.293)
& 0.670
& 5.915
\\

& CycleGAN
& 3.060
& 1.614
& (1.772, 3.046, 3.510)
& 0.873
& 1.660
\\

& \textbf{OTSynth}
& \textbf{3.895}
& \textbf{0.853}
& (3.385, \textbf{3.677}, 4.538)
& \textbf{0.404}
& \textbf{0.586}
\\

\cmidrule(lr){2-7}

& \textbf{Oracle}
& 3.500
& 0.840
& (2.889, 3.485, 4.254)
& ---
& ---
\\

\designrule


\multirow{6}{*}{BRCA:PDAC}
& TWFE
& 3.920
& 0.455
& (3.742, 3.902, 4.260)
& 0.518
& 1.039
\\

& Matching
& 3.785
& 1.006
& (3.228, 3.747, 4.348)
& 0.361
& 4.920
\\

& GenSynth
& 3.978
& \textbf{0.763}
& (3.355, 3.967, 4.178)
& 0.554
& 3.836
\\

& CycleGAN
& 2.367
& 0.692
& (1.874, 2.192, 2.772)
& 1.058
& 3.221
\\

& \textbf{OTSynth}
& \textbf{3.564}
& 0.739
& (\textbf{3.068}, \textbf{3.540}, \textbf{4.104})
& \textbf{0.206}
& \textbf{0.411}
\\

\cmidrule(lr){2-7}

& \textbf{Oracle}
& 3.424
& 0.855
& (2.894, 3.597, 4.012)
& ---
& ---
\\

\bottomrule
\end{tabular}
\end{adjustbox}

\vspace{2mm}

\begin{minipage}{0.98\textwidth}
\centering
\footnotesize
Closer to Oracle is better for mean/sd/quantiles; smaller is better for distance metrics.
\end{minipage}

\end{table}

Table~\ref{tab:realdata_results} reports the descriptive held-out reconstruction of the target-treated TTD distribution across cancer types. \model{OTSynth} achieves the lowest Wasserstein distance across all four targets and generally closely reproduces the Oracle mean and median. These results illustrate \model{OTSynth}'s ability to recover the target-treated outcome distribution in this application.

Table~\ref{tab:realdata_results} compares the competing methods across different cancer type pairs, on their estimated distribution of outcome, \emph{TTD}. \model{OTSynth} demonstrates strong alignment with Oracle distributions. It achieves the lowest Wasserstein distance and KL divergence across all pairs generally closely reproduces the Oracle mean and median, and quantiles. These results highlight \model{OTSynth}'s ability to recover the target-treated outcome distribution in this application.

\begin{figure}
    \centering
    \includegraphics[width=0.95\linewidth]{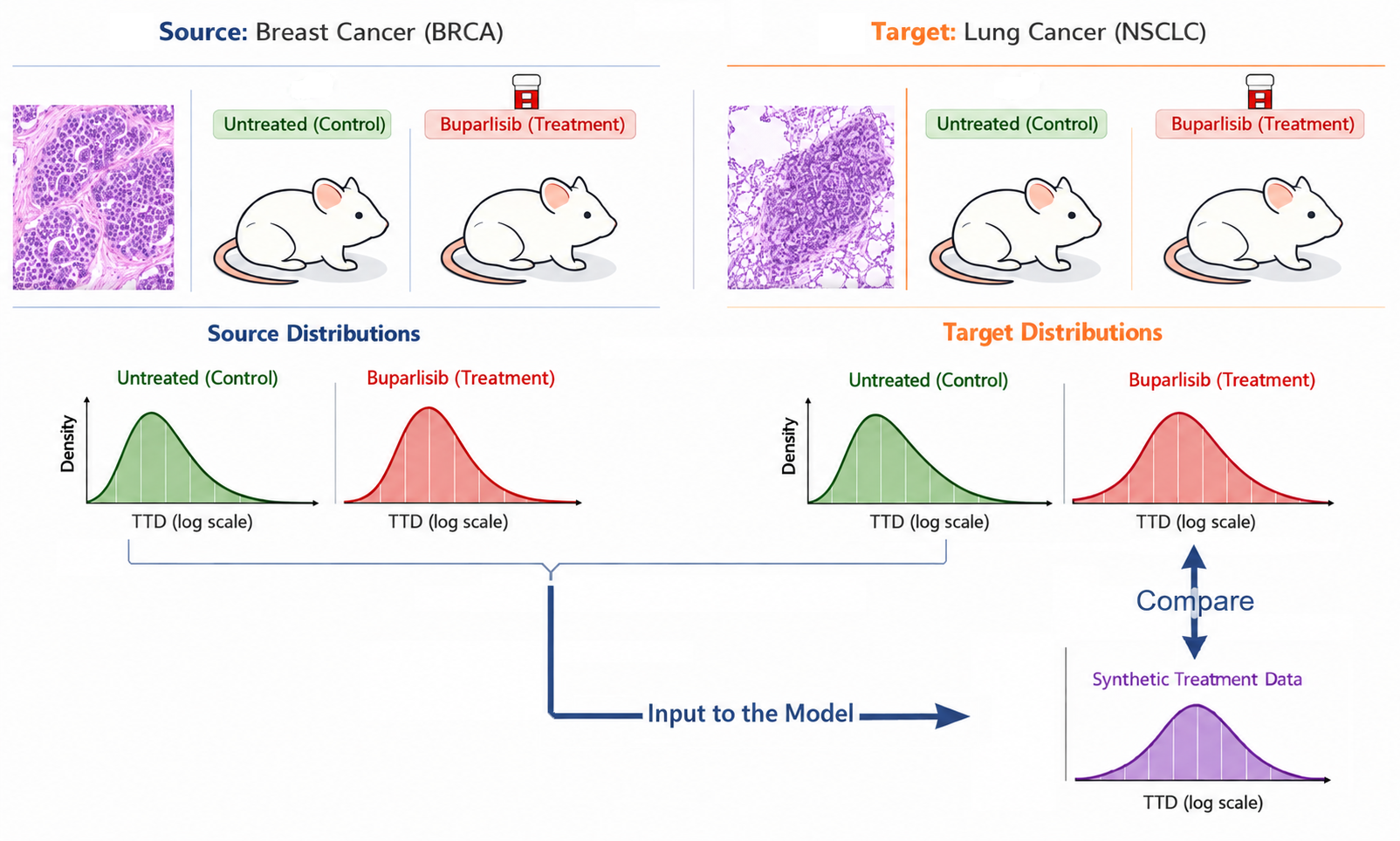}
    \caption{PDX application schematic. Using the source control, source treated, and target control samples, each method constructs a synthetic target-treated distribution, which is compared with the observed target treated distribution as an oracle benchmark.}
    \label{fig:placeholder}
\end{figure}

\section{Conclusion and Discussion} \label{sec:5}

This paper develops a framework for distributional transportability in the cross-site, one-armed target setting, where treated and control data are observed at a source site but only control data are observed at a target site. The approach addresses two limitations of existing transportability methods. First, rather than transporting only an average treatment effect, it aims to estimate the full target treated distribution. Second, rather than relying on conditional invariance or ``unconfounded location'' assumptions that reduce cross-site heterogeneity to differences in observed covariate distributions, it allows site heterogeneity to act on the joint feature--outcome distribution.

The proposed framework represents source and target sites as two feature--outcome spaces that may differ in covariate representation, measurement scale, outcome reporting, population composition, and similarity geometry. Identification is based on a shared transformation that links the source and target control distributions, as well as the source and target treated distributions. This transformation is learned from aligning the observed source and target control samples, through a regularized fused Gromov-Wasserstein criterion, and then applied to the source treated sample to synthesize the unobserved target treated distribution. The theoretical results establish weak convergence of the resulting synthetic treated empirical distribution to the target treated distribution. 

The simulations show that our method is competitive under simple linear cross-site shifts and provides substantially stronger distributional recovery as cross-site heterogeneity becomes nonlinear or the baseline design is made more challenging through observational treatment assignment, limited overlap, and heterogeneous care regimes. Performance deteriorates when the shared-transformation assumption is explicitly violated, highlighting the identifying role of this structure. The patient-derived xenograft application further demonstrates the ability of the method to recover distributional features of the target treated population and improve on competing approaches.

As in other transportability frameworks, the identifying assumptions are not verifiable from the observed data alone. The shared-transformation assumption requires the cross-site distributional shift learned from the control population to remain valid for the treated population. This assumption is most plausible when cross-site heterogeneity reflects environmental, measurement, reporting, or population-composition differences that affect treated and control units similarly. It may fail when treatment implementation differs substantially across sites, for example through dosing schedules, adherence patterns,  or treatment-specific follow-up protocols. In such cases, the observed  shift between the source and target control distributions may not reveal the  shift between the source and target treated distributions, potentially biasing reconstruction of the target treated distribution.

Future work could develop robustness solutions for  violations of the shared-transformation assumption, including partially shared or treatment-specific transformations. Other extensions include alternative anchor constructions, missing or partially observed covariates, and longitudinal or recurrent outcome settings.

\spacingset{1.2}

\bibliographystyle{chicago} 
\bibliography{refs} 

\end{document}


\def\spacingset#1{\renewcommand{\baselinestretch}%
{#1}\small\normalsize} \spacingset{1}

\makeatletter
\g@addto@macro\normalsize{%
  \setlength\abovedisplayskip{10pt plus 2pt minus 5pt}%
  \setlength\belowdisplayskip{10pt plus 2pt minus 5pt}%
  \setlength\abovedisplayshortskip{6pt plus 2pt minus 3pt}%
  \setlength\belowdisplayshortskip{6pt plus 2pt minus 3pt}%
  \setlength\jot{3pt}
}
\makeatother

\if1\blind
{
  \title{Supplementary Material for: \\ Distributional Treatment Effect Transportability across Heterogeneous Sites}
\author{ Borna Bateni\\
Department of Statistics \& Data Science, University of California, \\ Los Angeles, Los Angeles, CA, USA\\
Yubai Yuan\\
Department of Statistics, The Pennsylvania State University, \\ University Park, PA, USA\\
Qi Xu\\
School of Statistics, University of Minnesota, Minneapolis, MN, USA\\
and \\
Annie Qu\thanks{
    Corresponding author. E-mail: \texttt{aqu2@ucsb.edu}. 
    The authors gratefully acknowledge support from the U.S. National Science Foundation (NSF) under Grants DMS-2515275 and DMS-2515698.
  }\\
Department of Statistics \& Applied Probability, University of California, Santa Barbara, Santa Barbara, CA, USA
}

  \date{}
  \maketitle
  \vspace{-10mm}

} \fi

\if0\blind
{
  \bigskip
  \bigskip
  \bigskip
  \title{Supplementary Material for: \\ Distributional Treatment Effect Transportability across Heterogeneous Sites}
  \author{}
  \date{}
  \maketitle
  \medskip
} \fi

\spacingset{1.7}
\appendix
\section{Additional Theoretical Results and Proofs}

\subsection{pull-back metric Lemma and Proof}

We begin with a basic lemma establishing that the pull-back metric induced by a bijective mapping between the source and target spaces is indeed a proper metric.

\begin{lemma}\label{lem:1}
For any bijective mapping $\phi:\Z\to\Z'$ from the source to the target space, the pull-back metric define as: $$ d_\phi'(z_1',z_2') = d_Z(\phi^{-1}(z_1'),\phi^{-1}(z_2'))\foral z_1',z_2'\in\Z'.$$ constitutes a proper metric on $\Z'$.
\end{lemma}

\begin{proof}[Proof of Lemma \ref{lem:1}]
Symmetry and Triangle inequality of $d_\phi'$ follow directly from the corresponding properties of the original metric $d_Z$ and well-definedness of $\phi$. Positivity follows from the bijectivity of $\phi$: if $z'_1\neq z'_2\in\Z'$ then by injectivity of $\phi$, $\phi^{-1}(z'_1)\neq \phi^{-1}(z'_2)$, and by the positivity of the metric $d_Z$, we have $d'_\phi(z'_1,z'_2)=d_Z(\phi^{-1}(z'_1),\phi^{-1}(z'_2))>0$.
\end{proof}

\subsection{Proof of Lemma 3.1}

\subsubsection{Continuity and Boundedness of $d_{\mathcal Z}$ and $\mathcal K$}\label{subsec:cont}

Throughout the theoretical development, we work under the standing regularity condition that the source-space metric

\vspace{2.5mm}

\centerline{$\displaystyle d_{\mathcal Z}:\mathcal Z\times\mathcal Z\to\mathbb R_+\,,$}

\vspace{2.5mm}

\noindent and that the feature-alignment function

\vspace{2.5mm}

\centerline{$\displaystyle \mathcal K:\mathcal Z\times\mathcal Z'\to\mathbb R_+\,,$}

\vspace{2.5mm}

\noindent are continuous and bounded. Continuity of \(\mathcal K\) is already imposed in Assumption~2.2. The continuity of \(d_{\mathcal Z}\) is automatic, since every metric is continuous with respect to the product topology on \(\mathcal Z\times\mathcal Z\). The boundedness requirement is imposed without loss of generality. Indeed, if one starts from an unbounded metric \(d_{\mathcal Z}\), one may replace it by any bounded, continuous, strictly increasing transform of it, for instance

\vspace{2.5mm}

\centerline{$\displaystyle \bar d_{\mathcal Z}(z_1,z_2):=\tanh\!\bigl(d_{\mathcal Z}(z_1,z_2)\bigr),\qquad z_1,z_2\in\mathcal Z\,.$}

\vspace{2.5mm}

\noindent Then \(\bar d_{\mathcal Z}\) is again a metric on \(\mathcal Z\), induces the same topology as \(d_{\mathcal Z}\), and preserves the underlying notion of closeness of points while ensuring boundedness. Likewise, the alignment function \(\mathcal K\) may always be chosen bounded without affecting its role as a penalty quantifying cross-site feature mismatch.

\subsubsection{Regularity of the Admissible Transformation Class}
\label{sec:Phi_regular}

We impose the following regularity conditions on the admissible transformation
class \(\Phi\). In applications, \(\Phi\) is taken to be a parametrized class of neural networks with fixed architecture. Let
\(\Theta\subseteq\mathbb R^m\) denote the corresponding parameter space, where
\(m<\infty\) is the number of trainable parameters, and write

\vspace{2.5mm}

\centerline{$\displaystyle
\Phi=\{f_\theta:\theta\in\Theta\}.
$}

\vspace{2.5mm}

\noindent
Throughout the theoretical analysis, we identify \(\Phi\) with its parameter
space \(\Theta\) through the parametrization \(\theta\mapsto f_\theta\). We assume
that \(\Theta\) is compact and that, for every \(\theta\in\Theta\),
\(f_\theta:\mathcal Z\to\mathcal Z'\) is a bi-measurable bijection. Moreover, we
assume that the forward and inverse maps vary continuously with the parameter:
the maps

\vspace{2.5mm}

\centerline{$\displaystyle
(z,\theta)\longmapsto f_\theta(z),
\qquad
(z',\theta)\longmapsto f_\theta^{-1}(z')
$}

\vspace{2.5mm}

\noindent
are jointly continuous on \(\mathcal Z\times\Theta\) and
\(\mathcal Z'\times\Theta\), respectively.

The compactness of \(\Theta\) is imposed to preclude minimizing sequences from
escaping the admissible class and to guarantee the existence of subsequential
limits of empirical minimizers. This is a theoretical regularity condition, but
it is also compatible with common implementations. Since the neural-network
architecture is fixed, the transformation class is finite-dimensional; imposing
compactness amounts to restricting attention to bounded-weight networks. In
practice, this is naturally induced or approximated by standard procedures such
as weight decay, spectral normalization, gradient clipping, early stopping, or
explicit bounds on the parameter norm.

Under this parametrization, the empirical optimization problem is understood as
an optimization over

\vspace{2.5mm}

\centerline{$\displaystyle
(\pi,\theta)\in
\Pi(\widehat\mu_0,\widehat\mu_0')\times\Theta,
$}

\vspace{2.5mm}

\noindent
where \(\Pi(\widehat\mu_0,\widehat\mu_0')\) is equipped with the topology of weak
convergence on probability measures over \(\mathcal Z\times\mathcal Z'\). The
next lemma records the continuity, and hence measurability, of the RFGW
objective under these regularity conditions.

\begin{lemma}[Continuity of the RFGW objective]
\label{lem:RFGW_continuous}
Suppose \(d_{\mathcal Z}\) and \(\mathcal K\) are bounded and continuous, and
suppose the admissible class \(\Phi=\{f_\theta:\theta\in\Theta\}\) satisfies the
regularity conditions above. Then the map

\vspace{2.5mm}

\centerline{$\displaystyle
(\pi,\theta)
\longmapsto
\RFGW\!\left(\pi,d_{\mathcal Z},\mathcal K,f_\theta\right)
$}

\vspace{2.5mm}

\noindent
is continuous on \(\Pi(\mu_0,\mu_0')\times\Theta\), where
\(\Pi(\mu_0,\mu_0')\) is equipped with the topology of weak convergence. In
particular, it is Borel measurable.
\end{lemma}

\begin{proof}
Write

\vspace{2.5mm}

\centerline{$\displaystyle
F(\pi,\theta)
:=
\RFGW\!\left(\pi,d_{\mathcal Z},\mathcal K,f_\theta\right).
$}

\vspace{2.5mm}

\noindent
It suffices to verify continuity of each component of \(F\).

First, the alignment term is

\vspace{2.5mm}

\centerline{$\displaystyle
\mathcal A_{\mathcal K}(\pi)
=
\int_{\mathcal Z\times\mathcal Z'}
\mathcal K(z,z')\,d\pi(z,z').
$}

\vspace{2.5mm}

\noindent
Since \(\mathcal K\) is bounded and continuous, the map
\(\pi\mapsto\mathcal A_{\mathcal K}(\pi)\) is continuous under weak convergence
of \(\pi\).

Next, consider the graph-consistency term. By definition of the pull-back metric,

\vspace{2.5mm}

\centerline{$\displaystyle
d'_{f_\theta}\bigl(f_\theta(z),z'\bigr)
=
d_{\mathcal Z}\bigl(z,f_\theta^{-1}(z')\bigr).
$}

\vspace{2.5mm}

\noindent
Define

\vspace{2.5mm}

\centerline{$\displaystyle
G(z,z',\theta)
:=
d_{\mathcal Z}\bigl(z,f_\theta^{-1}(z')\bigr).
$}

\vspace{2.5mm}

\noindent
By the joint continuity of \((z',\theta)\mapsto f_\theta^{-1}(z')\) and the
continuity of \(d_{\mathcal Z}\), the function \(G\) is continuous on
\(\mathcal Z\times\mathcal Z'\times\Theta\). Since \(d_{\mathcal Z}\) is bounded,
\(G\) is bounded. Therefore, if \(\pi_n\longrightarrow\pi\) weakly, and
\(\theta_n\to\theta\), then

\vspace{2.5mm}

\centerline{$\displaystyle
\int G(z,z',\theta_n)\,d\pi_n(z,z')
\longrightarrow
\int G(z,z',\theta)\,d\pi(z,z').
$}

\vspace{2.5mm}

\noindent
Thus
\[
(\pi,\theta)\mapsto
\mathcal G_{f_\theta,d'_{f_\theta}}(\pi)
\]
is continuous.

Finally, consider the distortion term. Define

\vspace{2.5mm}

\centerline{$\displaystyle
H(z_1,z_1',z_2,z_2',\theta)
:=
\left|
d_{\mathcal Z}(z_1,z_2)
-
d_{\mathcal Z}\bigl(f_\theta^{-1}(z_1'),f_\theta^{-1}(z_2')\bigr)
\right|.
$}

\vspace{2.5mm}

\noindent
The joint continuity of the inverse parametrization and the continuity of
\(d_{\mathcal Z}\) imply that \(H\) is continuous on
\((\mathcal Z\times\mathcal Z')^2\times\Theta\). Moreover, \(H\) is bounded
because \(d_{\mathcal Z}\) is bounded. If \(\pi_n\longrightarrow\pi\) weakly, then
\(\pi_n\otimes\pi_n\longrightarrow\pi\otimes\pi\) weakly as well. Hence, if also
\(\theta_n\to\theta\),

\vspace{2.5mm}

\centerline{$\displaystyle
\int H(z_1,z_1',z_2,z_2',\theta_n)\,
d(\pi_n\otimes\pi_n)(z_1,z_1',z_2,z_2')
\longrightarrow
\int H(z_1,z_1',z_2,z_2',\theta)\,
d(\pi\otimes\pi)(z_1,z_1',z_2,z_2').
$}

\vspace{2.5mm}

\noindent
Therefore
\[
(\pi,\theta)\mapsto
\mathcal D_{d_{\mathcal Z},d'_{f_\theta}}(\pi)
\]
is continuous.

Combining the three terms,

\[
F(\pi,\theta)
=
(1-\lambda-\alpha)\mathcal D_{d_{\mathcal Z},d'_{f_\theta}}(\pi)
+
\alpha\mathcal A_{\mathcal K}(\pi)
+
\lambda\mathcal G_{f_\theta,d'_{f_\theta}}(\pi),
\]
and since \(\alpha,\lambda>0\) with \(\alpha+\lambda<1\), it follows that
\(F\) is continuous on \(\Pi(\mu_0,\mu_0')\times\Theta\). In particular, \(F\) is
Borel measurable.
\end{proof}

\subsubsection{Measurable Selection of Empirical Minimizers}
\label{subsec:measurable_selection_RFGW}

For each pair \((n_0,n_0')\), the empirical RFGW minimizer is chosen from the argmin correspondence
\[
\mathcal M_{n_0,n_0'}
:=
\operatorname*{arg\,min}_{(\pi,f)\in\Pi(\widehat\mu_0^{(n_0)},\widehat\mu_0'{}^{(n_0')})
\times\Phi}
\RFGW\!\left(\pi,d_{\mathcal Z},\mathcal K,f\right).
\]

Under the compactness assumption on \(\Theta\), the empirical coupling set
\(\Pi(\widehat\mu_0^{(n_0)},\widehat\mu_0'{}^{(n_0')})\) is compact in the weak
topology, and therefore \(\Pi\!\left(\widehat\mu_0^{(n_0)},\widehat\mu_0'{}^{(n_0')}\right)
\times\Phi\) is compact. Moreover, by
Lemma~\ref{lem:RFGW_continuous}, the RFGW objective is continuous in
\((\pi,\theta)\). Hence \(\mathcal M_{n_0,n_0'}\) is nonempty and compact-valued.

Since the empirical measures are measurable functions of the data and the
objective is Borel measurable, the argmin correspondence
\(\mathcal M_{n_0,n_0'}\) is a measurable closed-valued correspondence. By the measurable selection theorem
\citep{kuratowski1965selectors}, it admits a measurable selection. Throughout
the sequel, we fix one such selection and write
\[
(\widehat\pi_{n_0,n_0'},\widehat f_{n_0,n_0'})
\in
\mathcal M_{n_0,n_0'}\,.
\]
Thus \((\widehat\pi_{n_0,n_0'},\widehat f_{n_0,n_0'})\) is a well-defined
measurable empirical minimizer. 

\subsubsection{Population-Level: Characterization of the Zero-Level Set of the RFGW Objective}

We first consider the population-level case, where the distributions $\mu_0$ and $\mu'_0$ are observed in their entirety, and couplings are taken in $\Pi(\mu_0,\mu'_0)$.

The following lemma gives a complete characterization of the RFGW minimizer in the population case.

\begin{lemma}\label{lem:population}
A pair consisting of an admissible bijection $g\in\Phi$ and a coupling $\pi\in\Pi(\mu_0,\mu'_0)$ satisfies

\vspace{2.5mm}

\centerline{$\displaystyle 
\RFGW(\pi,d_{\mathcal Z},\mathcal K,g)=0
$}

\vspace{2.5mm}

\noindent if and only if

\vspace{2.5mm}

\centerline{$\displaystyle g_\#\mu_0=\mu'_0\,,\quad g_\#\mu_1=\mu'_1\,,\qquad \text{and \qquad} \mathcal K(z,g(z))=0\quad \text{for $\mu_0$-almost every}z\in\Z \,,$}

\vspace{2.5mm}

\noindent and the coupling is the deterministic coupling induced by $g$, 

\vspace{2.5mm}

\centerline{$\displaystyle \pi=\pi_g:=(\operatorname{id},g)_\#\mu_0\,.$}

\vspace{2.5mm}

\noindent
    
\end{lemma}

\begin{proof}

\noindent\textbf{I. Sufficiency}

Since \(g_\#\mu_0=\mu_0'\), the deterministic coupling according to $g$

\vspace{2.5mm}

\centerline{$\displaystyle \pi_g:A\times B\mapsto \mu_0(A\cap g^{-1}(B))\,,\qquad \text{for any measurable } A\subseteq \Z,\,B\subseteq \Z'\,,$}

\vspace{2.5mm}

\noindent is a valid coupling of $\mu_0$ and $\mu'_0$. Moreover, by definition of the pull-back metric,
\[
d'_g(g(z_1),g(z_2))=d_{\mathcal Z}(z_1,z_2)
\qquad\text{for all }z_1,z_2\in\mathcal Z,
\]
and therefore

\vspace{2.5mm}

\centerline{$\displaystyle \begin{aligned}
    \mathcal D_{d_{\mathcal Z},d'_g}(\pi_g) &= \int_{\Z\times \Z'}\int_{\Z\times \Z'} \Big|d_Z(z_1,z_2)-d'_g(z'_1,z'_2)\Big|\, d\pi_g(z_1,z'_1)\,d\pi_g(z_2,z'_2)\\
    &= \int_{\Z}\int_{\Z} \Big|d_Z(z_1,z_2)-d'_g(g(z_1),g(z_2))\Big|\, d\mu_0(z_1)\,d\mu_0(z_2)\\
    &=0\,.
\end{aligned}$}

\vspace{2.5mm}

\noindent Likewise, since \(\pi_g\) is supported on the graph of \(g\),
\[
\mathcal G_{g,d'_g}(\pi_g)
=
\int_{\mathcal Z\times\mathcal Z'}
d'_g\!\bigl(g(z),z'\bigr)\,d\pi_g(z,z')
=\int_\Z d'_g(g(z),g(z))\,d\mu_0(z)=0\,.
\]
Finally, we have

\vspace{2.5mm}

\centerline{$\displaystyle \mathcal K\bigl(z,g(z)\bigr)=0
\qquad\text{for $\mu_0$-almost every }z\in\mathcal Z\,,$}

\vspace{2.5mm}

\noindent so that

\vspace{2.5mm}

\centerline{$\displaystyle \mathcal A_{\mathcal K}(\pi_g)=\int_{\Z\times\Z'}\mathcal K(z,z')d\pi_g(z,z')=\int_Z\mathcal K(z,g(z))\,d\mu_0(z)=0\,.$}

\vspace{2.5mm}

\noindent Hence

\vspace{2.5mm}

\centerline{$\displaystyle \RFGW(\pi_g,d_{\mathcal Z},\mathcal K,g)=0\,.$}

\vspace{2.5mm}

\noindent\textbf{II. Necessity}

\noindent Let an arbitrary \(g\in\Phi\) and \(\pi\in\Pi(\mu_0,\mu_0')\) satisfy
\[
\RFGW(\pi,d_{\mathcal Z},\mathcal K,g)=0.
\]
Since each term in the definition of \(\RFGW\) is nonnegative and the weights
\((1-\lambda-\alpha)\), \(\alpha\), and \(\lambda\) are strictly positive, it follows that
\[
\mathcal D_{d_{\mathcal Z},d'_g}(\pi)=0,\qquad
\mathcal A_{\mathcal K}(\pi)=0,\qquad
\mathcal G_{g,d'_g}(\pi)=0.
\]

First, observe that the graph-consistency term

\vspace{2.5mm}

\centerline{$\displaystyle 0=\mathcal G_{g,d'_g}(\pi)
=
\int_{\mathcal Z\times\mathcal Z'}
d'_g\!\bigl(g(z),z'\bigr)\,d\pi(z,z')\,.$}

\vspace{2.5mm}

\noindent Since the integrand is nonnegative, it follows that

\vspace{2.5mm}

\centerline{$\displaystyle 
d'_g\!\bigl(g(z),z'\bigr)=0
\qquad\text{for }\pi\text{-almost every }(z,z')\in \Z\times\Z'\,.
$}

\vspace{2.5mm}

\noindent Equivalently, for every \(\varepsilon>0\), 

\vspace{2.5mm}

\centerline{$\displaystyle \pi\left(\Bigl\{(z,z')\in \Z\times\Z' : d'_g\!\bigl(g(z),z'\bigr)\ge \varepsilon\Bigr\}\right)=0\,.
$}

\vspace{2.5mm}

\noindent Now define the graph of \(g\) by

\vspace{2.5mm}

\centerline{$\displaystyle 
G_g:=\{(z,g(z)):z\in\Z\}\subseteq \Z\times\Z'\,.
$}

\vspace{2.5mm}

\noindent Since \(d'_g\) is a metric on \(\Z'\),

\vspace{2.5mm}

\centerline{$\displaystyle 
d'_g\!\bigl(g(z),z'\bigr)=0
\quad\Longleftrightarrow\quad
z'=g(z)\,,
$}

\vspace{2.5mm}

\noindent and hence

\vspace{2.5mm}

\centerline{$\displaystyle 
(\Z\times\Z')\setminus G_g
=
\Bigl\{(z,z')\in\Z\times\Z' : d'_g\!\bigl(g(z),z'\bigr)>0\Bigr\}
=
\bigcup_{m=1}^\infty
\Bigl\{(z,z') : d'_g\!\bigl(g(z),z'\bigr)\ge \tfrac1m\Bigr\}\,.
$}

\vspace{2.5mm}

\noindent Therefore,

\vspace{2.5mm}

\centerline{$\displaystyle 
\pi\bigl((\Z\times\Z')\setminus G_g\bigr)
\le
\sum_{m=1}^\infty
\pi\Bigl(\{(z,z') : d'_g\!\bigl(g(z),z'\bigr)\ge \tfrac1m\}\Bigr)
=0\,.
$}

\vspace{2.5mm}

\noindent Hence \(\pi\) is supported on \(G_g\).

\vspace{2.5mm}

\noindent Since \(\pi\in\Pi(\mu_0,\mu_0')\) has first marginal \(\mu_0\) and is supported on the graph of \(g\), for every measurable rectangle \(A\times A'\subseteq \Z\times\Z'\) we have

\vspace{2.5mm}

\centerline{$\displaystyle \begin{aligned}
\pi(A\times A')
=
\pi\bigl((A\times A')\cap G_g\bigr)
&=
\pi\bigl(\{(z,g(z)): z\in A\cap g^{-1}(A')\}\bigr)\\
&=\pi((A\cap g^{-1}(A'))\times \Z')
\\&=
\mu_0\bigl(A\cap g^{-1}(A')\bigr)\,.
\end{aligned}
$}

\vspace{2.5mm}

\noindent Thus

\vspace{2.5mm}

\centerline{$\displaystyle 
\pi=:\pi_g\,.
$}

\vspace{2.5mm}

\noindent Since $\pi \in \Pi(\mu_0,\mu'_0)$, we should have

\vspace{2.5mm}

\centerline{$\displaystyle 
g_{\#}\mu_0=\mu'_0\,.
$}

\vspace{2.5mm}

\noindent Also, substituting \(\pi=\pi_g\) into the alignment term yields
\[
0
=
\mathcal A_{\mathcal K}(\pi_g)
=
\int_{\mathcal Z\times\mathcal Z'} \mathcal K(z,z')\,d\pi_g(z,z')
=
\int_{\mathcal Z}\mathcal K\bigl(z,g(z)\bigr)\,d\mu_0(z).
\]
Since \(\mathcal K\ge 0\), this implies
\[
\mathcal K\bigl(z,g(z)\bigr)=0
\qquad \mu_0\text{-almost everywhere on }\mathcal Z.
\]

\noindent Therefore, every population zero-loss pair is of the form \((\pi_g,g)\), where \(g\in\Phi\) satisfies
\[
g_\#\mu_0=\mu_0'
\qquad\text{and}\qquad
\mathcal K\bigl(z,g(z)\bigr)=0
\quad \mu_0\text{-a.e.}\,,
\]

\noindent which, in turn, by Assumption 2.3. implies that 

\vspace{2.5mm}

\centerline{$\displaystyle g_{\#}\mu_1=\mu'_1\,.$}

\vspace{2.5mm}
    
\end{proof}

\subsubsection{Empirical-Level: Weak Convergence of the Empirical Measures}\label{subsec:Omega Star}

Next, we fix i.i.d.\ sequences \(\{z_{0i}\}_{i\ge 1}\sim\mu_0\) and \(\{z_{0j}'\}_{j\ge 1}\sim\mu_0'\), and define

\vspace{2.5mm}

\centerline{$\displaystyle 
\widehat\mu_0^{(n_0)}:=\frac{1}{n_0}\sum_{i=1}^{n_0}\delta_{z_{0i}}\,,\qquad 
\widehat\mu_0'^{(n_0')}:=\frac{1}{n_0'}\sum_{j=1}^{n_0'}\delta_{z'_{0j}}\,.
$}

\vspace{2.5mm}

\noindent Since \(\Z\) and \(\Z'\) are Polish, Varadarajan's theorem \citep[Theorem~11.4.1]{dudley2002real} provides an event \(\Omega^\star\) with \(\mathbb P(\Omega^\star)=1\) on which

\vspace{2.5mm}

\centerline{$\displaystyle 
\widehat\mu_0^{(n_0)}\xrightarrow{\text{weakly}}\mu_0\,,\qquad 
\widehat\mu_0'^{(n_0')}\xrightarrow{\text{weakly}}\mu_0'\,,
\qquad \text{as }n_0,n_0'\to\infty\,.
$}

\vspace{2.5mm}

\noindent We fix an arbitrary \(\omega\in\Omega^\star\) and suppress the dependence on \(\omega\); all subsequent limits are deterministic along this sample path.

\subsubsection{Empirical-Level: Asymptotic Feasibility of the Zero-Level Set}

We first show that the empirical minimizers of RFGW objective asymptotically  approach the zero-level set. 

\begin{lemma}\label{lem:A5}
Let \((\widehat\pi_{n_0,n_0'},\widehat f_{n_0,n_0'})\) belong to the set of empirical minimizers 

\vspace{2.5mm}

\centerline{$\displaystyle
 \operatorname*{arg\,min}_{\substack{
\pi \in \Pi(\widehat\mu_0,\widehat\mu_0')\\
f \in \Phi
}}
    \RFGW(\pi, d_{\Z}, \mathcal K, f)\,,
$}

\vspace{2.5mm}

\noindent based on control samples of sizes \((n_0,n_0')\). Under Assumptions~\textnormal{(2.1--2.2)}, with probability one,

\vspace{2.5mm}

\centerline{$\displaystyle
\RFGW\bigl(\widehat\pi_{n_0,n_0'},d_\Z,\mathcal K,\widehat f_{n_0,n_0'}\bigr)\longrightarrow 0
\qquad\text{as }n_0,n_0'\to\infty\,.
$}
\end{lemma}
\begin{proof}

Recall that by Assumption~2.2, there exists an admissible $\f:\Z\to\Z'$ with $\f_\#\mu_0=\mu'_0$. For each empirical $\widehat\mu_0^{(n_0)}$, define

\vspace{2.5mm}

\centerline{$\displaystyle 
\widehat\nu_0^{(n_0)}:=\f_\#\widehat\mu_0^{(n_0)}\,.
$}

\vspace{2.5mm}

\noindent Observe that \(\widehat\mu_0^{(n_0)}\xrightarrow{\text{weakly}}\mu_0\) and \(\f_\#\mu_0=\mu_0'\). Hence

\vspace{2.5mm}

\centerline{$\displaystyle 
\widehat\nu_0^{(n_0)}\xrightarrow{\text{weakly}}\mu_0'\,.
$}

\vspace{2.5mm}

\noindent Thus, the empirical distributions $\widehat\mu_0'^{(n'_0)}$ and $\widehat\nu_0^{(n_0)}$ are both defined on $\mathcal Z'$ and converge weakly to $\mu'_0$. 

For every measurable set $A\subseteq \Z'$, define the $\varepsilon$-enlargement:

\vspace{2.5mm}

\centerline{$\displaystyle
A^\varepsilon
:=
\{z'\in \mathcal Z':\ d'_\f(z',A)<\varepsilon\}\,,
$}

\vspace{2.5mm}

\noindent and define the \emph{Prokhorov metric} between $\widehat\mu_0'^{(n'_0)}$ and $\widehat\nu_0^{(n_0)}$ as

\vspace{2.5mm}

\centerline{$\displaystyle
\rho({\widehat\mu}_0^{\prime(n'_0)},\widehat\nu_0^{(n_0)})
:=
\inf\left\{\varepsilon>0:\;\begin{cases}
{{\widehat\mu}}_0^{\prime(n'_0)}(A)\le \widehat\nu_0^{(n_0)}(A^\varepsilon)+\varepsilon\;,
\\ \text{ and }\\
\widehat\nu_0^{(n_0)}(A)\le {{\widehat\mu}}_0^{\prime(n'_0)}(A^\varepsilon)+\varepsilon\;;\end{cases}
\ \text{for every measurable }A\subseteq \Z'
\right\}\;.
$}

\vspace{2.5mm}

\noindent The Prokhorov metric metrizes weak convergence on Polish spaces \cite[Theorem~6.8]{billingsley1999convergence}. Since \(\f\in\Phi\), the regularity conditions in
Subsection~\ref{sec:Phi_regular} imply that both \(\f\) and \(\f^{-1}\) are
continuous. Hence \(d'_\f\) generates the topology of \(\mathcal Z'\). Therefore,
the Prokhorov metric defined through \(d'_\f\)-enlargements metrizes weak
convergence on the Polish space \(\mathcal Z'\). Hence, on the Polish space $\Z'$, we have: 

\vspace{2.5mm}

\centerline{$\displaystyle 
\rho\!\left(\widehat\nu_0^{(n_0)},\mu_0'\right)\longrightarrow 0\,,\qquad\text{and}\qquad \rho\!\left(\widehat\mu_0^{\prime(n'_0)},\mu_0'\right)\longrightarrow 0\,.
$}

\vspace{2.5mm}

\noindent Hence

\vspace{2.5mm}

\centerline{$\displaystyle 
\rho\!\left(\widehat\nu_0^{(n_0)},\widehat\mu_0^{\prime(n_0')}\right)\leq \rho\!\left(\widehat\nu_0^{(n_0)},\mu_0'\right)\,+\, \rho\!\left(\widehat\mu_0^{\prime(n'_0)},\mu_0'\right) \longrightarrow 0\,.
$}

\vspace{2.5mm}

\noindent Accordingly, by the Strassen--Dudley theorem \citep[Theorem~11.6.2]{dudley2002real}, there exist a sequence of couplings
\[
\widetilde\varsigma_{n_0,n_0'}\in\Pi\!\bigl(\widehat\nu_0^{(n_0)},\widehat\mu_0'^{(n_0')}\bigr)
\]
such that

\vspace{2.5mm}

\centerline{$\displaystyle 
\widetilde\varsigma_{n_0,n_0'}\!\Bigl(\bigl\{(z_1',z_2')\in\Z'\times\Z' : d'_\f(z_1',z_2')>\rho(\widehat\nu_0^{(n_0)},\widehat\mu_0^{\prime (n_0')})\bigr\}\Bigr)\xrightarrow[n_0,n_0'\to\infty]{}0\,.
$}

\vspace{2.5mm}

\noindent Now define a sequence of couplings on \(\Z\times\Z'\) by pushing \(\widetilde\varsigma_{n_0,n_0'}\) through \(\f^{-1}\) on the first coordinate

\vspace{2.5mm}

\centerline{$\displaystyle 
\widetilde\pi_{n_0,n_0'}
:=
(\f^{-1}\times\mathrm{Id})_\#\widetilde\varsigma_{n_0,n_0'}
\in
\Pi\!\bigl(\widehat\mu_0^{(n_0)},\widehat\mu_0'^{(n_0')}\bigr)\,.
$}

\vspace{2.5mm}

\noindent The marginals of \(\widetilde\pi_{n_0,n_0'}\), \(\widehat\mu_0^{(n_0)}\) and \(\widehat\mu_0'^{(n_0')}\) converge weakly to \(\mu_0\) and \(\mu_0'\). Since weakly convergent sequences of
probability measures on Polish spaces are tight, both marginal sequences are
tight. Hence, for every \(\varepsilon>0\), there exist compact sets
\(K\subseteq\mathcal Z\) and \(K'\subseteq\mathcal Z'\) such that
\[
\widehat\mu_0^{(n_0)}(K)>1-\varepsilon/2,
\qquad
\widehat\mu_0'^{(n_0')}(K')>1-\varepsilon/2
\]
for all sufficiently large \(n_0,n_0'\). Since \(K\times K'\) is compact in
\(\mathcal Z\times\mathcal Z'\), and since \(\widetilde\pi_{n_0,n_0'}\) has
these two marginals,
\[
\widetilde\pi_{n_0,n_0'}(K\times K')
\ge
1-\widehat\mu_0^{(n_0)}(K^c)-\widehat\mu_0'^{(n_0')}(K'^c)
>1-\varepsilon.
\]
Therefore \(\{\widetilde\pi_{n_0,n_0'}\}\) is tight on
\(\mathcal Z\times\mathcal Z'\).

Moreover, because \(\widetilde\varsigma_{n_0,n_0'}\) concentrates asymptotically near the diagonal of \(\Z'\times\Z'\), the pushed-back coupling \(\widetilde\pi_{n_0,n_0'}\) concentrates asymptotically near the graph of \(\f\) in \(\Z\times\Z'\). Hence every weak limit point of \(\widetilde\pi_{n_0,n_0'}\) is supported on

\vspace{2.5mm}

\centerline{$\displaystyle 
G_\f:=\{(z,\f(z)):z\in\Z\}\,,
$}

\vspace{2.5mm}

\noindent and has first marginal \(\mu_0\). Therefore every such limit point must equal the deterministic coupling \(\pi_\f\), and consequently

\vspace{2.5mm}

\centerline{$\displaystyle 
\widetilde\pi_{n_0,n_0'}\xrightarrow{\text{weakly}}\pi_\f\,.
$}

\vspace{2.5mm}

\noindent

\noindent Now, observe that for the feasible pair \((\widetilde\pi_{n_0,n_0'},\f)\) we have

\vspace{2.5mm}

\centerline{$\displaystyle 
d'_\f\bigl(\f(z_1),\f(z_2)\bigr)=d_\Z(z_1,z_2)
\qquad\text{and}\qquad
\mathcal K\bigl(z,\f(z)\bigr)=0\,;
$}

\vspace{2.5mm}

\noindent and \(\widetilde\pi_{n_0,n_0'}\xrightarrow{\text{weakly}}\pi_\f\). Since, for the fixed map \(\f\), the distortion, alignment, and graph-consistency functionals are weakly continuous at \(\pi_\f\), and since each of these terms vanishes at \(\pi_\f\), it follows that

\vspace{2.5mm}

\centerline{$\displaystyle 
\RFGW\bigl(\widetilde\pi_{n_0,n_0'},d_\Z,\mathcal K,\f\bigr)\xrightarrow[n_0,n'_0\to \infty]{} 0\,.
$}

\vspace{2.5mm}

\noindent Since \((\widetilde\pi_{n_0,n_0'},\f)\) is feasible for the empirical optimization of $\operatorname{RFGW}$ objective,   while \((\widehat\pi_{n_0,n_0'},\widehat f_{n_0,n_0'})\) is an empirical minimizer, we obtain

\vspace{2.5mm}

\centerline{$\displaystyle 
0\le
\RFGW\bigl(\widehat\pi_{n_0,n_0'},d_\Z,\mathcal K,\widehat f_{n_0,n_0'}\bigr)
\le
\RFGW\bigl(\widetilde\pi_{n_0,n_0'},d_\Z,\mathcal K,\f\bigr)\xrightarrow[n_0,n_0'\to\infty]{}0\,.
$}
\end{proof}

\subsubsection{Empirical-Level: Identification of Weak Limit Points}

We now identify the weak limit points of the empirical minimizers. The next
lemma shows that any subsequential limit of the empirical minimizers must
 belong to the population zero-level set characterized in
Lemma~\ref{lem:population}.

\begin{lemma}\label{lem:weak_limit_points}
Let $
(\widehat\pi_{n_0,n_0'},\widehat f_{n_0,n_0'})$ 
be empirical minimizers of the RFGW objective based on control samples of sizes
\((n_0,n_0')\). On the event \(\Omega^\star\) defined in
Subsection~\ref{subsec:Omega Star}, every subsequential limit of $
(\widehat\pi_{n_0,n_0'},\widehat f_{n_0,n_0'})$ 
is of the form \((\pi_g,g)\), where \(g\in\Phi\) satisfies

\vspace{2.5mm}

\centerline{$\displaystyle
g_\#\mu_0=\mu_0'\,,
\qquad
\mathcal K(z,g(z))=0
\quad\mu_0\text{-almost everywhere}\,,
\qquad
g_\#\mu_1=\mu_1'\,,
$}

\vspace{2.5mm}

\noindent and $\pi_g$ is the deterministic coupling induced by $g$

\vspace{2.5mm}

\centerline{$\displaystyle
\pi_g:=(\operatorname{id},g)_\#\mu_0\,.
$}

\end{lemma}

\begin{proof}
Fix \(\omega\in\Omega^\star\), and suppress the dependence on \(\omega\). By the parametrization of
\(\Phi\) in Subsection~\ref{sec:Phi_regular}, we may write
\[
\widehat f_{n_0,n_0'}=f_{\widehat\theta_{n_0,n_0'}}
\qquad\text{for some }
\widehat\theta_{n_0,n_0'}\in\Theta.
\]

\noindent Since the marginals of \(\widehat\pi_{n_0,n_0'}\) are
\(\widehat\mu_0^{(n_0)}\) and \(\widehat\mu_0'^{(n_0')}\), and these
empirical measures converge weakly to \(\mu_0\) and \(\mu_0'\),
both marginal sequences are tight. Hence, for every \(\varepsilon>0\), there
exist compact sets \(K\subseteq\mathcal Z\) and
\(K'\subseteq\mathcal Z'\) such that
\[
\widehat\mu_0^{(n_0)}(K)>1-\varepsilon/2,
\qquad
\widehat\mu_0'^{(n_0')}(K')>1-\varepsilon/2
\]
for all sufficiently large \(n_0,n_0'\). Since \(K\times K'\) is compact in
\(\mathcal Z\times\mathcal Z'\), and since
\(\widehat\pi_{n_0,n_0'}\) has these two marginals,
\[
\widehat\pi_{n_0,n_0'}(K\times K')
\ge
1-\widehat\mu_0^{(n_0)}(K^c)
-\widehat\mu_0'^{(n_0')}(K'^c)
>1-\varepsilon.
\]
Therefore \(\{\widehat\pi_{n_0,n_0'}\}\) is tight on
\(\mathcal Z\times\mathcal Z'\), and by Prokhorov's theorem \citep[Theorem~8.6.2]{bogachev2007measureII}, has a weakly
convergent subsequence. Moreover, since \(\Theta\) is compact, the corresponding
parameter sequence has a convergent subsequence. Therefore, passing to a further
subsequence if necessary, there exist a probability measure
\(\pi_\ast\) on \(\mathcal Z\times\mathcal Z'\) and a parameter
\(\theta_\ast\in\Theta\) such that

\vspace{2.5mm}

\centerline{$\displaystyle
\widehat\pi_{n_0,n_0'}
\longrightarrow
\pi_\ast
\quad\text{weakly on }\mathcal Z\times\mathcal Z',
\qquad
\widehat\theta_{n_0,n_0'}\longrightarrow \theta_\ast.
$}

\vspace{2.5mm}

\noindent Define \(g:=f_{\theta_\ast}\). By the continuity of the parametrization
of \(\Phi\),

\vspace{2.5mm}

\centerline{$\displaystyle
\widehat f_{n_0,n_0'}(z)
=
f_{\widehat\theta_{n_0,n_0'}}(z)
\longrightarrow
f_{\theta_\ast}(z)
=
g(z)
\qquad\text{for every }z\in\mathcal Z.
$}

\vspace{2.5mm}

\noindent We first identify the marginals of \(\pi_\ast\). Let
\(\varphi\in C_b(\mathcal Z)\). Since the first marginal of
\(\widehat\pi_{n_0,n_0'}\) is \(\widehat\mu_0^{(n_0)}\), by Portmanteau theorem \citep[Theorem~2.1(v)]{billingsley1999convergence}

\vspace{2.5mm}

\centerline{$\displaystyle
\int_{\mathcal Z\times\mathcal Z'}\varphi(z)\,
d\widehat\pi_{n_0,n_0'}(z,z')
=
\int_{\mathcal Z}\varphi(z)\,
d\widehat\mu_0^{(n_0)}(z)
\longrightarrow
\int_{\mathcal Z}\varphi(z)\,d\mu_0(z).
$}

\vspace{2.5mm}

\noindent Passing to the weak limit on the left-hand side gives that the first
marginal of \(\pi_\ast\) is \(\mu_0\). Similarly, for every
\(\psi\in C_b(\mathcal Z')\),

\vspace{2.5mm}

\centerline{$\displaystyle
\int_{\mathcal Z\times\mathcal Z'}\psi(z')\,
d\widehat\pi_{n_0,n_0'}(z,z')
=
\int_{\mathcal Z'}\psi(z')\,
d\widehat\mu_0'^{(n_0')}(z')
\longrightarrow
\int_{\mathcal Z'}\psi(z')\,d\mu_0'(z'),
$}

\vspace{2.5mm}

\noindent so the second marginal of \(\pi_\ast\) is \(\mu_0'\). Therefore,

\vspace{2.5mm}

\centerline{$\displaystyle
\pi_\ast\in\Pi(\mu_0,\mu_0').
$}

\vspace{2.5mm}

\noindent By Lemma~\ref{lem:A5},

\vspace{2.5mm}

\centerline{$\displaystyle
\RFGW\bigl(
\widehat\pi_{n_0,n_0'},
d_{\mathcal Z},
\mathcal K,
\widehat f_{n_0,n_0'}
\bigr)
\longrightarrow 0.
$}

\vspace{2.5mm}

\noindent Since
\[
\widehat f_{n_0,n_0'}=f_{\widehat\theta_{n_0,n_0'}}
\qquad\text{and}\qquad
g=f_{\theta_\ast},
\]
and since the RFGW objective is continuous in \((\pi,\theta)\) by
Lemma~\ref{lem:RFGW_continuous}, we may pass to the limit along the extracted
subsequence to obtain

\vspace{2.5mm}

\centerline{$\displaystyle
\RFGW\bigl(
\pi_\ast,
d_{\mathcal Z},
\mathcal K,
g
\bigr)
=
\lim_{n_0,n_0'\to\infty}
\RFGW\bigl(
\widehat\pi_{n_0,n_0'},
d_{\mathcal Z},
\mathcal K,
\widehat f_{n_0,n_0'}
\bigr)
=
0.
$}

\vspace{2.5mm}

\noindent Thus \((\pi_\ast,g)\) belongs to the population zero-level set of the
RFGW objective. By Lemma~\ref{lem:population},

\vspace{2.5mm}

\centerline{$\displaystyle
\pi_\ast=\pi_g\,,
\qquad
g_\#\mu_0=\mu_0'\,,
\quad
\mathcal K(z,g(z))=0
\quad\mu_0\text{-almost everywhere}\,,\quad g_\#\mu_1=\mu_1'\,.
$}

\vspace{2.5mm}

\noindent Since the original subsequence was arbitrary, every subsequential limit
of the empirical minimizers is of the claimed form.
\end{proof}

\subsubsection{Conclusion}

We now conclude the proof of Lemma~3.1.
By Lemma~\ref{lem:A5}, the empirical RFGW value converges to
zero almost surely. Moreover, by tightness of the empirical couplings and
compactness of the parameter space \(\Theta\), every subsequence of empirical
minimizers has a further subsequence such that

\vspace{2.5mm}

\centerline{$\displaystyle
\widehat\pi_{n_0,n_0'}
\longrightarrow
\pi_g
\quad\text{weakly on }\mathcal Z\times\mathcal Z',
\qquad
\widehat f_{n_0,n_0'}(z)\longrightarrow g(z)
\quad\text{for every }z\in\mathcal Z,
$}

\vspace{2.5mm}

\noindent for some \(g\in\Phi\). Lemma~\ref{lem:weak_limit_points} identifies
every such limit point as satisfying

\vspace{2.5mm}

\centerline{$\displaystyle
\pi_g=(\operatorname{id},g)_\#\mu_0,
\qquad
g_\#\mu_0=\mu_0',
\qquad
\mathcal K(z,g(z))=0\quad\mu_0\text{-almost everywhere},
\qquad
g_\#\mu_1=\mu_1'.
$}

\vspace{2.5mm}

\noindent Therefore every asymptotic limit point of the empirical
control-alignment problem is a treatment-valid cross-site transport map,
together with its induced deterministic control coupling. This proves
Lemma~3.1.

\subsection{Proof of Theorem~3.1}

\subsubsection{Empirical Setup and Full-Probability Event}\label{emp_setup}

We work on the same probability space carrying the source-control, target-control,
and source-treatment samples. Let
\[
Z_0:=\{z_{0i}\}_{i\ge 1},\qquad
Z_0':=\{z'_{0j}\}_{j\ge 1},\qquad
Z_1:=\{z_{1k}\}_{k\ge 1},
\]
where
\[
z_{0i}\overset{\mathrm{i.i.d.}}{\sim}\mu_0,\qquad
z'_{0j}\overset{\mathrm{i.i.d.}}{\sim}\mu_0',
\qquad
z_{1k}\overset{\mathrm{i.i.d.}}{\sim}\mu_1,
\]
and the three sequences are mutually independent. Define the empirical measures
\[
\widehat\mu_0^{(n_0)}
:=
\frac{1}{n_0}\sum_{i=1}^{n_0}\delta_{z_{0i}},
\qquad
\widehat\mu_0'^{(n_0')}
:=
\frac{1}{n_0'}\sum_{j=1}^{n_0'}\delta_{z'_{0j}},
\qquad
\widehat\mu_1^{(n_1)}
:=
\frac{1}{n_1}\sum_{k=1}^{n_1}\delta_{z_{1k}}.
\]

Since \(\mathcal Z\) and \(\mathcal Z'\) are Polish, Varadarajan's theorem
\citep[Theorem~11.4.1]{dudley2002real} implies that there exists an event
\(\Omega^\star\) with \(\mathbb P(\Omega^\star)=1\) such that, on
\(\Omega^\star\),

\vspace{2.5mm}

\centerline{$\displaystyle
\widehat\mu_0^{(n_0)}
\xrightarrow{\text{weakly}}
\mu_0,
\qquad
\widehat\mu_0'^{(n_0')}
\xrightarrow{\text{weakly}}
\mu_0',
\qquad
\widehat\mu_1^{(n_1)}
\xrightarrow{\text{weakly}}
\mu_1,
$}

\vspace{2.5mm}

\noindent as \(n_0,n_0',n_1\to\infty\). Throughout the proof, we fix an arbitrary \(\omega\in\Omega^\star\) and omit the
dependence on \(\omega\). All limits below are therefore deterministic along this
fixed sample path.

\subsubsection{Measurable Selection of the Synthesis Map}
\label{subsec:measurable_selection_synthesis}

For a generic point \(z\in\mathcal Z\), define the empirical synthesis objective
\(Q_{n_0,n_0'}(\cdot\mid z):\mathcal Z'\to\mathbb R_+\) by

\vspace{2.5mm}

\centerline{$\displaystyle\begin{aligned}
Q_{n_0,n_0'}(z'\mid z)
:=
(1-\kappa-&\gamma)\;\;
\sum_{i=1}^{n_0}\sum_{j=1}^{n_0'}
\Big|
d_{\mathcal Z}(z_{0i},z)
-
d'_{\widehat f_{n_0,n_0'}}(z'_{0j},z')
\Big|
[\widehat p_{n_0,n'_0}]_{ij}\\
+\;&\kappa\,\;\; \mathcal K(z,z')
\\
+\;
&\gamma\,\;\;
d'_{\widehat f_{n_0,n_0'}}
\!\left(\widehat f_{n_0,n_0'}(z),z'\right)\,,\end{aligned}
$}

\vspace{2.5mm}

\noindent
where \(\kappa,\gamma\in(0,1)\) are tuning parameters with \(\kappa+\gamma<1\), and $[\widehat p_{n_0,n'_0}]_{ij}$ is the $(i,j)$th entry of the estimated coupling matrix $\widehat \pi_{n_0,n'_0}$. For each fixed pair \((n_0,n_0')\), the synthesis objective 
\(
(z,z')\longmapsto Q_{n_0,n_0'}(z'\mid z)
\) 
is jointly Borel measurable. Moreover, by the continuity of \(d_{\mathcal Z}\),
the regularity of the admissible class \(\Phi\), and the definition of the
pull-back metric, the map \(z'\mapsto Q_{n_0,n_0'}(z'\mid z)\) is lower
semicontinuous for every fixed \(z\in\mathcal Z\). Hence the synthesis argmin
correspondence
\[
\mathcal S_{n_0,n_0'}(z)
:=
\operatorname*{arg\,min}_{z'\in\mathcal Z'}
Q_{n_0,n_0'}(z'\mid z)
\]
is closed-valued whenever it is nonempty.  Under these conditions and assuming the argmin set is non-empty for
\(\mu_1\)-almost every \(z\in\mathcal Z\), the measurable selection theorem
\citep{kuratowski1965selectors} yields a measurable selection
\[
\widehat z'_{n_0,n_0'}:\mathcal Z\to\mathcal Z',
\qquad
\widehat z'_{n_0,n_0'}(z)\in \mathcal S_{n_0,n_0'}(z).
\]
Throughout the proof, \(\widehat z'_{n_0,n_0'}\) denotes one such measurable
selection:

\vspace{2.5mm}

\centerline{$\displaystyle
\widehat z'_{n_0,n_0'}(z)
\in
\operatorname*{arg\,min}_{z'\in\mathcal Z'}
Q_{n_0,n_0'}(z'\mid z)\,.
$}

\subsubsection{The Synthesis Objective and the Plug-In Comparison Point}

For the observed treated sample \(Z_1=\{z_{1k}\}_{k=1}^{n_1}\), the
synthetic observations are given by

\vspace{2.5mm}

\centerline{$\displaystyle
{z'_{1k}}^{\mathrm{(synth.)}}
=
\widehat z'_{n_0,n_0'}(z_{1k}),
\qquad k=1,\dots,n_1.
$}

\vspace{2.5mm}

\noindent
We also write the empirical graph-consistency term as

\vspace{2.5mm}

\centerline{$\displaystyle
\mathcal G_{\widehat f_{n_0,n_0'},d'_{\widehat f_{n_0,n_0'}}}
(\widehat\pi_{n_0,n_0'})
:=
\sum_{i=1}^{n_0}\sum_{j=1}^{n_0'}
d'_{\widehat f_{n_0,n_0'}}
\!\left(
\widehat f_{n_0,n_0'}(z_{0i}),z'_{0j}
\right)
[\widehat p_{n_0,n'_0}]_{ij}.
$}

\vspace{2.5mm}

\noindent
The following lemma shows that the synthetic point remains close, in the learned
pull-back metric, to the direct plug-in image
\(\widehat f_{n_0,n_0'}(z)\). This is the key finite-sample comparison that
replaces the need to analyze the full epi-limit of the synthesis objective.

\begin{lemma}[Plug-in comparison bound]
\label{lem:plugin_comparison}
For every \(z\in\mathcal Z\),

\vspace{2.5mm}

\centerline{$\displaystyle
d'_{\widehat f_{n_0,n_0'}}
\!\left(
\widehat z'_{n_0,n_0'}(z),
\widehat f_{n_0,n_0'}(z)
\right)
\le
\frac{1-\kappa-\gamma}{\gamma}\,
\mathcal G_{\widehat f_{n_0,n_0'},d'_{\widehat f_{n_0,n_0'}}}
(\widehat\pi_{n_0,n_0'})\;+\;
\frac{\kappa}{\gamma}\,
\mathcal K\!\left(z,\widehat f_{n_0,n'_0}(z)\right)\,.
$}

\vspace{2.5mm}

\noindent Consequently, on the event \(\Omega^\star\), , along any convergent subsequence
\[
(\widehat\pi_{n_0,n'_0},\widehat f_{n_0,n'_0})
\to
(\pi_g,g)
\]
as characterized in Lemma~3.1, we have

\vspace{2.5mm}

\centerline{$\displaystyle
d'_{\widehat f_{n_0,n_0'}}
\!\left(
\widehat z'_{n_0,n_0'}(z),
\widehat f_{n_0,n_0'}(z)
\right)
\longrightarrow 0
\qquad
\text{for \(\mu_1\)-almost every }z\in\mathcal Z\,.
$}

\vspace{2.5mm}

\end{lemma}

\begin{proof}
Fix \(z\in\mathcal Z\). Since \(\widehat z'_{n_0,n_0'}(z)\) minimizes the
empirical synthesis objective, we have

\vspace{2.5mm}

\centerline{$\displaystyle
\begin{aligned}\gamma\,
d'_{\widehat f_{n_0,n_0'}}
\!\left(
\widehat z'_{n_0,n_0'}(z),
\widehat f_{n_0,n_0'}(z)
\right)&\leq Q_{n_0,n_0'}
\!\left(\widehat z'_{n_0,n_0'}(z)\mid z\right)
\\&\le
Q_{n_0,n_0'}
\!\left(\widehat f_{n_0,n_0'}(z)\mid z\right)\\
&=(1-\kappa-\gamma)
\sum_{i=1}^{n_0}\sum_{j=1}^{n_0'}
\Big|
d_{\mathcal Z}(z_{0i},z)
-
d'_{\widehat f_{n_0,n_0'}}
\!\left(z'_{0j},\widehat f_{n_0,n_0'}(z)\right)
\Big|
[\widehat p_{n_0,n'_0}]_{ij}\\
&\quad + \kappa \; \mathcal K\Big(z, \widehat f_{n_0,n_0'}(z)\Big) + \gamma \; \cancelto{0}{d'_{\widehat f_{n_0,n_0'}}
\!\left(\widehat f_{n_0,n_0'}(z),\widehat f_{n_0,n_0'}(z)\right)}
\,.\end{aligned}
$}

\vspace{2.5mm}

\noindent
By the pull-back definition of \(d'_{\widehat f_{n_0,n_0'}}\),

\vspace{2.5mm}

\centerline{$\displaystyle
d'_{\widehat f_{n_0,n_0'}}
\!\left(z'_{0j},\widehat f_{n_0,n_0'}(z)\right)
=
d_{\mathcal Z}
\!\left(
\widehat f_{n_0,n_0'}^{-1}(z'_{0j}),z
\right).
$}

\vspace{2.5mm}

\noindent
Therefore, by the triangle inequality for \(d_{\mathcal Z}\),

\vspace{2.5mm}

\centerline{$\displaystyle
\begin{aligned}
\Big|
d_{\mathcal Z}(z_{0i},z)
-
d'_{\widehat f_{n_0,n_0'}}
\!\left(z'_{0j},\widehat f_{n_0,n_0'}(z)\right)
\Big|
&=
\Big|
d_{\mathcal Z}(z_{0i},z)
-
d_{\mathcal Z}
\!\left(
\widehat f_{n_0,n_0'}^{-1}(z'_{0j}),z
\right)
\Big|
\\
&\le
d_{\mathcal Z}
\!\left(
z_{0i},
\widehat f_{n_0,n_0'}^{-1}(z'_{0j})
\right)
\\
&=
d'_{\widehat f_{n_0,n_0'}}
\!\left(
\widehat f_{n_0,n_0'}(z_{0i}),
z'_{0j}
\right).
\end{aligned}
$}

\vspace{2.5mm}

\noindent
Substituting this bound into the plug-in objective gives

\vspace{2.5mm}

\centerline{$\displaystyle\begin{aligned}
\gamma\,
d'_{\widehat f_{n_0,n_0'}}
\!\left(
\widehat z'_{n_0,n_0'}(z),
\widehat f_{n_0,n_0'}(z)
\right)
&\le
(1-\kappa-\gamma)
\sum_{i=1}^{n_0}\sum_{j=1}^{n_0'}
d'_{\widehat f_{n_0,n_0'}}
\!\left(
\widehat f_{n_0,n_0'}(z_{0i}),
z'_{0j}
\right)
[\widehat p_{n_0,n'_0}]_{ij}+
\kappa\,\mathcal K\!\left(z,\widehat f_{n_0,n'_0}(z)\right)\\
&=(1-\kappa-\gamma)\,
\mathcal G_{\widehat f_{n_0,n_0'},d'_{\widehat f_{n_0,n_0'}}}
\!\!\!(\widehat\pi_{n_0,n_0'})\;\;+\;\;
\kappa\;\mathcal K\!\left(z,\widehat f_{n_0,n'_0}(z)\right)\,.\end{aligned}
$}

\vspace{2.5mm}

\noindent
which proves the comparison bound.

Now consider any convergent subsequence
\[
(\widehat\pi_{n_0,n'_0},\widehat f_{n_0,n'_0})
\to
(\pi_g,g)\,,
\]
as characterized in Lemma~3.1. By Lemma~\ref{lem:A5},

\vspace{2.5mm}

\centerline{$\displaystyle
\RFGW\bigl(
\widehat\pi_{n_0,n_0'},
d_{\mathcal Z},
d'_{\widehat f_{n_0,n_0'}},
\mathcal K,
\widehat f_{n_0,n_0'}
\bigr)
\longrightarrow 0\,,
$}

\vspace{2.5mm}

\noindent
and since the graph-consistency term is nonnegative and enters the RFGW objective
with strictly positive weight \(\lambda>0\), it follows that

\vspace{2.5mm}

\centerline{$\displaystyle
\mathcal G_{\widehat f_{n_0,n_0'},d'_{\widehat f_{n_0,n_0'}}}
(\widehat\pi_{n_0,n_0'})
\longrightarrow 0.
$}

\vspace{2.5mm}

\noindent Moreover, by the convergence \(\widehat f_{n_0,n'_0}(z)\to g(z)\) and the continuity of \(\mathcal K\),
\[
\mathcal K\!\left(z,\widehat f_{n_0,n'_0}(z)\right)
\longrightarrow
\mathcal K(z,g(z))\;=\;0
\qquad \mu_0\text{-almost everywhere}.
\]
Since \(\mu_1\ll\mu_0\) by Assumption, the same equality holds \(\mu_1\)-almost everywhere. Therefore,
\[
\mathcal K\!\left(z,\widehat f_{n_0,n'_0}(z)\right)
\longrightarrow 0
\qquad \mu_1\text{-almost everywhere}.
\]
The comparison bound then implies
\[
d'_{\widehat f_{n_0,n'_0}}
\!\left(
\widehat z'_{n_0,n'_0}(z),
\widehat f_{n_0,n'_0}(z)
\right)
\longrightarrow 0
\qquad
\mu_1\text{-almost everywhere}.
\]

\noindent
This completes the proof.
\end{proof}

\subsubsection{Subsequence Reduction and Convergence of the Synthetic Map}

We now pass to an arbitrary convergent subsequence of the empirical minimizers.
Let
\[
\left\{
(\widehat\pi_{n_{0,k},n'_{0,k}},
 \widehat f_{n_{0,k},n'_{0,k}})
\right\}_{k\ge 1}
\]
be any convergent subsequence of
\(
(\widehat\pi_{n_0,n_0'},\widehat f_{n_0,n_0'})\,.
\)
By Lemma~3.1, there exists an admissible
bijection \(g\in\Phi\) such that, along this subsequence,

\vspace{2.5mm}

\centerline{$\displaystyle
\widehat\pi_{n_{0,k},n'_{0,k}}
\longrightarrow
\pi_g
\quad\text{weakly on }\mathcal Z\times\mathcal Z',
\qquad
\widehat f_{n_{0,k},n'_{0,k}}(z)
\longrightarrow
g(z)
\quad\text{for every }z\in\mathcal Z.
$}

\vspace{2.5mm}

\noindent Moreover, the limiting map \(g\) satisfies

\vspace{2.5mm}

\centerline{$\displaystyle
g_\#\mu_0=\mu_0',
\qquad
g_\#\mu_1=\mu_1',
\qquad
\mathcal K(z,g(z))=0
\quad\mu_0\text{-almost everywhere},
$}

\vspace{2.5mm}

\noindent and the limiting coupling is the deterministic coupling induced by \(g\),

\vspace{2.5mm}

\centerline{$\displaystyle
\pi_g=(\operatorname{id},g)_\#\mu_0.
$}

\vspace{2.5mm}

\noindent For notational simplicity, throughout the next subsections we suppress
the subsequence index \(k\) and write

\vspace{2.5mm}

\centerline{$\displaystyle
(\widehat\pi_{n_0,n_0'},\widehat f_{n_0,n_0'})
\longrightarrow
(\pi_g,g),
$}

\vspace{2.5mm}

\noindent with the understanding that all limits are taken along the fixed
convergent subsequence above. 

By
Lemma~\ref{lem:plugin_comparison}, for $\mu_1$-almost every \(z\in\mathcal Z\),

\vspace{2.5mm}

\centerline{$\displaystyle
d'_{\widehat f_{n_0,n_0'}}
\!\left(
\widehat z'_{n_0,n_0'}(z),
\widehat f_{n_0,n_0'}(z)
\right)
\longrightarrow 0 \;\implies\;
d_{\mathcal Z}
\!\left(
\widehat f_{n_0,n_0'}^{-1}
\!\left(\widehat z'_{n_0,n_0'}(z)\right),
z
\right)
\longrightarrow 0.
$}

\vspace{2.5mm}

\noindent
Thus for the (sub)sequence {$
x_{n_0,n_0'}(z)
:=
\widehat f_{n_0,n_0'}^{-1}
\!\left(\widehat z'_{n_0,n_0'}(z)\right)
$} we have

\vspace{2.5mm}

\centerline{$\displaystyle
x_{n_0,n_0'}(z)\longrightarrow z
\qquad\text{in }\mathcal Z.
$}

\vspace{2.5mm}

\noindent
Moreover, by definition we have $
\widehat z'_{n_0,n_0'}(z)
=
\widehat f_{n_0,n_0'}
\!\left(x_{n_0,n_0'}(z)\right)
$. Along the fixed convergent subsequence, \(\widehat f_{n_0,n_0'}\to g\). By the
continuity of the admissible class \(\Phi\), convergence of
\(\widehat f_{n_0,n_0'}\to g\) together with
\(x_{n_0,n_0'}(z)\to z\) implies

\vspace{2.5mm}

\centerline{$\displaystyle
\widehat f_{n_0,n_0'}\!\left(x_{n_0,n_0'}(z)\right)
\longrightarrow
g(z).
$}

\vspace{2.5mm}

\noindent
Therefore,

\vspace{2.5mm}

\centerline{$\displaystyle
\widehat z'_{n_0,n_0'}(z)
\longrightarrow
g(z).
$}

\vspace{2.5mm}

\noindent
Since \(g^{-1}\) is continuous, it follows that

\vspace{2.5mm}

\centerline{$\displaystyle
g^{-1}\!\left(\widehat z'_{n_0,n_0'}(z)\right)
\longrightarrow
g^{-1}(g(z))=z.
$}

\vspace{2.5mm}

\noindent
Hence, by the definition of \(d'_g\),

\vspace{2.5mm}

\centerline{$\displaystyle
d'_g
\!\left(
\widehat z'_{n_0,n_0'}(z),
g(z)
\right)
=
d_{\mathcal Z}
\!\left(
g^{-1}\!\left(\widehat z'_{n_0,n_0'}(z)\right),
z
\right)
\longrightarrow 0.
$}

\subsubsection{Convergence of the Synthetic Push-Forward Law}

Since \(g\in\Phi\), the regularity conditions imply that \(g\) is a homeomorphism
between \(\mathcal Z\) and \(\mathcal Z'\). Therefore the pull-back metric
\[
d'_g(z_1',z_2')=d_{\mathcal Z}\bigl(g^{-1}(z_1'),g^{-1}(z_2')\bigr)
\]
generates the same topology on \(\mathcal Z'\). Accordingly
\[d'_g
\!\left(
\widehat z'_{n_0,n_0'}(z),
g(z)
\right)\longrightarrow 0\;\;\implies \;\;
\widehat z'_{n_0,n_0'}(z)\longrightarrow g(z)
\qquad\text{for }\mu_1\text{-almost every }z\in\mathcal Z\,,
\]

and for any \(\varphi\in C_b(\mathcal Z')\)
\[
\varphi\!\left(\widehat z'_{n_0,n_0'}(z)\right)
\longrightarrow
\varphi(g(z))
\qquad\text{for }\mu_1\text{-almost every }z\in\mathcal Z.
\]
Moreover,
\[
\left|\varphi\!\left(\widehat z'_{n_0,n_0'}(z)\right)\right|
\le
\|\varphi\|_\infty.
\]
Therefore, by the dominated convergence theorem,

\begin{equation}
    \label{eq: int varphi zhat}
    \int_{\mathcal Z}
\varphi\!\left(\widehat z'_{n_0,n_0'}(z)\right)\,d\mu_1(z)
\longrightarrow
\int_{\mathcal Z}
\varphi(g(z))\,d\mu_1(z).
\end{equation}

\noindent
By Lemma~3.1, the limiting map \(g\)
satisfies \(g_\#\mu_1=\mu_1'\). Hence

\vspace{2.5mm}

\centerline{$\displaystyle
\int_{\mathcal Z}
\varphi(g(z))\,d\mu_1(z)
=
\int_{\mathcal Z'}\varphi(z')\,d(g_\#\mu_1)(z')
=
\int_{\mathcal Z'}\varphi(z')\,d\mu_1'(z').
$}

\vspace{2.5mm}

\noindent
Since \(\varphi\in C_b(\mathcal Z')\) was arbitrary, we conclude that

\vspace{2.5mm}

\centerline{$\displaystyle
(\widehat z'_{n_0,n_0'})_\#\mu_1
\longrightarrow
\mu_1'
\qquad\text{weakly on }\mathcal Z'
$}

\vspace{2.5mm}

\noindent
along the fixed convergent subsequence.

\subsubsection{Empirical Approximation of the Synthetic Treatment Law}

Fix \(\varphi\in C_b(\mathcal Z')\). Recall that
\[
Z_0\overset{\mathrm{i.i.d.}}{\sim}\mu_0,
\qquad
Z_0'\overset{\mathrm{i.i.d.}}{\sim}\mu_0',
\qquad
Z_1\overset{\mathrm{i.i.d.}}{\sim}\mu_1,
\]
and that the treated sample \(Z_1\) is independent of the control samples
\((Z_0,Z_0')\). Conditionally on \(Z_0,Z_0'\), the measurable map
\[
\widehat z'_{n_0,n_0'}:\mathcal Z\to\mathcal Z'
\]
is fixed. Hence the random variables
\[
\varphi\!\left(\widehat z'_{n_0,n_0'}(z_{1k})\right),
\qquad k=1,\dots,n_1,
\]
are conditionally i.i.d., bounded by \(\|\varphi\|_\infty\), and have conditional
mean
\[
\int_{\mathcal Z}
\varphi\!\left(\widehat z'_{n_0,n_0'}(z)\right)\,d\mu_1(z).
\]

Therefore, by Hoeffding's inequality, for every \(\varepsilon>0\),

\vspace{2.5mm}

\centerline{$\displaystyle
\mathbb P\!\left(
\left|
\frac{1}{n_1}\sum_{k=1}^{n_1}
\varphi\!\left(\widehat z'_{n_0,n_0'}(z_{1k})\right)
-
\int_{\mathcal Z}
\varphi\!\left(\widehat z'_{n_0,n_0'}(z)\right)\,d\mu_1(z)
\right|
>\varepsilon
\;\middle|\; Z_0,Z_0'
\right)
\le
2\exp\!\left(
-\frac{n_1\varepsilon^2}{2\|\varphi\|_\infty^2}
\right).
$}

\vspace{2.5mm}

\noindent
Since \(n_1\to\infty\) along the relabeled subsequence under consideration, we
may pass, if necessary, to a further subsequence, again denoted by
\((n_0,n_0',n_1)\), such that

\vspace{2.5mm}

\centerline{$\displaystyle
\sum
\exp\!\left(
-\frac{n_1\varepsilon^2}{2\|\varphi\|_\infty^2}
\right)
<\infty .
$}

\vspace{2.5mm}

\noindent
By Borel--Cantelli, along this further subsequence,

\vspace{2.5mm}

\centerline{$\displaystyle
\frac{1}{n_1}\sum_{k=1}^{n_1}
\varphi\!\left(\widehat z'_{n_0,n_0'}(z_{1k})\right)
-
\int_{\mathcal Z}
\varphi\!\left(\widehat z'_{n_0,n_0'}(z)\right)\,d\mu_1(z)
\longrightarrow 0
\qquad\text{almost surely}.
$}

\vspace{2.5mm}

\noindent
Also, by Equation~\eqref{eq: int varphi zhat}, along the same fixed convergent subsequence,

\vspace{2.5mm}

\centerline{$\displaystyle
\int_{\mathcal Z}
\varphi\!\left(\widehat z'_{n_0,n_0'}(z)\right)\,d\mu_1(z)
\longrightarrow
\int_{\mathcal Z'}\varphi(z')\,d\mu_1'(z').
$}

\vspace{2.5mm}

\noindent
Combining the last two gives

\vspace{2.5mm}

\centerline{$\displaystyle \int_{\mathcal Z'}\varphi(z')\,
d\widehat\mu_1'{}^{\mathrm{(synth.)}}(z')=
\frac{1}{n_1}\sum_{k=1}^{n_1}
\varphi\!\left(\widehat z'_{n_0,n_0'}(z_{1k})\right)
\longrightarrow
\int_{\mathcal Z'}\varphi(z')\,d\mu_1'(z')
\qquad\text{almost surely}
$}

\subsubsection{Conclusion by the Subsequence Principle}

We started with an arbitrary subsequence of triples
\((n_0,n_0',n_1)\). From it, we extracted a further subsequence along which the
empirical minimizers converge to some \((\pi_g,g)\) as characterized in
Lemma~3.1. Along a still further subsequence,
the Borel--Cantelli argument above gives, for each bounded continuous
function \(\varphi\),

\vspace{2.5mm}

\centerline{$\displaystyle
\int_{\mathcal Z'}\varphi(z')\,
d\widehat\mu_1'{}^{\mathrm{(synth.)}}(z')
\longrightarrow
\int_{\mathcal Z'}\varphi(z')\,d\mu_1'(z').
$}

\vspace{2.5mm}

\noindent Hence every subsequence of
\(\widehat\mu_1'{}^{\mathrm{(synth.)}}\) has a further subsequence converging
weakly to \(\mu_1'\). By the subsequence principle,

\vspace{2.5mm}

\centerline{$\displaystyle
\widehat\mu_1'{}^{\mathrm{(synth.)}}
\longrightarrow
\mu_1'
\qquad\text{weakly on }\mathcal Z'.
$}

\vspace{2.5mm}

\noindent This proves Theorem~3.1.

\subsection{Sensitivity Theorem for the Shared Transformation Assumption}
\label{subsec:sensitivity_theorem_shared_transformation}

Theorem~3.1 establishes consistency under the exact shared-transformation condition, under which every admissible map \(f:\Z\to\Z'\) that satisfies the control-side transport, 

\vspace{2.5mm}

\centerline{$\displaystyle f_\#\mu_0=\mu'_0\,,$}

\vspace{2.5mm}

 \noindent and is compatible with the anchor alignment, 
 
 \vspace{2.5mm}

\centerline{$\displaystyle \mathcal K(z,f(z))=0\qquad\mu_0\,\text{-almost everywhere}\,,$}

\vspace{2.5mm}
 
 \noindent also transports the source treated distribution to the target treated distribution, 
 
 \vspace{2.5mm}

\centerline{$\displaystyle f_\#\mu_1=\mu'_1\,.$}

\vspace{2.5mm}
 
\noindent This condition is the central identification bridge in the cross-site, one-armed target setting: since \(\mu'_1\) is unobserved, the treated-side validity of the transformation learned from control distributions cannot be verified from the available data alone. We now record a sensitivity version of Theorem~3.1 for settings in which the shared-transformation condition holds only approximately. Specifically, we replace the exact constraint \(f_\#\mu_1=\mu'_1\) by the requirement that \(f_\#\mu_1\) belong to a neighborhood of \(\mu'_1\).

For an admissible limit map \(f\in\Phi\), let \(\operatorname{W}_f\) denote the Wasserstein distance induced by the pull-back metric \(d'_f\):
\[
\operatorname{W}_f(\nu,\nu')
:=
\inf_{\pi\in\Pi(\nu,\nu')}
\int_{\Z'\times\Z'}
d'_f(z'_1,z'_2)\,
d\pi(z'_1,z'_2)\,.
\]
The following result shows that if $f_\#\mu_1$ lies in a $\operatorname{W}_f$-neighborhood around $\mu'_1$, then the synthetic treated distribution is asymptotically within the same $\operatorname{W}_f$-neighborhood.

\begin{theorem}
\label{thm:sensitivity_shared_transformation}

Suppose the regularity conditions used in Theorem~3.1 hold. Let
\(
(\widehat\pi_{n_0,n'_0},\widehat f_{n_0,n'_0})
\)
be a sequence of empirical minimizers of the control-alignment problem, and consider any subsequence along which
\[
(\widehat\pi_{n_0,n'_0},\widehat f_{n_0,n'_0})
\xrightarrow[n_0,n'_0\to \infty]{}
(\pi_g,g),
\]
where $g:\Z\to\Z'$ is an admissible map that satisfies
\[
g_\#\mu_0=\mu'_0,
\qquad
\mathcal K(z,g(z))=0
\quad \mu_0\text{-almost everywhere}\,;
\]
however, the shared transformation condition of Assumption 2.3 holds only approximately, in the sense that for some \(\varepsilon> 0\),
\[
\operatorname{W}_g\bigl(g_\#\mu_1,\mu'_1\bigr)\le \varepsilon\,.
\]
Then, for the synthetic treated empirical distribution
\(
\widehat\mu_1'{}^{\mathrm{(synth.)}}\!\!
:=\!\!
\frac{1}{n_1}\sum_{k=1}^{n_1}
\delta_{\widehat z'_{n_0,n'_0}(z_{1k})}
\)
 generated by the proposed method we have, along the same subsequence,
\[
\operatorname W_g\!\left(
\widehat\mu_1'{}^{\mathrm{(synth.)}},
\mu'_1
\right)
\le
\varepsilon+o_{\mathbb P}(1).
\]

\end{theorem}

\subsubsection{Proof of Theorem~\ref{thm:sensitivity_shared_transformation}}

By Lemma~\ref{lem:1}, \(d'_g\) is a metric on \(\Z'\). Since \(d_{\Z}\) is bounded (see Appendix~\ref{subsec:cont}), \(d'_g\) is bounded as well, and therefore every probability measure on \(\Z'\) has finite first moment with respect to \(d'_g\). Hence \(\operatorname W_g\) is the usual Wasserstein distance induced by the metric \(d'_g\), and in particular satisfies the triangle inequality. The triangle inequality yields

\vspace{2.5mm}

\centerline{$\displaystyle \operatorname{W}_g\bigl(\widehat\mu_1'{}^{\mathrm{(synth.)}}, \mu'_1\bigr)\leq \operatorname{W}_g\bigl(\widehat\mu_1'{}^{\mathrm{(synth.)}},g_\#\mu_1\bigr)+\operatorname{W}_g\bigl(g_\#\mu_1,\mu'_1\bigr)\,.$}

\vspace{2.5mm}

\noindent It therefore remains to show that
\[
\operatorname{W}_g\bigl(
\widehat\mu_1'{}^{\mathrm{(synth.)}},
g_\#\mu_1
\bigr)
=
o_{\mathbb P}(1).
\]
We work on the full-probability event \(\Omega^\star\) defined in Appendix~\ref{emp_setup}, on which the source-control, target-control, and source-treated empirical measures converge weakly to their population limits. Fix an arbitrary \(\omega\in\Omega^\star\) and suppress the dependence on \(\omega\). All limits below are therefore deterministic along this fixed sample path. Along the subsequence under consideration,
\[
(\widehat\pi_{n_0,n'_0},\widehat f_{n_0,n'_0})
\longrightarrow
(\pi_g,g)\,,
\]
where $g:\Z\to\Z'$ satisfies
\[
g_\#\mu_0=\mu'_0,
\qquad
\mathcal K(z,g(z))=0
\quad \mu_0\text{-almost everywhere}\,.
\]
Lemma~\ref{lem:plugin_comparison}, applied along the subsequence in the theorem statement yields
\[\begin{aligned}
d'_{\widehat f_{n_0,n'_0}}\!\left(
\widehat z'_{n_0,n'_0}(z),
\widehat f_{n_0,n'_0}(z)
\right)
\;=\;
d_{\Z}\!\left(
\widehat f_{n_0,n'_0}^{-1}
\bigl(\widehat z'_{n_0,n'_0}(z)\bigr),
z
\right)
\longrightarrow 0
\qquad
\mu_1\text{-almost everywhere}\;,\\
\implies \widehat f_{n_0,n'_0}^{-1}
\bigl(\widehat z'_{n_0,n'_0}(z)\bigr)\longrightarrow
z\qquad
\mu_1\text{-almost everywhere}\,.\end{aligned}
\]
By the regularity conditions on the admissible transformation class imposed in Appendix~\ref{sec:Phi_regular}, we may write
\[
\widehat f_{n_0,n'_0}=f_{\widehat\theta_{n_0,n'_0}},
\qquad
g=f_{\theta_g},
\qquad
\widehat\theta_{n_0,n'_0}\to\theta_g,
\]
for parameters \(\widehat\theta_{n_0,n'_0},\theta_g\in\Theta\). The same regularity conditions impose joint continuity of the parametrization map
\[
(u,\theta)\longmapsto f_\theta(u).
\]
Thus we have, for \(\mu_1\)-almost every \(z\),
\[
\left(
\widehat f_{n_0,n'_0}^{-1}
\bigl(\widehat z'_{n_0,n'_0}(z)\bigr),
\widehat\theta_{n_0,n'_0}
\right)
\longrightarrow
(z,\theta_g).
\]
Therefore, by joint continuity,
\[ \begin{aligned}
\widehat z'_{n_0,n'_0}(z)=\widehat f_{n_0,n'_0}
\!\Big(
\widehat f_{n_0,n'_0}^{-1}&
\bigl(\widehat z'_{n_0,n'_0}(z)\bigr)
\Big)=
f_{\widehat\theta_{n_0,n'_0}}
\!\left(
\widehat f_{n_0,n'_0}^{-1}
\bigl(\widehat z'_{n_0,n'_0}(z)\bigr)
\right)
\longrightarrow
f_{\theta_g}(z)
=
g(z)
\\ &\implies
\widehat z'_{n_0,n'_0}(z)
\longrightarrow
g(z)
\qquad
\mu_1\text{-almost everywhere}\,.
\\\size{.75}{(\text{ \(d'_g\) metrizes the topology on \(\Z'\)})}
&\implies d'_g\!\left(
\widehat z'_{n_0,n'_0}(z),
g(z)
\right)
\longrightarrow 0
\qquad
\mu_1\text{-almost everywhere}.\end{aligned}
\]

Since \(d'_g\) is bounded, the dominated convergence theorem implies
\[
\int_{\Z}
d'_g\!\left(
\widehat z'_{n_0,n'_0}(z),
g(z)
\right)
\,d\mu_1(z)
\longrightarrow 0.
\]
Define the deterministic coupling $\varpi_{n_0,n'_0}:=(\widehat z'_{n_0,n'_0}, g)_\#\mu_1\in \Pi(\widehat z'_{n_0,n'_0}{}_\#\mu_1, g_\#\mu_1)$. By the definition of \(\operatorname W_g\),  
\begin{align*}
    \int_{\Z}
d'_g\!\left(
\widehat z'_{n_0,n'_0}(z),
g(z)
\right)
\,d\mu_1(z)&= \int_{\Z'\times \Z'}d'_g(z'_1,z'_2)\;d\varpi_{n_0,n'_0}(z'_1,z'_2)\geq \operatorname{W}_g(\widehat z'_{n_0,n'_0}{}_\#\mu_1, g_\#\mu_1)
\end{align*}
\begin{equation}\label{eq:Wg(z',g mu1)=0}
    \implies \operatorname{W}_g\!\left(
\widehat z'_{n_0,n'_0}{}_\#\mu_1,
g_\#\mu_1
\right)
\xrightarrow[n_0,n'_0\to\infty]{} 0\,.
\end{equation}

Next, define
\[
D_g:=\sup_{z'_1,z'_2\in\Z'} d'_g(z'_1,z'_2)<\infty.
\]
and define the space of 1-Lipschitz continuous functions on $\Z'$ with respect to $d'_g$
\[
\mathcal F_{1,d'_g}:=\big\{\psi:\Z'\to \R: \|\psi\|_{d'_g}:=\sup_{z'_1\neq z'_2}\dfrac{|\psi(z'_1)-\psi(z'_2)|}{d'_g(z'_1,z'_2)}\leq 1\big\}\,.
\]

By the Kantorovich--Rubinstein duality formula \cite[Remark~6.5]{villani2009optimal},
\begin{align*}
\operatorname W_g\!\left(
\widehat\mu_1'{}^{\mathrm{(synth.)}},
\widehat z'_{n_0,n'_0}{}_\#\mu_1
\right)
&=
\sup_{\psi\in\mathcal F_{1,d'_g}}
\left|
\int_{\Z'}\psi(z')\,d\widehat\mu_1'{}^{\mathrm{(synth.)}}(z')
-
\int_{\Z'}\psi(z')\,d\!\left(\widehat z'_{n_0,n'_0}{}_\#\mu_1\right)(z')
\right|\\
&= \sup_{\psi\in\mathcal F_{1,d'_g}}
\left|\;
\frac1{n_1}\;\sum_{k=1}^{n_1}\psi(\widehat z'_{n_0,n'_0}(z_{1k}))\quad
-
\int_{\Z}\psi(\widehat z'_{n_0,n'_0}(z))\,d\!\mu_1(z)
\right|
\end{align*}
Since adding a constant to \(\psi\) does not change the difference of the two
integrals, we may restrict the function space to
\[ \mathcal F_{1,d'_g}^0:=\big\{\psi\in \mathcal F_{1,d'_g}: \psi(z'_0)=0\big\}\,,
\]
for some fixed \(z'_0\in\Z'\). Thus
\begin{align*}
\operatorname W_g\!\left(
\widehat\mu_1'{}^{\mathrm{(synth.)}},
\widehat z'_{n_0,n'_0}{}_\#\mu_1
\right)
&= \sup_{\psi\in\mathcal F_{1,d'_g}^0}
\left|\;
\frac1{n_1}\;\sum_{k=1}^{n_1}\psi(\widehat z'_{n_0,n'_0}(z_{1k}))\quad
-
\int_{\Z}\psi(\widehat z'_{n_0,n'_0}(z))\,d\!\mu_1(z)
\right|
\end{align*}
Fix an arbitrary \(\eta>0\) and choose \(\delta>0\) small enough that
\(
\delta<\dfrac\eta{D_g}.
\)
 Since \(d'_g\) is bounded and metrizes the topology on \(\Z'\), \(\operatorname W_g\) convergence in \eqref{eq:Wg(z',g mu1)=0} implies weak convergence,
\(
\widehat z'_{n_0,n'_0}{}_\#\mu_1
\xrightarrow{\mathrm{weakly}}
g_\#\mu_1.
\) Also, 
because \(\Z'\) is Polish, every weakly convergent sequence of probability
measures on \(\Z'\) is tight, meaning for every \(\delta>0\), there exists a compact set
\(C_\delta\subseteq\Z'\) such that, for all sufficiently large \(n_0,n'_0\),
\[
\widehat z'_{n_0,n'_0}{}_\#\mu_1(C_\delta)=\mu_1(\widehat z_{n_0,n'_0}^{\prime-1}(C_\delta))>1-\delta.
\]

Observe that $(C_\delta, d'_g)$ is a compact metric space. Also, consider the restriction  of functions in $\mathcal F_{1,d'_g}^0$ to $C_\delta$, which is Lipschitz thus equicontinuous, and is bounded since
$$\sup_{z'\in\Z'}|\psi(z')|\leq \sup_{z'\in\Z'}d'_g(z',z'_0)\leq D_g\,.$$
Therefore, by the Arzelà--Ascoli theorem \cite[Theorem~4.43]{folland1999real}, the restricted class $\mathcal F_{1,d'_g}^0\big|_{C_\delta}$ is \emph{totally bounded} in the uniform metric on \(C_\delta\), meaning for any $\eta>0$ there exist finitely many $\psi_1,\ldots,\psi_M\in\mathcal F_{1,d'g}^0\big|_{C_\delta}, M<\infty$, that satisfy
$$\text{for any }\psi\in \mathcal F_{1,d'g}^0\big|_{C_\delta}: \text{ there exists }m\in 1,\cdots,M\text{ such that }\sup_{z'\in C_\delta}|\psi(z')-\psi_m(z')|<{\eta}\,.$$

For each fixed \(m\in\{1,\ldots,M\}\), conditional on the control samples \((Z_0,Z'_0)\), the random variables
\[
\psi_m\!\left(\widehat z'_{n_0,n'_0}(z_{1k})\right),
\qquad k=1,\ldots,n_1,
\]
are i.i.d. Moreover, since \(\psi_m\in\mathcal F_{1,d'_g}^0\), they are uniformly bounded by \(D_g\). Their conditional mean is
\[
\mathbb E\!\left[
\psi_m\!\left(\widehat z'_{n_0,n'_0}(z_{1k})\right)
\,\middle|\,
Z_0,Z'_0
\right]
=
\int_{\Z}
\psi_m\!\left(\widehat z'_{n_0,n'_0}(z)\right)\,d\mu_1(z).
\]
Therefore, by Hoeffding's inequality, for every \(t>0\),
\[
\mathbb P\!\left(
\left|
\frac1{n_1}\sum_{k=1}^{n_1}
\psi_m\!\left(\widehat z'_{n_0,n'_0}(z_{1k})\right)
-
\int_{\Z}
\psi_m\!\left(\widehat z'_{n_0,n'_0}(z)\right)\,d\mu_1(z)
\right|
>t
\,\middle|\,
Z_0,Z'_0
\right)
\le
2\exp\!\left(
-\frac{n_1t^2}{2D_g^2}
\right).
\]
Taking a union bound over \(m=1,\ldots,M\), we obtain
\[
\mathbb P\!\left(
\max_{1\le m\le M}
\left|
\frac1{n_1}\sum_{k=1}^{n_1}
\psi_m\!\left(\widehat z'_{n_0,n'_0}(z_{1k})\right)
-
\int_{\Z}
\psi_m\!\left(\widehat z'_{n_0,n'_0}(z)\right)\,d\mu_1(z)
\right|
>t
\,\middle|\,
Z_0,Z'_0
\right)
\le
2M\exp\!\left(
-\frac{n_1t^2}{2D_g^2}
\right).
\]
Since \(M<\infty\) is fixed once \(\eta\) and \(C_\delta\) are fixed, the right-hand side converges to zero as \(n_1\to\infty\). Hence
\[
\max_{1\le m\le M}
\left|
\frac1{n_1}\sum_{k=1}^{n_1}
\psi_m\!\left(\widehat z'_{n_0,n'_0}(z_{1k})\right)
-
\int_{\Z}
\psi_m\!\left(\widehat z'_{n_0,n'_0}(z)\right)\,d\mu_1(z)
\right|
=o_{\mathbb P}(1).
\]

Now, observe that 
\[
\begin{aligned}
\Big|
\frac1{n_1}\sum_{k=1}^{n_1}
\psi\!\left(\widehat z'_{n_0,n'_0}(z_{1k})\right)
-
\int_{\Z}
\psi\!&\left(\widehat z'_{n_0,n'_0}(z)\right)\,d\mu_1(z)
\Big|\leq
\\
&
\left|
\frac1{n_1}\sum_{k=1}^{n_1}
\psi_m\!\left(\widehat z'_{n_0,n'_0}(z_{1k})\right)
-
\int_{\Z}
\psi_m\!\left(\widehat z'_{n_0,n'_0}(z)\right)\,d\mu_1(z)
\right|
\\
&\qquad+
\frac1{n_1}\sum_{k=1}^{n_1}
\left|
\psi\!\left(\widehat z'_{n_0,n'_0}(z_{1k})\right)
-
\psi_m\!\left(\widehat z'_{n_0,n'_0}(z_{1k})\right)
\right|
\\
&\qquad+
\int_{\Z}
\left|
\psi\!\left(\widehat z'_{n_0,n'_0}(z)\right)
-
\psi_m\!\left(\widehat z'_{n_0,n'_0}(z)\right)
\right|\,d\mu_1(z)\,.
\end{aligned}
\]
On the set of points \(z\in\widehat z_{n_0,n'_0}^{\prime\, -1}(C_\delta)\), 
we have $|\psi(z')-\psi_m(z')|\le \eta/8$. Outside this set, since both
\(\psi\) and \(\psi_m\) belong to \(\mathcal F_{1,d'_g}^0\), we have $|\psi(z')-\psi_m(z')|\le 2D_g$. Therefore,

\[
\begin{aligned}
&\frac1{n_1}\sum_{k=1}^{n_1}
\left|
\psi\!\left(\widehat z'_{n_0,n'_0}(z_{1k})\right)
-
\psi_m\!\left(\widehat z'_{n_0,n'_0}(z_{1k})\right)
\right|\le
{\eta}
+
2D_g\,
\frac1{n_1}\sum_{k=1}^{n_1}
\1\!\left\{
z_{1k}\in
\widehat z_{n_0,n'_0}^{\prime\, -1}(C_\delta^c)
\right\},
\end{aligned}
\] and
\[
\begin{aligned}
\int_{\Z}
\left|
\psi\!\left(\widehat z'_{n_0,n'_0}(z)\right)
-
\psi_m\!\left(\widehat z'_{n_0,n'_0}(z)\right)
\right|\,d\mu_1(z)&\le
{\eta}
+
2D_g\;
\mu_1\!\left(
\widehat z_{n_0,n'_0}^{\prime\, -1}(C_\delta^c)
\right)\\
&\leq \eta +2D_g\;\delta\leq 3\eta\,.
\end{aligned}
\]

Conditional on \((Z_0,Z'_0)\), the indicators
\[
\1\!\left\{
z_{1k}\in
\widehat z_{n_0,n'_0}^{\prime\, -1}(C_\delta^c)
\right\},
\qquad k=1,\ldots,n_1,
\]
are i.i.d. Bernoulli random variables with mean less than \(\delta\). Hence, by Hoeffding's inequality,

\begin{align*}
&\mathbb P\left(\frac1{n_1}\sum_{k=1}^{n_1}
\1 \!\left\{
z_{1k}\in
\widehat z_{n_0,n'_0}^{\prime\, -1}(C_\delta^c)
\right\}-\delta>\delta\Big| Z_0,Z'_0\right)\\&=\mathbb P\left(\frac1{n_1}\sum_{k=1}^{n_1}
\1 \!\left\{
z_{1k}\in
\widehat z_{n_0,n'_0}^{\prime\, -1}(C_\delta^c)
\right\}>2\delta\Big| Z_0,Z'_0\right)
\le \exp(-2n_1\delta^2)\xrightarrow[n_1\to\infty]{0}\,0\\&\qquad
\implies \frac1{n_1}\sum_{k=1}^{n_1}
\1 \!\left\{
z_{1k}\in
\widehat z_{n_0,n'_0}^{\prime\, -1}(C_\delta^c)
\right\}\leq 2\delta + o_{\mathbb P}(1)\,,
\end{align*}
yielding
\[
\begin{aligned}
&\frac1{n_1}\sum_{k=1}^{n_1}
\left|
\psi\!\left(\widehat z'_{n_0,n'_0}(z_{1k})\right)
-
\psi_m\!\left(\widehat z'_{n_0,n'_0}(z_{1k})\right)
\right|\le
{\eta}
+
4D_g\;\delta + o_{\mathbb P}(1)\leq 5\eta\,+ o_{\mathbb P}(1).
\end{aligned}
\]
Since $\eta>0$ was chosen arbitrarily, setting $\eta\to0$ yields
\[
\begin{aligned}
&\left|
\frac1{n_1}\sum_{k=1}^{n_1}
\psi\!\left(\widehat z'_{n_0,n'_0}(z_{1k})\right)
-
\int_{\Z}
\psi\!\left(\widehat z'_{n_0,n'_0}(z)\right)\,d\mu_1(z)
\right|\le
o_{\mathbb P}(1)
+
8\eta=o_{\mathbb P}(1).
\end{aligned}
\] for any $\psi\in \mathcal F_{1,d'_g}^0$. Taking the supremum thus yields 
\begin{align*}
\operatorname W_g\!\left(
\widehat\mu_1'{}^{\mathrm{(synth.)}},
\widehat z'_{n_0,n'_0}{}_\#\mu_1
\right)
&= \sup_{\psi\in\mathcal F_{1,d'_g}^0}
\left|\;
\frac1{n_1}\;\sum_{k=1}^{n_1}\psi(\widehat z'_{n_0,n'_0}(z_{1k}))\quad
-
\int_{\Z}\psi(\widehat z'_{n_0,n'_0}(z))\,d\!\mu_1(z)
\right|=o_{\mathbb P}(1)\,,
\end{align*} which, in combination with \eqref{eq:Wg(z',g mu1)=0} proves the final result
\[
\operatorname W_g\!\left(
\widehat\mu_1'{}^{\mathrm{(synth.)}},
\mu'_1
\right)
\le
\varepsilon+o_{\mathbb P}(1).
\]

\section{Method Implementation Details}

\subsection{Construction of the Anchor Alignment Function}\label{app:anchor}

For each anchor coordinate \(j\in\{1,\ldots,|\mathcal H|\}\), we define a discrepancy \(\kappa_j\) according to the notion of cross-site comparability that is appropriate for that coordinate. To simplify notation, for \(z=(x,y)\in\mathcal Z\) and \(z'=(x',y')\in\mathcal Z'\), define
\[
a_j(z)
:=
\Big[\operatorname{Proj}^{\mathcal Z\to\mathcal H}(z)\Big]_j,
\qquad
a'_j(z')
:=
\Big[\operatorname{Proj}^{\mathcal Z'\to\mathcal H}(z')\Big]_j,
\]
where
\[
\operatorname{Proj}^{\mathcal Z\to\mathcal H}(z)
=
(x_{\ell_1},\ldots,x_{\ell_{|\mathcal H|}}),
\qquad
\operatorname{Proj}^{\mathcal Z'\to\mathcal H}(z')
=
(x'_{\ell_1},\ldots,x'_{\ell_{|\mathcal H|}}),
\]
are the canonical projections from $\mathcal Z$ and $\mathcal Z'$ into the common anchor space $\mathcal H$, and \(\{\ell_1,\ldots,\ell_{|\mathcal H|}\}\) denote the indices of the anchor features.

When an anchor is measured on a common scale at both sites, $\kappa_j$ may be defined on the original scale by directly comparing the coordinate values:
\vspace{2.5mm}

\centerline{$\displaystyle
\kappa_j(z,z')
=
\left|a_j(z)-a'_j(z')\right|\,.
$}

\vspace{2.5mm}

\noindent In this case, the condition \(\kappa_j(z,\f(z))=0\) requires the transformation to preserve the value of the \(j\)th anchor coordinate.

By contrast, when comparability holds up to a monotone distortion, $\kappa_j$ may instead be defined through comparing the marginal quantiles. Let 
\[
G_j(t)
=
\mu_0\!\left(\left\{
\widetilde z\in\mathcal Z:
a_j(\widetilde z)\le t
\right\}\right),
\qquad
G'_j(t)
=
\mu'_0\!\left(\left\{
\widetilde z'\in\mathcal Z':
a'_j(\widetilde z')\le t
\right\}\right)
\]
denote the corresponding source and target control marginal distribution functions of anchor covariates. For such \emph{quantile-comparable} anchors, a natural choice of $\kappa_j$ would be:
\[
\kappa_j(z,z')
=
\left|
G_j\!\left(a_j(z)\right)
-
G'_j\!\left(a'_j(z')\right)
\right|\,.
\]
\noindent Hence, \(\kappa_j(z,\f(z))=0\) requires that \(\f\) preserve the marginal \emph{rank} of the \(j\)th anchor across sites. Equivalently, for continuous marginals, we may define
\[
h_j^\star(t)
=
{G'_j}^{-1}\!\left(G_j(t)\right).
\]
Then
\[\kappa_j(z,z')=0\quad\Longleftrightarrow\quad
G_j\!\left(a_j(z)\right)
=
G'_j\!\left(a'_j(z')\right)
\quad\Longleftrightarrow\quad
h_j^\star\!\left(a_j(z)\right)
=
a'_j(z').
\]
Thus we may equivalently define $\kappa_j$ through $$\kappa_j(z,z')=\left|h^\star_j(a_j(z))-a'_j(z')\right|\,.$$
\noindent In either case, $\mathcal K$ takes the unified form 

\vspace{2.5mm}

\centerline{$\displaystyle
\mathcal K\!\left(z,z'\right)\,=\, \sum_{j=1}^{|\mathcal H|}\alpha_j\, \kappa_j(z,z')\,,\qquad \alpha_j\geq 0\,,   
$}

\vspace{2.5mm}

\noindent with the choice of $\kappa_j$ reflecting which notion of comparability is scientifically appropriate for each anchor. The distance-based form is natural for variables such as age or binary gender indicator that are recorded under common definitions, and we expect matched units across sites to remain close on those coordinates. The quantile-based form arises naturally for measurements such as blood pressure or biomarker concentrations, where absolute scales may differ across institutions due to calibration or assay differences, but the relative ordering of patients is preserved; for instance, a patient near the upper tail of the source distribution should still correspond to a patient near the upper tail of the target distribution. We use equal $\alpha_j$ weights across anchor coordinates in the numerical experiments, while directly comparable anchors are standardized before their absolute differences are computed.

In the numerical implementation, we replace \(G_j\) and \(G'_j\) by the empirical distribution functions computed from the observed control samples $Z_0$ and $Z'_0$
\[
\widehat G_j(t)
=
\frac{1}{n_0}
\sum_{i=1}^{n_0}
\1\!\left\{
a_j(z_{0i})\le t
\right\},
\qquad
\widehat G'_j(t)
=
\frac{1}{n'_0}
\sum_{i=1}^{n'_0}
\1\!\left\{
a'_j(z'_{0i})\le t
\right\}\,,\qquad z_{0i}\in Z_0,\;z'_{0i}\in Z'_0
\]
and approximate $h^\star=G_j^{\prime-1}\circ G_j$ by a minimizer of the 1-Wasserstein distance between the empirical distributions
$$\widehat h_j\in \arg\min_{h \in H^{\uparrow}} \operatorname{W}_1\left(h_\#\widehat G_j, \widehat G'_j\right)\,,$$
where $H^{\uparrow}$ is a class of monotone one-dimensional neural networks. The resulting $\kappa_j$ is then defined on the target scale as $$\widehat \kappa_j=\left|\widehat h_j(a_j(z))-a'_j(z')\right|$$ 

Note that the requirement $\mathcal K(z,\f(z))=0$ holds exactly at the population level for both direct- and quantile-comparable anchors: for the distance-based form, $\kappa_j(z,\f(z))=0$ requires that $\f$ acts as the identity on the $j^{\mathrm {th}}$ coordinate of $\mathcal H$. For the quantile-based form, it requires that $\f$ is rank-preserving on the $j^{\mathrm {th}}$ coordinate. In practice, estimating $h^\star$ using the empirical quantile functions $\widehat G_j, \widehat G'_j$ makes the alignment condition $\mathcal K(z,\f(z))=0$ hold up to a sampling error.

\subsection{Parametrization of the Transformation Class}
\label{app:Phi_implementation}

In computation, we represent the cross-site transformation by a finite-dimensional neural map \(f_\theta:\mathcal Z\to\mathcal Z'\). All coordinates are first standardized separately using the source- and target-control samples, so \(f_\theta\) is estimated on the standardized scale and its output is subsequently transformed back to the original target scale.

To simplify notation, for each \(z\in\mathcal Z\), write
\[
z=(x_D,x_Q,x_U,y),
\]
where \(x_D\in\mathcal X_D\) denotes the directly comparable anchors,
\(x_Q\in\mathcal X_Q\) the quantile-comparable anchors,
\(x_U\in\mathcal X_U\) the remaining non-comparable covariates, and
\(y\in\mathcal Y\) the outcome. The corresponding target-space coordinates are denoted by
\(z'=(x'_D,x'_Q,x'_U,y')\), where $(x'_D,x'_Q)\in \mathcal X_D\times \mathcal X_Q$ since the source and target spaces share the same anchor space.

We use a gated residual parametrization of the transformation. We define an anchor-based scalar gate \(g_\eta(z)\in(0,1)\):
\[
g_\eta(z)
=
\operatorname{sigmoid}\!\left\{N_\eta(x_D,x_Q)\right\},
\]
where \(N_\eta\) is a feed-forward neural network (the gate may alternatively depend on the full input \(z\)). The basic parametrization is
\[
f_\theta(z)
=
z+
\{1-g_\eta(z)\}r_0(z)
+
g_\eta(z)r_1(z)
+
g_\eta(z)b,
\]
where \(r_0\) and \(r_1\) are residual neural networks and \(b\) is a learned shift vector. Thus, the transformation interpolates smoothly between two nonlinear residual maps, while the gate determines the relative contribution of each branch.

For the reported implementation, we use a factorized version of $f_\theta$ that utilizes the known partition
\(z=(x_D,x_Q,x_U,y)\). For the directly comparable coordinates, the factorized map preserves
\[
x'_D=x_D.
\]
For the quantile-comparable coordinates, it uses the collection of coordinate-wise monotone maps, $\widehat h_Q=(\widehat h_1, \widehat h_2,\dots)$ estimated for each $x_Q$, as described in Appendix~\ref{app:anchor}:
\[
\widehat h_Q(x_Q)
=
\big(
\widehat h_j(x_{Q,j})
\big)_{j\in Q}\,.
\]
Thus, the anchor transformation is fixed at
\[
(x'_D, x_Q'\big)= \big(x_D,\widehat h_Q(x_Q)\big)\,,
\]
throughout the subsequent optimization of \(f_\theta\). 

The non-anchor covariates are transformed through
\[
x'_U
=
x_U+
\{1-g_\eta(z)\}\,r_{U,0}(x_D,\widehat h_Q(x_Q),x_U)
+
g_\eta(z)r_{U,1}(x_D,\widehat h_Q(x_Q),x_U)
+
g_\eta(z)b_U,
\]
Hence, the transformation of \(X_U\) may depend jointly on all of $(x_D,x_Q,x_U)$. Finally, the outcome is transformed through:
\[
y'
=
y+
\{1-g_\eta(z)\}r_{Y,0}(x_D,\widehat h_Q(x_Q),x'_U,y)
+
g_\eta(z)r_{Y,1}(x_D,\widehat h_Q(x_Q),x'_U,y)
+
g_\eta(z)b_Y\,.
\]
thus the transformation of the outcome may depend on both the original outcome and the transformed covariate profile. 

The residual networks $r_{U,0},r_{U,1},r_{Y,0},r_{Y,1}$ are multilayer perceptrons with two hidden ReLU layers. In the numerical experiments, the hidden width is 128. The gate consists of two hidden ReLU layers followed by a sigmoid output. To initialize the two branches near one another, the second branch is initialized as a copy of the first with a small random perturbation. This provides a flexible nonlinear parametrization while beginning optimization close to a single residual transformation.

\subsection{Implementation and Tuning Details}
\label{app:implementation_details}

We estimate the coupling and transformation through a sequence of pretraining and alternating optimization steps. For the factorized specification described in Appendix~\ref{app:Phi_implementation}, the coordinate-wise transformations of the quantile-comparable anchors are estimated first as described in Appendix~\ref{app:anchor} and subsequently held fixed. The remaining parameters of \(f_\theta\) are then initialized using the observed source and target control samples.

\subsubsection{Pretraining}

Before optimizing the RFGW criterion, we pretrain the neural transformation to approximately align the source and target control distributions. At this stage, the coordinate-wise transformations of the anchors, \((x'_D,x'_Q) = (x_D, \widehat h_Q(x_Q))\) have already been estimated and fixed as described in Appendices~\ref{app:anchor}--\ref{app:Phi_implementation}.

For notational simplicity, let $f_{\theta,U}(z):=\operatorname{Proj}^{\mathcal Z'\to \mathcal X'_U}\circ f_\theta$ and  $f_{\theta,Y}(z):= \operatorname{Proj}^{\mathcal Z'\to \mathcal Y'}\circ f_\theta$ denote the transformed non-anchor covariates and outcome component of $f_\theta(z)$, respectively. Also, let \(\widehat\mu_{0}{}_{\#}f_{\theta,U}\) and \(\widehat\mu'_{0,U}\) denote the empirical distribution of non-anchor variables in the transformed source control sample, $f_{\theta,U}(Z_0)$, and the target control sample, $X'_{0,U}$, respectively. For each coordinate \(j\) of $\mathcal X'_U$, denote the corresponding one-dimensional marginal distributions by
\((\widehat\mu_{0}{}_{\#}f_{\theta,U})_j\) and \((\widehat\mu'_{0,U})_j\). Similarly, define the marginal outcome distributions as
\(\widehat\mu_{0}{}_{\#}f_{\theta,Y}\) and \(\widehat\mu'_{0,Y}\). Note, that although \(X_U\) and \(X'_U\) are not assumed to be directly comparable across sites, their comparison becomes well-defined after transformation: \(f_\theta\) maps the source observations into the target space \(\mathcal Z'\), making $f_{\theta,U}(Z_0)$ and $X'_{0,U}$ coordinatewise-comparable.  Consequently, coordinate-wise discrepancies between
\((\widehat\mu_{0}{}_{\#}f_{\theta,U})_j\) and
\((\widehat\mu'_{0,U})_j\)
may be used to assess and encourage marginal alignment after transformation.

We define the non-anchor marginal alignment loss as the average coordinate-wise 1-Wasserstein discrepancy
\[
\mathscr L_U(\theta)
=
\frac{1}{|\mathcal X_U'|}
\sum_{j=1}^{|\mathcal X_U'|}
W_1\!\left(
\big(\widehat\mu_{0}{}_{\#}f_{\theta,U}\big)_j,\,
\big(\widehat\mu'_{0,U}\big)_j
\right),
\]
and for the outcome 
\[
\mathscr L_Y(\theta)
=
W_1\!\left(
\widehat\mu_{0}{}_{\#}f_{\theta,Y},\,
\widehat\mu'_{0,Y}
\right).
\]
In addition to marginal alignment, we encourage the transformed source controls to reproduce the dependence structure of the target controls through correlation-alignment Frobenius norm loss
\[
\mathscr L_{\mathrm{corr}}(\theta)
=
\left\|
\operatorname{Corr}\!\Big(
(X_{0,D}, \widehat h_Q(Z_0), f_{\theta,U}(Z_0),f_{\theta,Y}(Z_0)
\Big)
-
\operatorname{Corr}\!\Big(
(X'_{0,D},X'_{0,Q},X'_{0,U}, Y')
\Big)
\right\|_F,
\]
Thus,
\(\mathscr L_U\) and \(\mathscr L_Y\) encourage marginal distributional agreement after transformation, while
\(\mathscr L_{\mathrm{corr}}\) provides a constraint on the dependence structure of the mapped control distribution. The pretraining criterion is therefore
\[
\mathscr L_{\mathrm{pre}}(\theta)
=
w_U(\theta)\mathscr L_U(\theta)
+
w_Y(\theta)\mathscr L_Y(\theta)
+
w_{\mathrm{corr}}(\theta)\mathscr L_{\mathrm{corr}}(\theta),
\]
where the weights are assigned based on gradient balancing:
\[
w_r(\theta)
\propto
\frac{c_r}{\|\nabla_\theta \mathscr L_r(\theta)\|_2}\,.
\]
In the numerical experiments, we use relative contributions $(c_U,c_Y,c_{\mathrm{corr}})=(1,1.5,0.3)$,and optimize the pretraining criterion using Adam for 300 iterations with learning rate \(0.01\). The resulting parameters initialize the subsequent RFGW optimization.

\subsubsection{Alternating Optimization}

After pretraining, estimation alternates between updating the coupling \(\pi\) and the transformation parameters \(\theta\). The empirical geometry, anchor, and graph losses follow from the corresponding components of the RFGW criterion
\[
\mathcal D_\theta(\pi)
=
\sum_{i,k=1}^{n_0}\sum_{j,m=1}^{n'_0}
\Big|
\|f_\theta(z_{0i})-f_\theta(z_{0k})\|_2
-
\|z'_{0j}-z'_{0m}\|_2
\Big|
p_{ij}p_{km},
\]
\[
\mathcal A(\pi)
=
\sum_{i=1}^{n_0}\sum_{j=1}^{n'_0}
\mathcal K(z_{0i},z'_{0j})p_{ij},
\qquad
\mathcal G_\theta(\pi)
=
\sum_{i=1}^{n_0}\sum_{j=1}^{n'_0}
\|f_\theta(z_{0i})-z'_{0j}\|_2p_{ij}.
\]

Given the current \(f_\theta\), the coupling is updated using the discrete fused Gromov--Wasserstein solver in the Python Optimal Transport package. The update takes the form
\[
\pi
\leftarrow
\arg\min_{\pi\in\Pi(\widehat\mu_0,\widehat\mu'_0)}
\left\{
\rho\,\mathcal D_\theta(\pi)
+
(1-\rho)\,\mathcal A(\pi)
+
\lambda\,\mathcal G_\theta(\pi)
\right\}.
\]
The first coupling update omits \(\mathcal G_\theta\), providing an initial geometry- and anchor-based alignment; subsequent updates include the graph term. In the numerical experiments, we set the mixing parameters to $\rho=0.7$ and $\lambda=0.2$ for the subsequent updates. 

Conditional on the current coupling, \(\theta\) is updated by Adam while \(\pi\) is held fixed, using
\[
\mathscr L_{\mathrm{map}}(\theta\mid\pi)
=
(1-\lambda)\mathcal D_\theta(\pi)
+
\lambda\mathcal G_\theta(\pi).
\]
Because the anchor transformations are fixed before this stage, \(\mathcal A(\pi)\) does not vary directly with \(\theta\), but continues to affect the transformation through the coupling \(\pi\). After a block of neural-network updates, the coupling is recomputed and the two steps are repeated.

In the numerical experiments, we additionally include the gradient-balanced calibration term
\[
\mathscr L_{\mathrm{aid}}(\theta)
=
w_U\mathscr L_U(\theta)
+
w_Y\mathscr L_Y(\theta)
+
w_{\mathrm{corr}}\mathscr L_{\mathrm{corr}}(\theta)
\]
during neural-network optimization. This term is used only as a numerical aid and is not part of the defining RFGW criterion; simulations with and without this aid produced nearly indistinguishable results.

\subsubsection{Synthetic Treatment Generation}

Once \((\widehat\pi,\widehat f)\) has been estimated from the control samples, the general synthesis criterion can be written compactly for each \(z_{1k}\in Z_1\) as
\[
\mathscr L_{\mathrm{synth}}(z';z_{1k})
=
(1-\kappa-\gamma)\,
\mathcal D_{1k}(z')
+
\kappa\,\mathcal K(z_{1k},z')
+
\gamma\,d'_{\widehat f}\!\left(\widehat f(z_{1k}),z'\right),
\]
where
\[
\mathcal D_{1k}(z')
=
\sum_{i=1}^{n_0}\sum_{j=1}^{n'_0}
\left|
d_{\mathcal Z}(z_{0i},z_{1k})
-
d'_{\widehat f}(z'_{0j},z')
\right|
\widehat p_{ij}
\]
is the control-relative geometry loss.

In the reported numerical experiments, we use the direct-map specialization
\[
\kappa=0,\qquad \gamma=1,
\]
for which
\[
{z'_{1k}}^{\mathrm{(synth.)}}
=
\arg\min_{z'\in\mathcal Z'}
d'_{\widehat f}\!\left(\widehat f(z_{1k}),z'\right)
=
\widehat f(z_{1k}).
\]
Thus, each source treated observation is standardized using the source-control normalization, passed directly through \(\widehat f\), and then transformed back using the target-control normalization.

We also evaluated synthesis using the additional geometry and anchor terms above and obtained nearly indistinguishable synthetic distributions once \((\widehat\pi,\widehat f)\) had converged. We therefore use the direct-map specification in the reported experiments, since evaluating \(\widehat f(z_{1k})\) is immediate whereas repeatedly evaluating the control-relative geometry loss for every treated observation is substantially more computationally expensive.

\section{Simulation Details}
\subsection{Data-Generating Process and Cross-Site Transformations}
\label{app:simulation_dgp}

We construct a hospital-transfer simulation in which the source and target sites observe related patient biology but partially encode the resulting clinical state through different variables. Each observation is written as
\[
Z=(X_Q,X_U,Y),
\]
where \(X_Q\) contains quantile-comparable features, \(X_U\) contains site-specific clinical features, and \(Y\) is the outcome. Treatment is denoted by \(D\in\{0,1\}\).

\subsubsection{Patient Biology and Source-Site Variables}

We first generate a common underlying patient population. Let the age be distributed according to a truncated normal distribution
\[
\operatorname{Age}\sim \operatorname{TN}(65,15^2;18,95)\,,
\] and define a latent comorbidity burden according to a Gamma distribution
\[
C=\operatorname{Gamma}(2,1)+0.02(\operatorname{Age}-65).
\]
Conditional on these quantities, we generate latent physiological components
\[
P,M,N,K,V,L,O\stackrel{\mathrm{iid}}{\sim}N(0,1),
\]
representing perfusion, metabolic dysfunction, neurologic status, cardiac status,
volume status, pulmonary dysfunction, and Organ reserve, respectively. Renal dysfunction and
frailty additionally depend on comorbidity:
\[
R=0.3\,\frac{C}{\operatorname{sd}(C)}+\varepsilon_R,
\qquad
F=0.5\,\frac{C}{\operatorname{sd}(C)}+\varepsilon_F,
\qquad
\varepsilon_R,\varepsilon_F\sim N(0,1).
\]
We define an overall latent severity index
\[
S=0.4P+0.4M+0.2L.
\]

The quantile-comparable covariates \(X_Q\) are then generated as noisy nonlinear
functions of these latent physiological states. Writing \([u]_+=\max(u,0)\),
representative equations are
\begin{align*}
\operatorname{SBP}
    &=120-13P+\varepsilon_{\mathrm{SBP}},
        &\varepsilon_{\mathrm{SBP}}&\sim N(0,8^2),\\
H&=\left[\frac{110-\operatorname{SBP}}{15}\right]_+,\\
\operatorname{HR}
    &=85+5S+4H+7K+\varepsilon_{\mathrm{HR}},
        &\varepsilon_{\mathrm{HR}}&\sim N(0,8^2),\intertext{}
\operatorname{Lactate}
    &=\exp(0.4+0.48M+\varepsilon_L),
        &\varepsilon_L&\sim N(0,0.30^2),\\
\operatorname{Creatinine}
    &=\exp(0.42R+\varepsilon_C),
        &\varepsilon_C&\sim N(0,0.33^2),\\
\operatorname{BUN}
    &=12+3\,\operatorname{Creatinine}+9V+\varepsilon_B,
        &\varepsilon_B&\sim N(0,6^2),\\
\operatorname{GCS}
    &=15-G-1.6N,
        &G&\sim\Gamma(1.3,1),\\
\operatorname{PF}
    &=\exp\!\left\{\log(350)-0.35\max(L,-0.5)-0.10S+\varepsilon_P\right\},
        &\varepsilon_P&\sim N(0,0.18^2).
\end{align*}
All measurements are truncated to clinically plausible ranges. Together with
age, these variables form
\[
X_Q=(\operatorname{Age},\operatorname{SBP},\operatorname{HR},
\operatorname{Lactate},\operatorname{Creatinine},
\operatorname{GCS},\operatorname{BUN},\operatorname{PF}).
\]

The source-site non-anchor variables \(X_U\) are four deterministic clinical
scores constructed from these measurements and latent patient characteristics,
chosen to resemble severity, comorbidity,  acuity, and respiratory scores:
\begin{align*}
X_U
&=
\big(
U_{\mathrm{sev}},
U_{\mathrm{com}},
U_{\mathrm{acuity}},
U_{\mathrm{resp}}
\big)\,;\\
U_{\mathrm{sev}}
&=
\Big[
\operatorname{round}\!\Big(
1.6\Big\{
\Big[\frac{\operatorname{Creatinine}-1}{0.5}\Big]_{0}^{4}
+
[H]_{0}^{4}
+
\Big[\frac{15-\operatorname{GCS}}{3}\Big]_{0}^{4}
+
0.3[\operatorname{Lactate}-2]_+
\Big\}
\Big)
\Big]_{0}^{24},
\\[2mm]
U_{\mathrm{com}}
&=
\Big[
\operatorname{round}\!\Big(
C
+
1.2[F]_+
+
0.04(\operatorname{Age}-50)
\Big)
\Big]_{0}^{17},
\\[2mm]
U_{\mathrm{acuity}}
&=
\Big[
\operatorname{round}\!\Big(
0.08|\operatorname{HR}-80|
+
0.05|\operatorname{SBP}-120|
+
3[\operatorname{Creatinine}-1]_+ \\
&\qquad\qquad\qquad\qquad\qquad\quad\;
+
0.6(15-\operatorname{GCS})
+
0.06[\operatorname{Age}-45]_+
+
1.5[O]_+
\Big)
\Big]_{0}^{71},
\\[2mm]
U_{\mathrm{resp}}
&=
\Big[
4\,\operatorname{Sigmoid}\!\Big(
-\frac{\operatorname{PF}-250}{60}
\Big)
+
0.4[O]_+
\Big]_{0}^{4}\,.
\end{align*}

Potential outcomes are generated on a latent severity scale. We set
\[
Y(0)=\ell_0(X_Q,X_U,F,L),
\qquad
Y(1)=Y(0)+\tau(X_U),
\]
where \(\ell_0\) is a nonlinear risk function 
\begin{align*}
Y(0)=\ell_0
={}&-5.9
+0.02[\operatorname{Age}-70]_+
+0.02\operatorname{Age}
-0.006\,\operatorname{SBP}
+0.005\,\operatorname{HR}
\\
&+0.5L_{\mathrm{nl}}
+0.08\,\operatorname{Creatinine}
\\
&+0.25\,
\operatorname{Sigmoid}\!\left(
\frac{\operatorname{Lactate}-2}{0.5}
\right)
\operatorname{Sigmoid}\!\left(
\frac{100-\operatorname{SBP}}{8}
\right)
\\
&+I
+1.15\,O_{\mathrm{occ}}
+M
\\
&+0.05\,U_{\mathrm{sev}}
+0.08\,U_{\mathrm{com}}
-\frac{0.12}{3}(\operatorname{GCS}-15)
+0.15\,U_{\mathrm{resp}}\,;
\end{align*}
containing the following unobserved components
\begin{align*}
    L_{\mathrm{nl}}&=
0.18(\operatorname{Lactate}-1.4)
+
0.15[\operatorname{Lactate}-2]_+
+
0.15[\operatorname{Lactate}-4]_+,\\
O_{\mathrm{occ}}
&=
0.45F
+
0.45O
+
0.30[L]_+,\\
I
&=
2.6\Bigg\{
\tanh\!\left(
\frac{
[\operatorname{Lactate}-2]_+
[110-\operatorname{SBP}]_+
}{60}
\right)
\\
&\qquad\qquad\qquad+
\tanh\!\left(
\frac{
[\operatorname{Creatinine}-1]_+
[\operatorname{BUN}-20]_+
}{40}
\right)
\\
&\qquad\qquad\qquad+
\tanh\!\left(
\frac{
[\operatorname{Lactate}-2]_+
[12-\operatorname{GCS}]_+
}{25}
\right)
\Bigg\},\\
M
&=
0.80\,\widetilde U_{\mathrm{sev}}[O]_+
+
0.80\,\widetilde U_{\mathrm{com}}[F]_++
0.65\,\widetilde U_{\mathrm{acuity}}[O]_+
+
0.55\,\widetilde U_{\mathrm{resp}}[L]_+,
\\
\widetilde U_r
&=
\frac{U_r-\mathbb E[U_r]}
{\operatorname{sd}(U_r)},
\qquad
r\in\{\mathrm{sev},\mathrm{com},\mathrm{acuity},\mathrm{resp}\},
\end{align*}
and the treatment effects are heterogeneous:
\[
\tau(X_U)
=
\log(0.87)
-
0.05\left(
U_{\mathrm{sev}}-\mathbb E[U_{\mathrm{sev}}]
\right)\,,
\]
making treatment more beneficial for patients with greater baseline severity.

\subsubsection{Cross-Site Transformation}

The target representation is obtained by applying a common source-to-target transformation to the patient variables. The transformation acts separately on the quantile-comparable block \(X_Q\), the site-specific score block \(X_U\), and the outcome.

\paragraph{Linear transformation.}

For the quantile-comparable block, the linear design uses the coordinate-wise affine map
\[
X'_Q=A_Q\odot X_Q+B_Q,
\]
where
\[
A_Q=(1.05,0.97,1.06,1.15,1.20,1.00,1.10,0.95),
\qquad
B_Q=(2,4,-2,0,0,0,3,0).
\]

The target non-anchor variables are then generated from the source scores and the source- and target-scale clinical measurements. Writing
\[
X_U=
\left(
U_{\mathrm{sev}},
U_{\mathrm{com}},
U_{\mathrm{acuity}},
U_{\mathrm{resp}}
\right),
\]
the target score block 
\( \displaystyle
X'_U=
\left(
U'_{\mathrm{sev}},
U'_{\mathrm{com}},
U'_{\mathrm{acuity}},
U'_{\mathrm{resp}}
\right)
\) 
is given by
\begin{align*}
U'_{\mathrm{sev}}
&=
\Big[
U_{\mathrm{sev}}
-2[\operatorname{Creatinine}-1]_+
+0.09|\operatorname{HR}'-80|
+0.05[\operatorname{Age}'-45]_+
+3
\Big]_{0}^{90},
\\
U'_{\mathrm{com}}
&=
\Big[
0.35\,U_{\mathrm{com}}
-0.4[U_{\mathrm{com}}-6]_+
+0.22(\operatorname{Age}'-60)
\\
&\hspace{18mm}
+0.004(\operatorname{Age}'-60)^2
\operatorname{sign}(\operatorname{Age}'-60)
\Big]_{-15}^{60},
\\
U'_{\mathrm{acuity}}
&=
\Big[
0.72\,U_{\mathrm{acuity}}
-0.04|\operatorname{HR}-80|
+0.32(\operatorname{BUN}'-20)
\Big]_{0}^{120},
\\
U'_{\mathrm{resp}}
&=
\Big[
2.2\,\operatorname{Sigmoid}\!\left(
-\frac{\operatorname{PF}'-200}{40}
\right)
+0.15\,U_{\mathrm{resp}}
\\&\qquad\qquad\qquad+1.3\,\operatorname{Sigmoid}\!\left(
\frac{\operatorname{Age}'-70}{10}
\right)
-0.5[U_{\mathrm{resp}}-2]_+
\Big]_{0}^{10}.
\end{align*}
Thus, the source and target score blocks represent related clinical dimensions but are defined through different scoring systems.

Finally, under the shared-transformation design, outcomes are mapped according to
\[
Y'
=
0.95\,Y
-
\left(
0.25+0.02\,U'_{\mathrm{sev}}
\right).
\]

\paragraph{Nonlinear transformation.}

In the nonlinear design, each coordinate of \(X_Q\) receives an additional nonlinear perturbation. Writing
\[
\widetilde X_{Qj}
=
\frac{X_{Qj}-\mathbb E[X_{Qj}]}
{\operatorname{sd}(X_{Qj})},
\]
we set
\[
X'_{Qj}
=
A_{Qj}X_{Qj}+B_{Qj}
+
0.30\,\operatorname{sd}(X_{Qj})
\left\{
1.1\tanh(\widetilde X_{Qj})
+
0.25\,\operatorname{sign}(\widetilde X_{Qj})
\widetilde X_{Qj}^{\,2}
\right\}.
\]

For the non-anchor block, let
\[
\bar U'_{\mathrm{sev}},\quad
\bar U'_{\mathrm{com}},\quad
\bar U'_{\mathrm{acuity}},\quad
\bar U'_{\mathrm{resp}}
\]
denote the four target scores obtained from the linear formulas above before truncation, and define
\[
H=\left[\frac{110-\operatorname{SBP}}{15}\right]_+,
\qquad
Z_L=[\operatorname{Lactate}-2]_+,
\qquad
Z_C=[\operatorname{Creatinine}-1]_+,
\]
together with
\[
R_H=
\left(\frac{\operatorname{HR}'-80}{25}\right)^2,
\qquad
R_{BC}
=
\frac{\operatorname{BUN}'}
{\operatorname{Creatinine}'+0.5},
\qquad
R_{LP}
=
\frac{100\,\operatorname{Lactate}}
{\max(\operatorname{SBP}',60)}.
\]
The nonlinear target scores are
\begin{align*}
U'_{\mathrm{sev}}
&=
\Big[
0.65\,\bar U'_{\mathrm{sev}}
+
30\tanh\!\left(
\frac{8Z_LH}{30}
\right)
+
3.61\,R_H
+
1.22\,R_{LP}
\Big]_{0}^{90},
\\[1.5mm]
U'_{\mathrm{com}}
&=
\Big[
0.36\,\bar U'_{\mathrm{com}}
+
1.15
\left(\frac{\operatorname{Age}'-60}{10}\right)
\left(\frac{U_{\mathrm{com}}}{4}\right)
+
8
\left(\frac{\operatorname{Age}'-60}{10}\right)
Z_C
\Big]_{-15}^{60},
\\[1.5mm]
U'_{\mathrm{acuity}}
&=
\Big[
0.45\,\bar U'_{\mathrm{acuity}}
+
0.23\,R_{BC}
+
8\,R_H
+
7.41\,Z_C
\left(\frac{\operatorname{BUN}'-20}{10}\right)
\Big]_{0}^{120},
\\[1.5mm]
U'_{\mathrm{resp}}
&=
\Big[
0.55\,\bar U'_{\mathrm{resp}}
+
1.5\,
\operatorname{Sigmoid}\!\left(
-\frac{\operatorname{PF}'-180}{30}
\right)
\frac{\operatorname{Lactate}}{2}
\\
&\hspace{18mm}
+
\left(\frac{\operatorname{Age}'-65}{10}\right)
\operatorname{Sigmoid}\!\left(
-\frac{\operatorname{PF}'-200}{40}
\right)
\Big]_{0}^{10}.
\end{align*}
Thus, the nonlinear design introduces interactions among the source and target clinical measurements in addition to the site-specific redefinition of the score variables. 

The latent target potential outcomes are then mapped according to
\begin{align*}
\widetilde Y'
&=
0.95\,Y
-
\left(
0.25+0.02\,U'_{\mathrm{sev}}
\right)
-
0.6\tanh\!\left(
\frac{Y+2.7}{1.5}
\right)
-
0.25\tanh\!\left(
\frac{Y+2.7}{0.6}
\right)\\
Y'
&=
\operatorname{sign}(\widetilde Y')
|\widetilde Y'|^{1.8}.
\end{align*}

\subsubsection{Departure Scenarios from the Baseline Design}

The departure scenarios are constructed from two modifications of the baseline data-generating process: a severity-dependent treatment propensity and an alternative target-care outcome map. All departures are obtained by introducing one or both of these components, while leaving the remaining elements of the baseline design unchanged. 

\paragraph{Severity-dependent propensity score.}

To induce treatment selection and define the severity index used in several departure scenarios, we construct a propensity score from clinically relevant markers of patient severity. Let
\[
\begin{aligned}
S(X)
={}&
0.9\,\1\{\operatorname{Lactate}\ge 2\}
+
1.1\,\1\{\operatorname{Lactate}\ge 4\}
+
0.5[\operatorname{Lactate}-2]_+
\\
&+
\1\{\operatorname{SBP}<90\}
+
0.4\left[\frac{90-\operatorname{SBP}}{10}\right]_+
+
0.15\,U_{\mathrm{sev}}
+
0.05\,U_{\mathrm{com}}
-
\frac{0.20}{3}(\operatorname{GCS}-15).
\end{aligned}
\]
We define
\[
e(X)
=
\operatorname{Sigmoid}\!\Big(
a+\{S(X)-\E[S(X)]\}
\Big),
\]
where \(a\) is chosen so that \(\E[e(X)]=0.5\), and truncate \(e(X)\) to the interval \([0.05,0.95]\).

\paragraph{Alternative target-care outcome maps.}

To generate departures from the shared cross-site outcome transformation, we introduce two alternative target-care outcome maps. For the linear cross-site design, these are defined as
\[
\f_{Y}^{1}(Y;s)
=
0.50\,Y+4.0+0.4[s]_+,
\qquad
\f_{Y}^{2}(Y;s)
=
0.95\,Y+0.4+0.8[s]_+.
\]
Here, \(s\) denotes a patient-severity index. In the heterogeneous-care-regime scenario we define $s$ using the propensity score $e(X)$ defined above
\[
s=6\{e(X)-0.62\},
\]
whereas in the arm-specific-transformation scenario \(s\) is the standardized target severity score,
\[
s=
\frac{U'_{\mathrm{sev}}-\E[U'_{\mathrm{sev}}]}
{\operatorname{sd}(U'_{\mathrm{sev}})}.
\]

For the nonlinear cross-site design, the corresponding maps are augmented as
\[
\f_{Y}^{i}(Y;s)
\leftarrow
\f_{Y}^{i}(Y;s)
+
0.5\tanh\!\left(\frac{Y+2.7}{1.5}\right),\qquad i\in\{1,2\}\,.
\]

\paragraph{Simulation departures.}

We consider four modifications of the baseline design. Under \emph{observational treatment assignment},
\[
D\mid X\sim \operatorname{Bernoulli}\{e(X)\},
\]
instead of \(D\sim\operatorname{Bernoulli}(0.5)\). Under \emph{non-random site membership}, we form
\[
R=e(X)+\varepsilon,\qquad \varepsilon\sim N(0,0.05^2),
\]
and assign observations in the upper half of \(R\) to the target site and those in the lower half to the source site, while retaining randomized treatment within each site.

For the \emph{arm-specific transformation} design, we set
\[
Y'=
\begin{cases}
\f_Y^2(Y;s), & \text{if }D_i=0\,,\\[1mm]
\f_Y(Y), & \text{if }D_i=1\,,
\end{cases}
\]
where \(\f_Y\) denotes the baseline target outcome map. Finally, for the \emph{heterogeneous-care-regime} design, both treatment arms follow
\[
Y'=
\begin{cases}
\f_Y(Y), &\text{if } e(X)>0.62\,,\\[1mm]
\f_Y^{1}(Y;s), &\text{if } e(X)\le 0.62\,.
\end{cases}\
\]

\subsection{Competing Methods and Simulation Implementation}
\label{app:baseline_details}

For \model{TWFE}, \model{MatchSynth}, and \model{GenSynth}, we considered both a full specification using all available covariates and restricted specifications based on the comparable anchor coordinates, with quantile-comparable anchors represented either on their original scale or through empirical ranks. For each method, we conducted the full Monte Carlo study using both the full and restricted specifications and report results for the variant with the lowest overall evaluation error (the restricted variants for \model{TWFE} and \model{Matching}, and the full variant for \model{GenSynth}). Thus, the comparison
does not advantage \model{OTSynth} by providing it with anchor information that is withheld from the competing methods.

\paragraph{TWFE. }The Two-way Fixed Effects (\model{TWFE}) baseline is a straightforward extension of linear regression that accounts for treatment and target-site effects by introducing treatment and site indicators
\[
D_i=\1\{z_i\in Z_1\},
\qquad
T_i=\1\{z_i\in Z'_0\},
\]
together with their interactions with the included covariates. The full version of the model is estimated jointly using the observed datasets \(Z_0,Z_1,Z'_0\):
\[
[\widehat\alpha,\widehat\beta,\widehat\tau,\widehat\gamma,
\widehat\nu,\widehat\lambda]
=
\argmin_{[\alpha,\beta,\tau,\gamma,\nu,\lambda]}
\sum_{z_i\in Z_0\cup Z_1\cup Z'_0}
\left[
y_i-
\left\{
\alpha+x_i^\top\beta
+D_i(\tau+x_i^\top\gamma)
+T_i(\nu+x_i^\top\lambda)
\right\}
\right]^2.
\]
Using the fitted coefficients, synthetic target-treated outcomes are generated on the target-control covariates:
\[
y_{1i}^{\mathrm{synth}}
=
\widehat\alpha
+{x'}_{0i}^{\top}\widehat\beta
+\widehat\tau
+{x'}_{0i}^{\top}\widehat\gamma
+\widehat\nu
+{x'}_{0i}^{\top}\widehat\lambda\,,
\qquad \text{for each }
z'_{0i}=({x'}_{0i},y'_{0i})\in Z'_0.
\]
Thus, \model{TWFE} leaves the target-control feature vectors unchanged and synthesizes only their treated outcomes.

In its restricted version, \model{TWFE} exploits the partition
\(X=(X_D,X_Q,X_U)\). Let
\[
R(x)
=
\left(
x_D,\,
\widehat G_{Q,1}(x_{Q,1}),\ldots,
\widehat G_{Q,|X_Q|}(x_{Q,|X_Q|})
\right)
\]
denote the representation containing the directly comparable covariates and the empirical ranks of the quantile-comparable covariates, where the empirical CDFs $G_{Q,i}$ are computed separately using the source and target control samples. The restricted \model{TWFE} model is
\[
\begin{aligned}
y_i
={}&
\alpha
+R(x_i)^\top\beta
+x_{U,i}^\top\delta
+D_i\{\tau+R(x_i)^\top\gamma\}\\
&+
T_i\{\nu+R(x_i)^\top\lambda+x_{U,i}^\top\xi\}
+\epsilon_i .
\end{aligned}
\]
Thus, $X_Q$ affects the outcome through its rank, and \(X_U\) may affect the baseline outcome and have a target-specific effect, but not the treatment-specific effect. 

\paragraph{Matching. }Inspired by \cite{hotz2005predicting}, the \model{Matching} baseline begins by identifying, for each target-control observation, the nearest source-treated observation in the feature representation:
\[
z^{\mathrm m}_{1i}
=
\argmin_{z_1=[x_1,y_1]\in Z_1}
\left\|
R(x_1)-R(x'_{0i})
\right\|_2,
\qquad\text{for each }z'_{0i}=
[x'_{0i},y'_{0i}]\in Z'_0,
\]
where, for the full specification of the model, $$R(x) = \left(\frac{x_1-\mu_1}{\sigma_1},\dots, \frac{x_{|X|}-\mu_{|X|}}{\sigma_{|X|}}\right)\,,$$ is the standardized vector of features. The matched outcomes are then regressed on the same representation,
\[
[\widehat\alpha,\widehat\beta]
=
\argmin_{[\alpha,\beta]}
\sum_{i}
\left[
y_{1i}^{\mathrm m}
-\left\{\alpha+R(x_{1i}^{\mathrm m})^\top\beta\right\}
\right]^2,
\]
and the synthetic responses are regression-adjusted according to
\[
y_{1i}^{\mathrm{synth}}
=
y_{1i}^{\mathrm m}
+
\widehat\beta^\top
\left\{
R(x'_{0i})-R(x_{1i}^{\mathrm m})
\right\}.
\]
As with \model{TWFE}, the resulting synthetic observations retain the target-control covariates.

The restricted version of \model{Matching} uses $X_D$ and the ranks of $X_Q$ as the feature representation for matching:
$$R(x) = \left(\frac{x_{D,1}-\mu_{D,1}}{\sigma_{D,1}},\dots,\frac{x_{D,|X_D|}-\mu_{D,|X_D|}}{\sigma_{D,|X_D|}},\widehat G_{Q,1}(x_{Q,1}),\ldots,
\widehat G_{Q,|X_Q|}(x_{Q,|X_Q|})\right)\,.$$

\paragraph{GenSynth. }The \model{GenSynth} baseline combines \model{Matching} with the Generalized Synthetic Control framework of \cite{xu2017generalized}. Each source control and source treated observation is first matched to a target-control observation in the selected feature representation:
\begin{align*}
    {z'}^{\text{m}}_{0i} &= \argmin_{z'_0=[x'_{0},y'_0]\in Z_0'}\|R(x_0') - R(x_{0i})\|_2\qquad\text{ for each }z_{0i}=[x_{0i},y_{0i}]\in Z_0,&\text{ and}\intertext{}
    {z'}^{\text{m}}_{1i} &= \argmin_{z'_0=[x'_{0},y'_0]\in Z_0'}\|R(x_0') - R(x_{1i})\|_2\qquad\text{ for each }z_{1i}=[x_{1i},y_{1i}]\in Z_1,
\end{align*}
For the matched source--target control pairs, the method assumes the rank-\(r\) factor model
\[
\begin{cases}
y_{0i}
=
\alpha+x_{0i}^{\top}\beta
+\lambda_{0i}^{\top}f+\epsilon_{0i}\,,\\
{y'}_{0i}^{\mathrm m}
=
\alpha+{{x'}_{0i}^{\mathrm m}}^{\top}\beta
+\lambda_{0i}^{\top}f'+\epsilon'_{0i}\,,
\end{cases} \qquad \text{for each }z_{0i}=[x_{0i},y_{0i}]\in Z_0\,,
\]
\[
\begin{cases}
y_{1i}
=
\alpha+x_{1i}^{\top}\beta
+\lambda_{1i}^{\top}f+\tau+\epsilon_{1i}\,,\\
{y'}_{1i}^{\mathrm m}
=
\alpha+{{x'}_{1i}^{\mathrm m}}^{\top}\beta
+\lambda_{1i}^{\top}f'+\epsilon'_{0i}\,,
\end{cases} \qquad \text{for each }z_{1i}=[x_{1i},y_{1i}]\in Z_1\,,
\]
Where factors $f=(f_1,\dots,f_r)$ and $f'=(f'_1,\dots,f'_r)$ represent the source and target membership, respectively. The estimation of parameters  are according to
\begin{align*}
[\widehat\alpha,\widehat\beta,\widehat f,\widehat f',\widehat\Lambda_0]
&=
\argmin_{[\alpha,\beta, f, f',\Lambda_0]}
\sum_{z_{0i}\in Z_0}
\left\|
\begin{bmatrix}
y_{0i}\\
{y'}_{0i}^{\mathrm m}
\end{bmatrix}
-
\begin{bmatrix}
\alpha+\mathbf{x}_{0i}^\top\beta+\lambda_{0i}^\top f\\
\alpha+{{\mathbf{x}'}_{0i}^{\mathrm m}}^\top\beta+\lambda_{0i}^\top f'
\end{bmatrix}
\right\|_2^2,
\\
[\widehat\Lambda_1,\widehat\tau]
&=
\argmin_{[\Lambda_1,\tau]}
\sum_{z_{1i}\in Z_1}
\left(
y_{1i}
-(\widehat\alpha
+\mathbf{x}_{1i}^\top\widehat\beta
+\lambda_{1i}^\top\widehat f
+\tau)
\right)^2,\\ &
\qquad\qquad\qquad
\text{subject to }
\sum_{z_{1i}\in Z_1}\lambda_{1i}=0,\quad
\lambda_{1i}\in\operatorname{span}(\widehat f),
\\
\intertext{which has the closed-form solution}
\widehat\tau
&=
\frac{1}{|Z_1|}
\sum_{z_{1i}\in Z_1}
\left(
y_{1i}
-(\widehat\alpha
+\mathbf{x}_{1i}^\top\widehat\beta
)\right),
\\
\widehat\lambda_{1i}
&=
\frac{
y_{1i}
-(\widehat\alpha
+\mathbf{x}_{1i}^\top\widehat\beta
+
\widehat\tau)
}{
\widehat f^\top\widehat f
}
\,\widehat f,
\intertext{and the synthetic responses are then computed as}
y_{1i}^{\mathrm{synth}}
&=
y_{1i}
+
\widehat\beta^\top
\left(
{\mathbf{x}'}_{1i}^{\mathrm m}
-\mathbf{x}_{1i}
\right)
+
\widehat\lambda_{1i}^\top
(\widehat f-\widehat f').
\end{align*}
We use $r=2$ in the numerical experiments.

Similar to \model{Matching}, the restricted version of \model{GenSynth} uses the anchor-aware feature respresentation 
$$R(x) = \left(\frac{x_{D,1}-\mu_{D,1}}{\sigma_{D,1}},\dots,\frac{x_{D,|X_D|}-\mu_{D,|X_D|}}{\sigma_{D,|X_D|}},\widehat G_{Q,1}(x_{Q,1}),\ldots,
\widehat G_{Q,|X_Q|}(x_{Q,|X_Q|})\right)\,.$$

\paragraph{CycleGAN.}
The \model{CycleGAN} baseline adapts the framework of \cite{zhu2017unpaired} to learn mappings between the source and target control distributions. It consists of two generators,
\[
\Gamma:\mathcal Z\to\mathcal Z',
\qquad
\Phi:\mathcal Z'\to\mathcal Z,
\]
mapping source control sample $Z_0$ to the target space, and target control sample $Z'_0$ to the source space, respectively. Additionally, there are two discriminators \(D_\Gamma\) and \(D_\Phi\). All coordinates of \(Z_0\) and \(Z'_0\) are standardized separately using the corresponding control samples. The generators operate jointly on the complete \((X,Y)\) vector.

We use least-squares adversarial losses. Writing \(z\sim Z_0\) and \(z'\sim Z'_0\), the discriminator objective is
\[
\begin{aligned}
\mathcal L_{\mathrm{adv},D}
=
\frac{1}{2}\Big[
&\E\{D_\Gamma(z')-1\}^2
+\E D_\Gamma\{\Gamma(z)\}^2\\
&+\E\{D_\Phi(z)-1\}^2
+\E D_\Phi\{\Phi(z')\}^2
\Big],
\end{aligned}
\]
whereas the generator adversarial loss is
\[
\mathcal L_{\mathrm{adv},G}
=
\E\{D_\Gamma(\Gamma(z))-1\}^2
+
\E\{D_\Phi(\Phi(z'))-1\}^2.
\]
Cycle consistency is imposed through
\[
\mathcal L_{\mathrm{cycle}}
=
\E\left\|\Phi\{\Gamma(z)\}-z\right\|_2^2
+
\E\left\|\Gamma\{\Phi(z')\}-z'\right\|_2^2,
\]
and we add an identity loss
\[
\mathcal L_{\mathrm{id}}
=
\E\left\|\Gamma(z')-z'\right\|_2^2
+
\E\left\|\Phi(z)-z\right\|_2^2.
\]
Accordingly, the generators are trained to minimize
\[
\mathcal L_G
=
\mathcal L_{\mathrm{adv},G}
+
c_1\mathcal L_{\mathrm{cycle}}
+
c_2\mathcal L_{\mathrm{id}},
\]
while the discriminators minimize \(\mathcal L_{\mathrm{adv},D}\). In the numerical experiments, we set $c_1=10$ and $c_2=5$. Each generator is a residual neural network with two hidden layers of width \(128\), using layer normalization and LeakyReLU activations. Each discriminator has two hidden layers of width \(128\) with LeakyReLU activations and spectral normalization. The networks are trained with Adam for \(300\) epochs. After training, the source-to-target generator is applied to the standardized source-treated sample and its output is transformed back to the target scale, yielding
\[
{Z'_1}^{\mathrm{synth}}=\Gamma(Z_1).
\]

\subsection{Additional Simulation Results}
\label{app:additional_results}

We supplement the performance measures with additional diagnostics of the outcome distribution, feature distribution, and feature--outcome dependence. As in the main text, all comparisons are between the synthetic target-treated sample and the corresponding oracle target-treated sample, and smaller values indicate better recovery. The reported entries are Monte Carlo means over \(R=50\) replications, with standard errors in parentheses.

\paragraph{Additional outcome-distribution metrics.}
For
\[
\tau\in\{.01,.05,.25,.50,.75,.95,.99\},
\]
we report the standardized quantile error
\[
E_{q_\tau}
=
\frac{
\left|
Q_\tau\!\left(Y_1^{\prime\mathrm{(synth.)}}\right)-Q_{\tau}\!\left(Y_1^{\prime\mathrm{(oracle)}}\right)
\right|
}{\operatorname{sd}\!\left(Y_1^{\prime\mathrm{(oracle)}}\right)}.
\]
We additionally report the Kolmogorov--Smirnov distance \[ D_{\mathrm{KS}} = \sup_y \left| F_{\mathrm{synth}}(y)-F_{\mathrm{oracle}}(y) \right|, \] the Hellinger distance \[ D_{\mathrm{H}} = \left\{ \frac{1}{2} \int \left( \sqrt{p_{\mathrm{synth}}(y)} - \sqrt{p_{\mathrm{oracle}}(y)} \right)^2dy \right\}^{1/2}, \] and the Jensen--Shannon divergence \[ D_{\mathrm{JS}} = \frac{1}{2} D_{\mathrm{KL}}(P_{\mathrm{synth}}\|M) + \frac{1}{2} D_{\mathrm{KL}}(P_{\mathrm{oracle}}\|M), \qquad M=\frac{1}{2}(P_{\mathrm{synth}}+P_{\mathrm{oracle}})\,, \]
where \(F\), \(p\), and \(P\) denote the corresponding CDF, density, and probability distribution, respectively.

\paragraph{Additional feature-distribution metrics.}
Let
\(\widetilde X'_{1,Q}, \widetilde X'_{1,U}\) denote the corresponding coordinates after standardization by the coordinate-wise mean and standard deviation of the oracle target-treated sample. In addition to the aggregate feature metrics reported in the main text, we separately report the sliced 1-Wasserstein distance for the quantile-comparable and non-comparable feature blocks,
\[
SW_1(X_B)
=
\E_{\theta\sim\mathrm{Unif}(\mathbb S^{p_B-1})}
W_1\!\left(
\theta^\top\widetilde X_{1,B}^{\prime\mathrm{(synth.)}},
\theta^\top\widetilde X_{1,B}^{\prime\mathrm{(oracle)}}
\right),
\qquad B\in\{Q,U\},
\]
where the expectation is approximated using random projections. We also report discrepancies in the first two moments,
\[
E_{\mu,B}
=
\frac{1}{\sqrt{p_B}}
\left\|
\overline{\widetilde X}_{1,B}^{\prime\mathrm{(synth.)}}
-
\overline{\widetilde X}_{1,B}^{\prime\mathrm{(oracle)}}
\right\|_2,
\]
and
\[
E_{\Sigma,B}
=
\frac{1}{p_B}
\left\|
\widehat\Sigma_{B}^{\mathrm{(synth.)}}
-
\widehat\Sigma_{B}^{\mathrm{(oracle)}}
\right\|_F,
\qquad B\in\{Q,U\},
\]where \(p_B\) denotes the dimension of the corresponding feature block.

Finally, for each of the four non-comparable features, $X_U=($severity, comorbidity, acuity, respiratory$)$, we report the coordinate-wise 1-Wasserstein distance between its standardized synthetic and oracle distributions. 

\paragraph{Additional dependence metrics.}
To distinguish dependence recovery across the two feature blocks, we report the feature--outcome correlation errors separately for \(X_Q\) and \(X_U\). For \(B\in\{Q,U\}\), 
\[
E_{\mathrm{corr},B}
=
\frac{1}{\sqrt{p_B}}
\left\|
\operatorname{Corr}
\!\left(
Y_1^{\prime\mathrm{(synth.)}},
X_{1,B}^{\prime\mathrm{(synth.)}}
\right)
-
\operatorname{Corr}
\!\left(
Y_1^{\prime\mathrm{(oracle)}},
X_{1,B}^{\prime\mathrm{(oracle)}}
\right)
\right\|_2.
\]
We additionally report the Spearman correlation error
\[\begin{aligned}
E_{\mathrm{Sp},B}
&=
\frac{1}{\sqrt{p_B}}
\Big\|
\operatorname{Corr}
\!\left(
\operatorname{rank}\!\left(Y_1^{\prime\mathrm{(synth.)}}\right),
\operatorname{rank}\!\left(X_{1,B}^{\prime\mathrm{(synth.)}}\right)
\right)\\ &\qquad\qquad\qquad
-
\operatorname{Corr}
\!\left(
\operatorname{rank}\!\left(Y_1^{\prime\mathrm{(oracle)}}\right),
\operatorname{rank}\!\left(X_{1,B}^{\prime\mathrm{(oracle)}}\right)
\right)
\Big\|_2,
\end{aligned}
\]
where ranks are computed coordinate-wise for the feature vectors. The former measures recovery of linear feature--outcome dependence, whereas the latter measures recovery of rank-based monotone dependence.
\newpage

\definecolor{Yshade}{RGB}{244,248,244}
\definecolor{Xshade}{RGB}{240,242,248}
\definecolor{Dshade}{RGB}{250,248,238}

\newcommand{\estold}[2]{%
\begingroup
\renewcommand{\arraystretch}{0.75}%
\begin{tabular}[c]{@{}c@{}}
#1\\[-2.5pt]
{\scriptsize\color{black!55}#2}
\end{tabular}%
\endgroup
}

\begin{table}[H]
\centering
\captionsetup{justification=centering}

\caption{Additional diagnostics for the baseline design.}
\label{tab:dgp_baseline_additional_1}

\setlength{\tabcolsep}{3.2pt}
\renewcommand{\arraystretch}{1.55}

\begin{adjustbox}{max width=\textwidth}
\begin{tabular}{
ll
*{10}{>{\columncolor{Yshade}}c}
*{2}{>{\columncolor{Xshade}}c}
}
\toprule
&
&
\multicolumn{10}{c}{\cellcolor{Yshade}Outcome \(Y\)}
&
\multicolumn{2}{c}{\cellcolor{Xshade}Features \(X\)}
\\
\cmidrule(lr){3-12}
\cmidrule(lr){13-14}

DGP type
& Model
& \(E_{q_{.01}}\)
& \(E_{q_{.05}}\)
& \(E_{q_{.25}}\)
& \(E_{q_{.50}}\)
& \(E_{q_{.75}}\)
& \(E_{q_{.95}}\)
& \(E_{q_{.99}}\)
& \(D_{\mathrm{KS}}\)
& \(D_{\mathrm{H}}\)
& \(D_{\mathrm{JS}}\)
& \(SW_1(X_Q)\)
& \(SW_1(X_U)\)
\\
\midrule

\multirow{5}{*}{Linear}
& TWFE
& \estold{0.0992}{(0.0109)}
& \estold{0.1503}{(0.0078)}
& \estold{0.2112}{(0.0059)}
& \estold{0.1669}{(0.0083)}
& \estold{\textbf{0.0752}}{(0.0072)}
& \estold{0.6892}{(0.0202)}
& \estold{1.6863}{(0.0620)}
& \estold{0.1519}{(0.0038)}
& \estold{0.2151}{(0.0031)}
& \estold{0.0414}{(0.0013)}
& \estold{\textbf{0.0585}}{(0.0010)}
& \estold{\textbf{0.0590}}{(0.0018)}
\\

& Matching
& \estold{\textbf{0.0545}}{(0.0069)}
& \estold{0.1177}{(0.0065)}
& \estold{0.2871}{(0.0064)}
& \estold{0.3924}{(0.0104)}
& \estold{0.5041}{(0.0162)}
& \estold{0.7608}{(0.0459)}
& \estold{1.1964}{(0.1075)}
& \estold{0.2046}{(0.0037)}
& \estold{0.2035}{(0.0025)}
& \estold{0.0381}{(0.0009)}
& \estold{\textbf{0.0585}}{(0.0010)}
& \estold{\textbf{0.0590}}{(0.0018)}
\\

& GenSynth
& \estold{1.4551}{(0.0263)}
& \estold{1.0556}{(0.0178)}
& \estold{0.5604}{(0.0119)}
& \estold{0.2534}{(0.0088)}
& \estold{0.1108}{(0.0105)}
& \estold{0.9842}{(0.0388)}
& \estold{2.1321}{(0.1082)}
& \estold{0.2720}{(0.0032)}
& \estold{0.3628}{(0.0032)}
& \estold{0.1057}{(0.0018)}
& \estold{0.2299}{(0.0019)}
& \estold{0.3786}{(0.0037)}
\\

& CycleGAN
& \estold{0.0561}{(0.0071)}
& \estold{\textbf{0.0368}}{(0.0052)}
& \estold{\textbf{0.0386}}{(0.0039)}
& \estold{\textbf{0.0693}}{(0.0072)}
& \estold{0.1476}{(0.0154)}
& \estold{0.3466}{(0.0365)}
& \estold{0.6881}{(0.0726)}
& \estold{\textbf{0.0621}}{(0.0033)}
& \estold{\textbf{0.1304}}{(0.0023)}
& \estold{\textbf{0.0147}}{(0.0005)}
& \estold{0.0740}{(0.0016)}
& \estold{0.0906}{(0.0031)}
\\

& \textbf{OTSynth}
& \estold{0.2363}{(0.0210)}
& \estold{0.0696}{(0.0070)}
& \estold{0.0672}{(0.0070)}
& \estold{0.0943}{(0.0085)}
& \estold{0.1095}{(0.0115)}
& \estold{\textbf{0.1859}}{(0.0208)}
& \estold{\textbf{0.4338}}{(0.0569)}
& \estold{0.0732}{(0.0038)}
& \estold{0.1447}{(0.0019)}
& \estold{0.0179}{(0.0005)}
& \estold{0.0712}{(0.0016)}
& \estold{0.0929}{(0.0028)}
\\

\addlinespace[2pt]
\midrule
\addlinespace[2pt]

\multirow{5}{*}{Nonlinear}
& TWFE
& \estold{0.5949}{(0.0227)}
& \estold{0.3889}{(0.0148)}
& \estold{0.1329}{(0.0075)}
& \estold{0.0348}{(0.0045)}
& \estold{0.0725}{(0.0080)}
& \estold{0.1408}{(0.0178)}
& \estold{1.6253}{(0.1071)}
& \estold{0.1384}{(0.0039)}
& \estold{0.2368}{(0.0051)}
& \estold{0.0474}{(0.0019)}
& \estold{\textbf{0.0591}}{(0.0010)}
& \estold{\textbf{0.0610}}{(0.0018)}
\\

& Matching
& \estold{1.3621}{(0.0436)}
& \estold{1.1106}{(0.0363)}
& \estold{0.7536}{(0.0241)}
& \estold{0.5421}{(0.0178)}
& \estold{0.4967}{(0.0180)}
& \estold{0.2939}{(0.0236)}
& \estold{1.4761}{(0.1083)}
& \estold{0.6751}{(0.0026)}
& \estold{0.6039}{(0.0046)}
& \estold{0.2942}{(0.0044)}
& \estold{\textbf{0.0591}}{(0.0010)}
& \estold{\textbf{0.0610}}{(0.0018)}
\\

& GenSynth
& \estold{0.2956}{(0.0163)}
& \estold{0.1838}{(0.0105)}
& \estold{\textbf{0.0341}}{(0.0037)}
& \estold{0.1090}{(0.0059)}
& \estold{\textbf{0.0569}}{(0.0059)}
& \estold{\textbf{0.1108}}{(0.0107)}
& \estold{1.4300}{(0.1067)}
& \estold{0.1066}{(0.0033)}
& \estold{0.1669}{(0.0039)}
& \estold{0.0254}{(0.0012)}
& \estold{0.3040}{(0.0015)}
& \estold{0.4920}{(0.0022)}
\\

& CycleGAN
& \estold{0.2558}{(0.0205)}
& \estold{0.1120}{(0.0098)}
& \estold{0.0897}{(0.0072)}
& \estold{0.1162}{(0.0071)}
& \estold{0.1857}{(0.0160)}
& \estold{0.9036}{(0.0529)}
& \estold{0.8122}{(0.0843)}
& \estold{0.1207}{(0.0031)}
& \estold{0.2129}{(0.0039)}
& \estold{0.0416}{(0.0014)}
& \estold{0.0780}{(0.0017)}
& \estold{0.1321}{(0.0024)}
\\

& \textbf{OTSynth}
& \estold{\textbf{0.1220}}{(0.0136)}
& \estold{\textbf{0.0819}}{(0.0081)}
& \estold{0.0499}{(0.0057)}
& \estold{\textbf{0.0319}}{(0.0038)}
& \estold{0.1066}{(0.0105)}
& \estold{0.2986}{(0.0307)}
& \estold{\textbf{0.8002}}{(0.0929)}
& \estold{\textbf{0.0901}}{(0.0042)}
& \estold{\textbf{0.1375}}{(0.0040)}
& \estold{\textbf{0.0171}}{(0.0010)}
& \estold{0.0709}{(0.0016)}
& \estold{0.1025}{(0.0033)}
\\

\bottomrule
\end{tabular}
\end{adjustbox}
\end{table}

\vspace{-6mm}

\begin{table}[H]
\centering
\setlength{\tabcolsep}{3.8pt}
\renewcommand{\arraystretch}{1.55}

\begin{adjustbox}{max width=\textwidth}
\begin{tabular}{
ll
*{8}{>{\columncolor{Xshade}}c}
*{4}{>{\columncolor{Dshade}}c}
}
\toprule
&
&
\multicolumn{8}{c}{\cellcolor{Xshade}Features \(X\)}
&
\multicolumn{4}{c}{\cellcolor{Dshade}Dependence}
\\
\cmidrule(lr){3-10}
\cmidrule(lr){11-14}

DGP type
& Model
& \(E_{\mu,Q}\)
& \(E_{\mu,U}\)
& \(E_{\Sigma,Q}\)
& \(E_{\Sigma,U}\)
& \shortstack{\(W_1\)\\Agg.\ Sev.}
& \shortstack{\(W_1\)\\Comorb.}
& \shortstack{\(W_1\)\\Acuity}
& \shortstack{\(W_1\)\\Resp.}
& \(E_{\mathrm{corr},Q}\)
& \(E_{\mathrm{corr},U}\)
& \(E_{\mathrm{Sp},Q}\)
& \(E_{\mathrm{Sp},U}\)
\\
\midrule

\multirow{5}{*}{Linear}
& TWFE
& \estold{\textbf{0.0439}}{(0.0016)}
& \estold{\textbf{0.0444}}{(0.0028)}
& \estold{\textbf{0.0527}}{(0.0011)}
& \estold{\textbf{0.0522}}{(0.0017)}
& \estold{\textbf{0.0574}}{(0.0035)}
& \estold{\textbf{0.0541}}{(0.0027)}
& \estold{\textbf{0.0586}}{(0.0036)}
& \estold{\textbf{0.0560}}{(0.0030)}
& \estold{0.1303}{(0.0022)}
& \estold{0.2006}{(0.0048)}
& \estold{0.0968}{(0.0025)}
& \estold{0.1207}{(0.0041)}
\\

& Matching
& \estold{\textbf{0.0439}}{(0.0016)}
& \estold{\textbf{0.0444}}{(0.0028)}
& \estold{\textbf{0.0527}}{(0.0011)}
& \estold{\textbf{0.0522}}{(0.0017)}
& \estold{\textbf{0.0574}}{(0.0035)}
& \estold{\textbf{0.0541}}{(0.0027)}
& \estold{\textbf{0.0586}}{(0.0036)}
& \estold{\textbf{0.0560}}{(0.0030)}
& \estold{0.0641}{(0.0023)}
& \estold{\textbf{0.0604}}{(0.0031)}
& \estold{0.0623}{(0.0020)}
& \estold{\textbf{0.0601}}{(0.0029)}
\\

& GenSynth
& \estold{0.2032}{(0.0025)}
& \estold{0.3687}{(0.0040)}
& \estold{0.1561}{(0.0009)}
& \estold{0.2924}{(0.0020)}
& \estold{0.6650}{(0.0064)}
& \estold{0.4142}{(0.0031)}
& \estold{0.2194}{(0.0038)}
& \estold{0.1537}{(0.0041)}
& \estold{0.1559}{(0.0023)}
& \estold{0.1608}{(0.0027)}
& \estold{0.1630}{(0.0025)}
& \estold{0.1410}{(0.0021)}
\\

& CycleGAN
& \estold{0.0644}{(0.0023)}
& \estold{0.0653}{(0.0043)}
& \estold{0.0700}{(0.0019)}
& \estold{0.1191}{(0.0028)}
& \estold{0.0950}{(0.0053)}
& \estold{0.1057}{(0.0040)}
& \estold{0.1181}{(0.0049)}
& \estold{0.0838}{(0.0032)}
& \estold{\textbf{0.0560}}{(0.0019)}
& \estold{0.0842}{(0.0031)}
& \estold{0.0637}{(0.0023)}
& \estold{0.1026}{(0.0044)}
\\

& \textbf{OTSynth}
& \estold{0.0618}{(0.0023)}
& \estold{0.0640}{(0.0037)}
& \estold{0.0655}{(0.0024)}
& \estold{0.1063}{(0.0041)}
& \estold{0.0810}{(0.0047)}
& \estold{0.0873}{(0.0044)}
& \estold{0.1270}{(0.0050)}
& \estold{0.0965}{(0.0046)}
& \estold{0.0656}{(0.0021)}
& \estold{0.0690}{(0.0036)}
& \estold{\textbf{0.0571}}{(0.0020)}
& \estold{0.0644}{(0.0036)}
\\

\addlinespace[2pt]
\midrule
\addlinespace[2pt]

\multirow{5}{*}{Nonlinear}
& TWFE
& \estold{\textbf{0.0439}}{(0.0016)}
& \estold{\textbf{0.0428}}{(0.0024)}
& \estold{\textbf{0.0578}}{(0.0019)}
& \estold{\textbf{0.0861}}{(0.0032)}
& \estold{\textbf{0.0587}}{(0.0033)}
& \estold{\textbf{0.0501}}{(0.0022)}
& \estold{\textbf{0.0567}}{(0.0030)}
& \estold{\textbf{0.0544}}{(0.0025)}
& \estold{0.1839}{(0.0039)}
& \estold{0.2940}{(0.0064)}
& \estold{0.1425}{(0.0027)}
& \estold{0.1728}{(0.0052)}
\\

& Matching
& \estold{\textbf{0.0439}}{(0.0016)}
& \estold{\textbf{0.0428}}{(0.0024)}
& \estold{\textbf{0.0578}}{(0.0019)}
& \estold{\textbf{0.0861}}{(0.0032)}
& \estold{\textbf{0.0587}}{(0.0033)}
& \estold{\textbf{0.0501}}{(0.0022)}
& \estold{\textbf{0.0567}}{(0.0030)}
& \estold{\textbf{0.0544}}{(0.0025)}
& \estold{0.1029}{(0.0035)}
& \estold{0.1067}{(0.0059)}
& \estold{0.0745}{(0.0020)}
& \estold{\textbf{0.0561}}{(0.0031)}
\\

& GenSynth
& \estold{0.2312}{(0.0021)}
& \estold{0.4275}{(0.0023)}
& \estold{0.2086}{(0.0009)}
& \estold{0.4499}{(0.0013)}
& \estold{0.6083}{(0.0046)}
& \estold{0.4723}{(0.0021)}
& \estold{0.5024}{(0.0035)}
& \estold{0.1562}{(0.0027)}
& \estold{0.1632}{(0.0031)}
& \estold{0.1669}{(0.0039)}
& \estold{0.1989}{(0.0032)}
& \estold{0.1914}{(0.0045)}
\\

& CycleGAN
& \estold{0.0677}{(0.0025)}
& \estold{0.0650}{(0.0036)}
& \estold{0.0781}{(0.0025)}
& \estold{0.1667}{(0.0045)}
& \estold{0.1317}{(0.0042)}
& \estold{0.2021}{(0.0039)}
& \estold{0.1362}{(0.0045)}
& \estold{0.1845}{(0.0038)}
& \estold{\textbf{0.0752}}{(0.0029)}
& \estold{0.1857}{(0.0060)}
& \estold{0.0690}{(0.0022)}
& \estold{0.0892}{(0.0026)}
\\

& \textbf{OTSynth}
& \estold{0.0614}{(0.0022)}
& \estold{0.0619}{(0.0040)}
& \estold{0.0649}{(0.0022)}
& \estold{0.1296}{(0.0058)}
& \estold{0.1104}{(0.0056)}
& \estold{0.1275}{(0.0041)}
& \estold{0.1334}{(0.0057)}
& \estold{0.1375}{(0.0046)}
& \estold{\textbf{0.0752}}{(0.0034)}
& \estold{\textbf{0.0926}}{(0.0065)}
& \estold{\textbf{0.0616}}{(0.0022)}
& \estold{0.0641}{(0.0040)}
\\

\bottomrule
\end{tabular}
\end{adjustbox}

\vspace{2mm}

\begin{minipage}{0.98\textwidth}
\centering
\footnotesize
Entries report means over \(R=50\) Monte Carlo runs, with standard errors in parentheses.
\end{minipage}

\end{table}

\begin{table}[H]
\centering
\captionsetup{justification=centering}

\caption{Additional diagnostics for the Observational treatment assignment design.}
\label{tab:dgp_observational_additional_1}

\setlength{\tabcolsep}{3.2pt}
\renewcommand{\arraystretch}{1.55}

\begin{adjustbox}{max width=\textwidth}
\begin{tabular}{
ll
*{10}{>{\columncolor{Yshade}}c}
*{2}{>{\columncolor{Xshade}}c}
}
\toprule
&
&
\multicolumn{10}{c}{\cellcolor{Yshade}Outcome \(Y\)}
&
\multicolumn{2}{c}{\cellcolor{Xshade}Features \(X\)}
\\
\cmidrule(lr){3-12}
\cmidrule(lr){13-14}

DGP type
& Model
& \(E_{q_{.01}}\)
& \(E_{q_{.05}}\)
& \(E_{q_{.25}}\)
& \(E_{q_{.50}}\)
& \(E_{q_{.75}}\)
& \(E_{q_{.95}}\)
& \(E_{q_{.99}}\)
& \(D_{\mathrm{KS}}\)
& \(D_{\mathrm{H}}\)
& \(D_{\mathrm{JS}}\)
& \(SW_1(X_Q)\)
& \(SW_1(X_U)\)
\\
\midrule

\multirow{5}{*}{Linear}
& TWFE
& \estold{0.1320}{(0.0133)}
& \estold{0.0785}{(0.0067)}
& \estold{0.0691}{(0.0060)}
& \estold{\textbf{0.0542}}{(0.0047)}
& \estold{0.2980}{(0.0084)}
& \estold{1.1234}{(0.0210)}
& \estold{2.2700}{(0.0398)}
& \estold{0.1315}{(0.0026)}
& \estold{0.2273}{(0.0031)}
& \estold{0.0434}{(0.0011)}
& \estold{0.1926}{(0.0018)}
& \estold{0.2063}{(0.0036)}
\\

& Matching
& \estold{0.0670}{(0.0077)}
& \estold{\textbf{0.0314}}{(0.0038)}
& \estold{0.1419}{(0.0061)}
& \estold{0.1867}{(0.0099)}
& \estold{0.1941}{(0.0136)}
& \estold{\textbf{0.2704}}{(0.0304)}
& \estold{0.5532}{(0.0851)}
& \estold{0.1132}{(0.0041)}
& \estold{0.1538}{(0.0026)}
& \estold{0.0211}{(0.0007)}
& \estold{0.1926}{(0.0018)}
& \estold{0.2063}{(0.0036)}
\\

& GenSynth
& \estold{1.3163}{(0.0198)}
& \estold{0.9903}{(0.0127)}
& \estold{0.5462}{(0.0080)}
& \estold{0.2660}{(0.0092)}
& \estold{\textbf{0.0940}}{(0.0103)}
& \estold{0.9992}{(0.0378)}
& \estold{2.1003}{(0.0932)}
& \estold{0.2738}{(0.0030)}
& \estold{0.3621}{(0.0027)}
& \estold{0.1057}{(0.0015)}
& \estold{0.2572}{(0.0019)}
& \estold{0.4040}{(0.0033)}
\\

& CycleGAN
& \estold{\textbf{0.0473}}{(0.0048)}
& \estold{0.0349}{(0.0034)}
& \estold{\textbf{0.0425}}{(0.0043)}
& \estold{0.0827}{(0.0084)}
& \estold{0.1440}{(0.0130)}
& \estold{0.3102}{(0.0319)}
& \estold{\textbf{0.5185}}{(0.0606)}
& \estold{\textbf{0.0619}}{(0.0034)}
& \estold{\textbf{0.1296}}{(0.0019)}
& \estold{\textbf{0.0144}}{(0.0004)}
& \estold{0.0776}{(0.0017)}
& \estold{\textbf{0.1092}}{(0.0038)}
\\

& \textbf{OTSynth}
& \estold{0.2576}{(0.0193)}
& \estold{0.0863}{(0.0087)}
& \estold{0.0553}{(0.0060)}
& \estold{0.0959}{(0.0096)}
& \estold{0.1120}{(0.0143)}
& \estold{0.3088}{(0.0324)}
& \estold{0.6497}{(0.1042)}
& \estold{0.0714}{(0.0041)}
& \estold{0.1530}{(0.0020)}
& \estold{0.0198}{(0.0005)}
& \estold{\textbf{0.0760}}{(0.0019)}
& \estold{0.1203}{(0.0039)}
\\

\addlinespace[2pt]
\midrule
\addlinespace[2pt]

\multirow{5}{*}{Nonlinear}
& TWFE
& \estold{0.4192}{(0.0187)}
& \estold{0.2330}{(0.0106)}
& \estold{\textbf{0.0249}}{(0.0028)}
& \estold{0.1011}{(0.0043)}
& \estold{0.1566}{(0.0072)}
& \estold{0.3578}{(0.0263)}
& \estold{2.5302}{(0.0933)}
& \estold{0.1439}{(0.0040)}
& \estold{0.2320}{(0.0049)}
& \estold{0.0463}{(0.0019)}
& \estold{0.1900}{(0.0018)}
& \estold{0.2142}{(0.0037)}
\\

& Matching
& \estold{1.0900}{(0.0306)}
& \estold{0.8656}{(0.0235)}
& \estold{0.5552}{(0.0157)}
& \estold{0.4045}{(0.0120)}
& \estold{0.2792}{(0.0101)}
& \estold{\textbf{0.1090}}{(0.0112)}
& \estold{2.2094}{(0.0886)}
& \estold{0.6214}{(0.0026)}
& \estold{0.5583}{(0.0054)}
& \estold{0.2516}{(0.0051)}
& \estold{0.1900}{(0.0018)}
& \estold{0.2142}{(0.0037)}
\\

& GenSynth
& \estold{\textbf{0.1200}}{(0.0129)}
& \estold{\textbf{0.0538}}{(0.0068)}
& \estold{0.1194}{(0.0065)}
& \estold{0.1482}{(0.0061)}
& \estold{0.1176}{(0.0069)}
& \estold{0.1393}{(0.0134)}
& \estold{1.9636}{(0.0864)}
& \estold{0.1689}{(0.0058)}
& \estold{0.1758}{(0.0036)}
& \estold{0.0285}{(0.0012)}
& \estold{0.3362}{(0.0015)}
& \estold{0.5295}{(0.0024)}
\\

& CycleGAN
& \estold{0.2253}{(0.0169)}
& \estold{0.0939}{(0.0085)}
& \estold{0.0936}{(0.0062)}
& \estold{0.0658}{(0.0048)}
& \estold{0.1460}{(0.0153)}
& \estold{0.7148}{(0.0610)}
& \estold{\textbf{0.9348}}{(0.0944)}
& \estold{0.1241}{(0.0038)}
& \estold{0.2107}{(0.0039)}
& \estold{0.0408}{(0.0015)}
& \estold{0.0786}{(0.0017)}
& \estold{0.1617}{(0.0025)}
\\

& \textbf{OTSynth}
& \estold{0.1485}{(0.0226)}
& \estold{0.0639}{(0.0067)}
& \estold{0.0406}{(0.0042)}
& \estold{\textbf{0.0359}}{(0.0036)}
& \estold{\textbf{0.0656}}{(0.0076)}
& \estold{0.3163}{(0.0376)}
& \estold{1.2584}{(0.1153)}
& \estold{\textbf{0.0898}}{(0.0033)}
& \estold{\textbf{0.1477}}{(0.0036)}
& \estold{\textbf{0.0194}}{(0.0009)}
& \estold{\textbf{0.0758}}{(0.0019)}
& \estold{\textbf{0.1178}}{(0.0036)}
\\

\bottomrule
\end{tabular}
\end{adjustbox}
\end{table}

\vspace{-6mm}

\begin{table}[H]
\centering
\setlength{\tabcolsep}{3.8pt}
\renewcommand{\arraystretch}{1.55}

\begin{adjustbox}{max width=\textwidth}
\begin{tabular}{
ll
*{8}{>{\columncolor{Xshade}}c}
*{4}{>{\columncolor{Dshade}}c}
}
\toprule
&
&
\multicolumn{8}{c}{\cellcolor{Xshade}Features \(X\)}
&
\multicolumn{4}{c}{\cellcolor{Dshade}Dependence}
\\
\cmidrule(lr){3-10}
\cmidrule(lr){11-14}

DGP type
& Model
& \(E_{\mu,Q}\)
& \(E_{\mu,U}\)
& \(E_{\Sigma,Q}\)
& \(E_{\Sigma,U}\)
& \shortstack{\(W_1\)\\Agg.\ Sev.}
& \shortstack{\(W_1\)\\Comorb.}
& \shortstack{\(W_1\)\\Acuity}
& \shortstack{\(W_1\)\\Resp.}
& \(E_{\mathrm{corr},Q}\)
& \(E_{\mathrm{corr},U}\)
& \(E_{\mathrm{Sp},Q}\)
& \(E_{\mathrm{Sp},U}\)
\\
\midrule

\multirow{5}{*}{Linear}
& TWFE
& \estold{0.2165}{(0.0021)}
& \estold{0.2123}{(0.0035)}
& \estold{0.1120}{(0.0010)}
& \estold{\textbf{0.0929}}{(0.0024)}
& \estold{0.3789}{(0.0065)}
& \estold{\textbf{0.0711}}{(0.0047)}
& \estold{0.1729}{(0.0075)}
& \estold{\textbf{0.0767}}{(0.0043)}
& \estold{0.1359}{(0.0022)}
& \estold{0.2280}{(0.0050)}
& \estold{0.0996}{(0.0022)}
& \estold{0.1453}{(0.0048)}
\\

& Matching
& \estold{0.2165}{(0.0021)}
& \estold{0.2123}{(0.0035)}
& \estold{0.1120}{(0.0010)}
& \estold{\textbf{0.0929}}{(0.0024)}
& \estold{0.3789}{(0.0065)}
& \estold{\textbf{0.0711}}{(0.0047)}
& \estold{0.1729}{(0.0075)}
& \estold{\textbf{0.0767}}{(0.0043)}
& \estold{0.0596}{(0.0024)}
& \estold{\textbf{0.0594}}{(0.0038)}
& \estold{\textbf{0.0585}}{(0.0023)}
& \estold{\textbf{0.0574}}{(0.0034)}
\\

& GenSynth
& \estold{0.2350}{(0.0024)}
& \estold{0.3994}{(0.0037)}
& \estold{0.1651}{(0.0011)}
& \estold{0.2908}{(0.0022)}
& \estold{0.7375}{(0.0071)}
& \estold{0.4085}{(0.0043)}
& \estold{0.2256}{(0.0047)}
& \estold{0.1369}{(0.0038)}
& \estold{0.1445}{(0.0021)}
& \estold{0.1545}{(0.0026)}
& \estold{0.1493}{(0.0025)}
& \estold{0.1313}{(0.0023)}
\\

& CycleGAN
& \estold{0.0694}{(0.0023)}
& \estold{\textbf{0.0858}}{(0.0054)}
& \estold{\textbf{0.0680}}{(0.0013)}
& \estold{0.1510}{(0.0038)}
& \estold{0.1184}{(0.0082)}
& \estold{0.1225}{(0.0052)}
& \estold{\textbf{0.1159}}{(0.0048)}
& \estold{0.0937}{(0.0038)}
& \estold{\textbf{0.0520}}{(0.0019)}
& \estold{0.0916}{(0.0033)}
& \estold{0.0596}{(0.0022)}
& \estold{0.0733}{(0.0025)}
\\

& \textbf{OTSynth}
& \estold{\textbf{0.0660}}{(0.0025)}
& \estold{0.0928}{(0.0044)}
& \estold{0.0708}{(0.0025)}
& \estold{0.1622}{(0.0120)}
& \estold{\textbf{0.1116}}{(0.0078)}
& \estold{0.1080}{(0.0054)}
& \estold{0.1514}{(0.0090)}
& \estold{0.1280}{(0.0066)}
& \estold{0.0780}{(0.0033)}
& \estold{0.0916}{(0.0047)}
& \estold{0.0688}{(0.0027)}
& \estold{0.0813}{(0.0043)}
\\

\addlinespace[2pt]
\midrule
\addlinespace[2pt]

\multirow{5}{*}{Nonlinear}
& TWFE
& \estold{0.2129}{(0.0021)}
& \estold{0.1983}{(0.0031)}
& \estold{0.1143}{(0.0012)}
& \estold{0.2097}{(0.0037)}
& \estold{0.3273}{(0.0059)}
& \estold{0.1537}{(0.0051)}
& \estold{\textbf{0.1348}}{(0.0074)}
& \estold{\textbf{0.1084}}{(0.0052)}
& \estold{0.1990}{(0.0032)}
& \estold{0.3133}{(0.0052)}
& \estold{0.1468}{(0.0021)}
& \estold{0.2018}{(0.0065)}
\\

& Matching
& \estold{0.2129}{(0.0021)}
& \estold{0.1983}{(0.0031)}
& \estold{0.1143}{(0.0012)}
& \estold{0.2097}{(0.0037)}
& \estold{0.3273}{(0.0059)}
& \estold{0.1537}{(0.0051)}
& \estold{\textbf{0.1348}}{(0.0074)}
& \estold{\textbf{0.1084}}{(0.0052)}
& \estold{0.1039}{(0.0030)}
& \estold{0.1131}{(0.0053)}
& \estold{0.0658}{(0.0021)}
& \estold{\textbf{0.0569}}{(0.0030)}
\\

& GenSynth
& \estold{0.2740}{(0.0020)}
& \estold{0.4588}{(0.0024)}
& \estold{0.2186}{(0.0009)}
& \estold{0.4558}{(0.0010)}
& \estold{0.6514}{(0.0062)}
& \estold{0.5107}{(0.0042)}
& \estold{0.5155}{(0.0065)}
& \estold{0.1681}{(0.0028)}
& \estold{0.1711}{(0.0031)}
& \estold{0.1792}{(0.0050)}
& \estold{0.2007}{(0.0033)}
& \estold{0.1895}{(0.0037)}
\\

& CycleGAN
& \estold{0.0690}{(0.0023)}
& \estold{\textbf{0.0672}}{(0.0044)}
& \estold{0.0709}{(0.0016)}
& \estold{0.2086}{(0.0030)}
& \estold{0.2006}{(0.0051)}
& \estold{0.1983}{(0.0037)}
& \estold{0.1660}{(0.0053)}
& \estold{0.2037}{(0.0037)}
& \estold{\textbf{0.0771}}{(0.0031)}
& \estold{0.2283}{(0.0055)}
& \estold{\textbf{0.0607}}{(0.0024)}
& \estold{0.1138}{(0.0036)}
\\

& \textbf{OTSynth}
& \estold{\textbf{0.0654}}{(0.0025)}
& \estold{0.0837}{(0.0046)}
& \estold{\textbf{0.0696}}{(0.0028)}
& \estold{\textbf{0.1776}}{(0.0117)}
& \estold{\textbf{0.1225}}{(0.0056)}
& \estold{\textbf{0.1412}}{(0.0046)}
& \estold{0.1688}{(0.0072)}
& \estold{0.1663}{(0.0071)}
& \estold{0.0848}{(0.0030)}
& \estold{\textbf{0.1110}}{(0.0063)}
& \estold{0.0735}{(0.0027)}
& \estold{0.0739}{(0.0041)}
\\

\bottomrule
\end{tabular}
\end{adjustbox}

\vspace{2mm}

\begin{minipage}{0.98\textwidth}
\centering
\footnotesize
Entries report means over \(R=50\) Monte Carlo runs, with standard errors in parentheses.
\end{minipage}

\end{table}

\begin{figure}[H]
\centering
\includegraphics[width=0.95\textwidth]{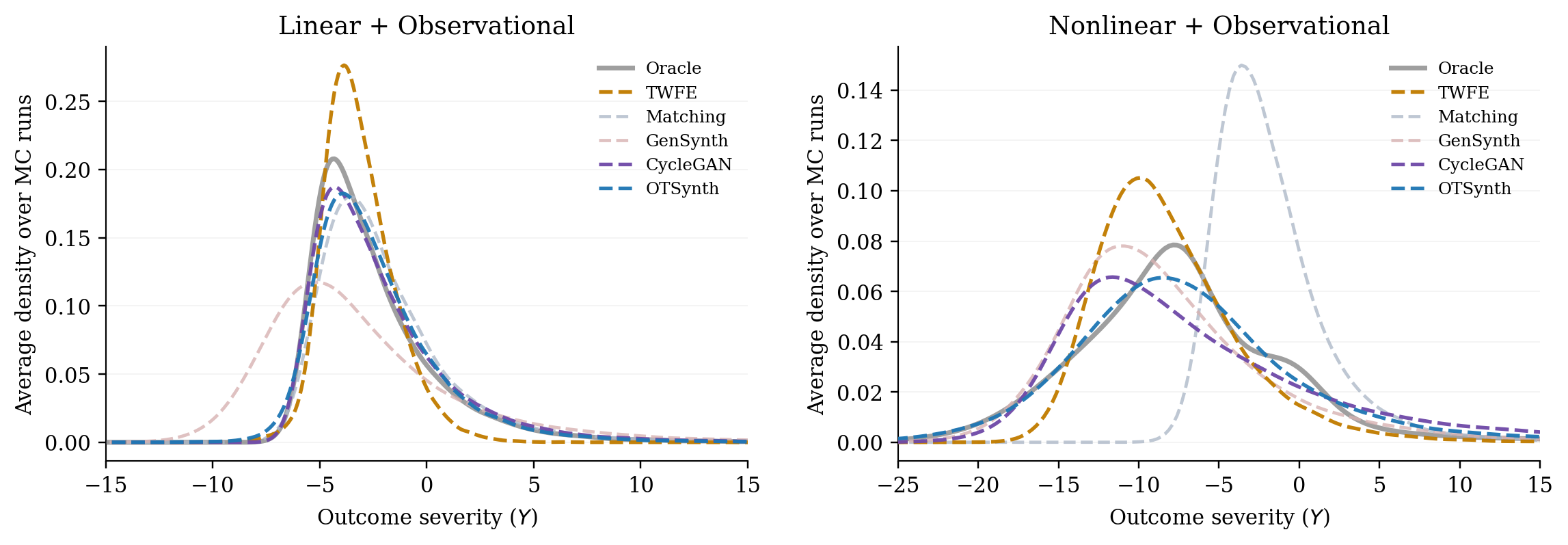}
\caption{Estimated outcome densities under the Observational treatment assignment design.}
\label{fig:y_density_obs}
\end{figure}

\newpage

\begin{table}[H]
\centering
\captionsetup{justification=centering}

\caption{Additional diagnostics for the Limited cross-site overlap design.}
\label{tab:dgp_overlap_additional_1}

\setlength{\tabcolsep}{3.2pt}
\renewcommand{\arraystretch}{1.55}

\begin{adjustbox}{max width=\textwidth}
\begin{tabular}{
ll
*{10}{>{\columncolor{Yshade}}c}
*{2}{>{\columncolor{Xshade}}c}
}
\toprule
&
&
\multicolumn{10}{c}{\cellcolor{Yshade}Outcome \(Y\)}
&
\multicolumn{2}{c}{\cellcolor{Xshade}Features \(X\)}
\\
\cmidrule(lr){3-12}
\cmidrule(lr){13-14}

DGP type
& Model
& \(E_{q_{.01}}\)
& \(E_{q_{.05}}\)
& \(E_{q_{.25}}\)
& \(E_{q_{.50}}\)
& \(E_{q_{.75}}\)
& \(E_{q_{.95}}\)
& \(E_{q_{.99}}\)
& \(D_{\mathrm{KS}}\)
& \(D_{\mathrm{H}}\)
& \(D_{\mathrm{JS}}\)
& \(SW_1(X_Q)\)
& \(SW_1(X_U)\)
\\
\midrule

\multirow{5}{*}{Linear}
& TWFE
& \estold{\textbf{0.0926}}{(0.0081)}
& \estold{0.2313}{(0.0076)}
& \estold{0.2785}{(0.0060)}
& \estold{0.2087}{(0.0074)}
& \estold{\textbf{0.0655}}{(0.0068)}
& \estold{0.7887}{(0.0234)}
& \estold{1.7963}{(0.0564)}
& \estold{0.1886}{(0.0040)}
& \estold{0.2532}{(0.0032)}
& \estold{0.0579}{(0.0015)}
& \estold{\textbf{0.0575}}{(0.0008)}
& \estold{\textbf{0.0560}}{(0.0016)}
\\

& Matching
& \estold{0.2165}{(0.0278)}
& \estold{0.1642}{(0.0177)}
& \estold{0.1633}{(0.0174)}
& \estold{0.1718}{(0.0206)}
& \estold{0.2121}{(0.0209)}
& \estold{0.5742}{(0.0421)}
& \estold{1.2396}{(0.0876)}
& \estold{0.1317}{(0.0085)}
& \estold{0.1987}{(0.0048)}
& \estold{0.0348}{(0.0018)}
& \estold{\textbf{0.0575}}{(0.0008)}
& \estold{\textbf{0.0560}}{(0.0016)}
\\

& GenSynth
& \estold{0.9864}{(0.0216)}
& \estold{0.6822}{(0.0146)}
& \estold{0.3436}{(0.0143)}
& \estold{0.2311}{(0.0147)}
& \estold{0.2194}{(0.0163)}
& \estold{0.2806}{(0.0231)}
& \estold{\textbf{0.4497}}{(0.0506)}
& \estold{0.2049}{(0.0057)}
& \estold{0.2933}{(0.0042)}
& \estold{0.0696}{(0.0018)}
& \estold{0.3885}{(0.0021)}
& \estold{0.6496}{(0.0036)}
\\

& CycleGAN
& \estold{0.1157}{(0.0091)}
& \estold{\textbf{0.0537}}{(0.0060)}
& \estold{\textbf{0.0335}}{(0.0038)}
& \estold{\textbf{0.0532}}{(0.0060)}
& \estold{0.1320}{(0.0136)}
& \estold{0.2664}{(0.0290)}
& \estold{0.5225}{(0.0576)}
& \estold{\textbf{0.0535}}{(0.0027)}
& \estold{\textbf{0.1361}}{(0.0019)}
& \estold{\textbf{0.0158}}{(0.0004)}
& \estold{0.0796}{(0.0015)}
& \estold{0.0939}{(0.0022)}
\\

& \textbf{OTSynth}
& \estold{0.2976}{(0.0237)}
& \estold{0.0886}{(0.0081)}
& \estold{0.0783}{(0.0059)}
& \estold{0.1099}{(0.0079)}
& \estold{0.1083}{(0.0108)}
& \estold{\textbf{0.2208}}{(0.0198)}
& \estold{0.5257}{(0.0598)}
& \estold{0.0771}{(0.0030)}
& \estold{0.1562}{(0.0025)}
& \estold{0.0210}{(0.0007)}
& \estold{0.0962}{(0.0017)}
& \estold{0.1032}{(0.0030)}
\\

\addlinespace[2pt]
\midrule
\addlinespace[2pt]

\multirow{5}{*}{Nonlinear}
& TWFE
& \estold{0.4374}{(0.0153)}
& \estold{0.2908}{(0.0105)}
& \estold{0.1067}{(0.0068)}
& \estold{0.1175}{(0.0078)}
& \estold{0.0729}{(0.0092)}
& \estold{\textbf{0.2187}}{(0.0234)}
& \estold{2.4751}{(0.1125)}
& \estold{0.1686}{(0.0058)}
& \estold{0.2434}{(0.0046)}
& \estold{0.0495}{(0.0018)}
& \estold{\textbf{0.0581}}{(0.0008)}
& \estold{\textbf{0.0603}}{(0.0017)}
\\

& Matching
& \estold{0.8796}{(0.0243)}
& \estold{0.6992}{(0.0191)}
& \estold{0.4228}{(0.0114)}
& \estold{0.3343}{(0.0091)}
& \estold{0.1267}{(0.0086)}
& \estold{0.3954}{(0.0218)}
& \estold{3.0408}{(0.0952)}
& \estold{0.5674}{(0.0044)}
& \estold{0.5262}{(0.0061)}
& \estold{0.2268}{(0.0053)}
& \estold{\textbf{0.0581}}{(0.0008)}
& \estold{\textbf{0.0603}}{(0.0017)}
\\

& GenSynth
& \estold{0.2583}{(0.0485)}
& \estold{0.1656}{(0.0322)}
& \estold{0.0745}{(0.0085)}
& \estold{0.0926}{(0.0107)}
& \estold{0.2095}{(0.0223)}
& \estold{0.4892}{(0.0323)}
& \estold{2.8295}{(0.1156)}
& \estold{0.1609}{(0.0086)}
& \estold{0.2150}{(0.0082)}
& \estold{0.0428}{(0.0040)}
& \estold{0.4173}{(0.0022)}
& \estold{0.6100}{(0.0023)}
\\

& CycleGAN
& \estold{0.1104}{(0.0154)}
& \estold{0.0962}{(0.0107)}
& \estold{0.1142}{(0.0069)}
& \estold{\textbf{0.0270}}{(0.0033)}
& \estold{0.2109}{(0.0182)}
& \estold{0.8003}{(0.0688)}
& \estold{\textbf{1.0199}}{(0.0887)}
& \estold{0.1422}{(0.0038)}
& \estold{0.2108}{(0.0046)}
& \estold{0.0420}{(0.0018)}
& \estold{0.0818}{(0.0016)}
& \estold{0.1456}{(0.0020)}
\\

& \textbf{OTSynth}
& \estold{\textbf{0.0826}}{(0.0075)}
& \estold{\textbf{0.0420}}{(0.0056)}
& \estold{\textbf{0.0313}}{(0.0049)}
& \estold{0.0656}{(0.0051)}
& \estold{\textbf{0.0682}}{(0.0076)}
& \estold{0.2837}{(0.0290)}
& \estold{1.2182}{(0.1232)}
& \estold{\textbf{0.1044}}{(0.0045)}
& \estold{\textbf{0.1388}}{(0.0033)}
& \estold{\textbf{0.0170}}{(0.0008)}
& \estold{0.0949}{(0.0018)}
& \estold{0.1060}{(0.0036)}
\\

\bottomrule
\end{tabular}
\end{adjustbox}
\end{table}

\vspace{-6mm}

\begin{table}[H]
\centering
\setlength{\tabcolsep}{3.8pt}
\renewcommand{\arraystretch}{1.55}

\begin{adjustbox}{max width=\textwidth}
\begin{tabular}{
ll
*{8}{>{\columncolor{Xshade}}c}
*{4}{>{\columncolor{Dshade}}c}
}
\toprule
&
&
\multicolumn{8}{c}{\cellcolor{Xshade}Features \(X\)}
&
\multicolumn{4}{c}{\cellcolor{Dshade}Dependence}
\\
\cmidrule(lr){3-10}
\cmidrule(lr){11-14}

DGP type
& Model
& \(E_{\mu,Q}\)
& \(E_{\mu,U}\)
& \(E_{\Sigma,Q}\)
& \(E_{\Sigma,U}\)
& \shortstack{\(W_1\)\\Agg.\ Sev.}
& \shortstack{\(W_1\)\\Comorb.}
& \shortstack{\(W_1\)\\Acuity}
& \shortstack{\(W_1\)\\Resp.}
& \(E_{\mathrm{corr},Q}\)
& \(E_{\mathrm{corr},U}\)
& \(E_{\mathrm{Sp},Q}\)
& \(E_{\mathrm{Sp},U}\)
\\
\midrule

\multirow{5}{*}{Linear}
& TWFE
& \estold{\textbf{0.0421}}{(0.0015)}
& \estold{\textbf{0.0386}}{(0.0024)}
& \estold{\textbf{0.0517}}{(0.0012)}
& \estold{\textbf{0.0591}}{(0.0024)}
& \estold{\textbf{0.0564}}{(0.0026)}
& \estold{\textbf{0.0550}}{(0.0029)}
& \estold{\textbf{0.0558}}{(0.0029)}
& \estold{\textbf{0.0505}}{(0.0029)}
& \estold{0.1533}{(0.0027)}
& \estold{0.2238}{(0.0053)}
& \estold{0.1199}{(0.0029)}
& \estold{0.1545}{(0.0046)}
\\

& Matching
& \estold{\textbf{0.0421}}{(0.0015)}
& \estold{\textbf{0.0386}}{(0.0024)}
& \estold{\textbf{0.0517}}{(0.0012)}
& \estold{\textbf{0.0591}}{(0.0024)}
& \estold{\textbf{0.0564}}{(0.0026)}
& \estold{\textbf{0.0550}}{(0.0029)}
& \estold{\textbf{0.0558}}{(0.0029)}
& \estold{\textbf{0.0505}}{(0.0029)}
& \estold{0.1194}{(0.0050)}
& \estold{0.0928}{(0.0049)}
& \estold{0.1067}{(0.0049)}
& \estold{\textbf{0.0735}}{(0.0041)}
\\

& GenSynth
& \estold{0.3823}{(0.0026)}
& \estold{0.6353}{(0.0038)}
& \estold{0.2168}{(0.0009)}
& \estold{0.3655}{(0.0022)}
& \estold{1.0888}{(0.0082)}
& \estold{0.4755}{(0.0049)}
& \estold{0.5220}{(0.0061)}
& \estold{0.2768}{(0.0068)}
& \estold{0.1665}{(0.0029)}
& \estold{0.2049}{(0.0042)}
& \estold{0.1554}{(0.0034)}
& \estold{0.1555}{(0.0040)}
\\

& CycleGAN
& \estold{0.0657}{(0.0024)}
& \estold{0.0664}{(0.0036)}
& \estold{0.0761}{(0.0014)}
& \estold{0.1424}{(0.0024)}
& \estold{0.1032}{(0.0045)}
& \estold{0.1173}{(0.0036)}
& \estold{0.1387}{(0.0038)}
& \estold{0.0886}{(0.0032)}
& \estold{0.0717}{(0.0023)}
& \estold{0.0776}{(0.0038)}
& \estold{0.0720}{(0.0023)}
& \estold{0.0766}{(0.0038)}
\\

& \textbf{OTSynth}
& \estold{0.0846}{(0.0025)}
& \estold{0.0630}{(0.0032)}
& \estold{0.0974}{(0.0012)}
& \estold{0.1354}{(0.0056)}
& \estold{0.0950}{(0.0053)}
& \estold{0.1081}{(0.0046)}
& \estold{0.1242}{(0.0053)}
& \estold{0.0974}{(0.0048)}
& \estold{\textbf{0.0635}}{(0.0023)}
& \estold{\textbf{0.0776}}{(0.0048)}
& \estold{\textbf{0.0625}}{(0.0024)}
& \estold{0.0761}{(0.0048)}
\\

\addlinespace[2pt]
\midrule
\addlinespace[2pt]

\multirow{5}{*}{Nonlinear}
& TWFE
& \estold{\textbf{0.0420}}{(0.0015)}
& \estold{\textbf{0.0427}}{(0.0024)}
& \estold{\textbf{0.0547}}{(0.0015)}
& \estold{\textbf{0.0882}}{(0.0043)}
& \estold{\textbf{0.0589}}{(0.0027)}
& \estold{\textbf{0.0510}}{(0.0029)}
& \estold{\textbf{0.0594}}{(0.0031)}
& \estold{\textbf{0.0484}}{(0.0028)}
& \estold{0.2153}{(0.0035)}
& \estold{0.3326}{(0.0058)}
& \estold{0.1663}{(0.0034)}
& \estold{0.1807}{(0.0050)}
\\

& Matching
& \estold{\textbf{0.0420}}{(0.0015)}
& \estold{\textbf{0.0427}}{(0.0024)}
& \estold{\textbf{0.0547}}{(0.0015)}
& \estold{\textbf{0.0882}}{(0.0043)}
& \estold{\textbf{0.0589}}{(0.0027)}
& \estold{\textbf{0.0510}}{(0.0029)}
& \estold{\textbf{0.0594}}{(0.0031)}
& \estold{\textbf{0.0484}}{(0.0028)}
& \estold{0.1662}{(0.0032)}
& \estold{0.1775}{(0.0056)}
& \estold{0.1127}{(0.0038)}
& \estold{0.0927}{(0.0045)}
\\

& GenSynth
& \estold{0.3651}{(0.0022)}
& \estold{0.5301}{(0.0021)}
& \estold{0.2418}{(0.0010)}
& \estold{0.4629}{(0.0012)}
& \estold{0.7516}{(0.0058)}
& \estold{0.5509}{(0.0054)}
& \estold{0.5882}{(0.0056)}
& \estold{0.2332}{(0.0059)}
& \estold{0.1894}{(0.0048)}
& \estold{0.2032}{(0.0101)}
& \estold{0.1736}{(0.0050)}
& \estold{0.1718}{(0.0080)}
\\

& CycleGAN
& \estold{0.0669}{(0.0024)}
& \estold{0.0599}{(0.0032)}
& \estold{0.0775}{(0.0015)}
& \estold{0.1809}{(0.0031)}
& \estold{0.1655}{(0.0040)}
& \estold{0.2129}{(0.0031)}
& \estold{0.1694}{(0.0048)}
& \estold{0.2301}{(0.0046)}
& \estold{0.0892}{(0.0025)}
& \estold{0.1910}{(0.0049)}
& \estold{0.0882}{(0.0025)}
& \estold{\textbf{0.0760}}{(0.0025)}
\\

& \textbf{OTSynth}
& \estold{0.0833}{(0.0026)}
& \estold{0.0627}{(0.0041)}
& \estold{0.0981}{(0.0013)}
& \estold{0.1459}{(0.0076)}
& \estold{0.1208}{(0.0049)}
& \estold{0.1378}{(0.0049)}
& \estold{0.1336}{(0.0059)}
& \estold{0.1484}{(0.0056)}
& \estold{\textbf{0.0677}}{(0.0026)}
& \estold{\textbf{0.1041}}{(0.0050)}
& \estold{\textbf{0.0658}}{(0.0022)}
& \estold{0.0799}{(0.0039)}
\\

\bottomrule
\end{tabular}
\end{adjustbox}

\vspace{2mm}

\begin{minipage}{0.98\textwidth}
\centering
\footnotesize
Entries report means over \(R=50\) Monte Carlo runs, with standard errors in parentheses.
\end{minipage}

\end{table}

\begin{figure}[H]
\centering
\includegraphics[width=0.95\textwidth]{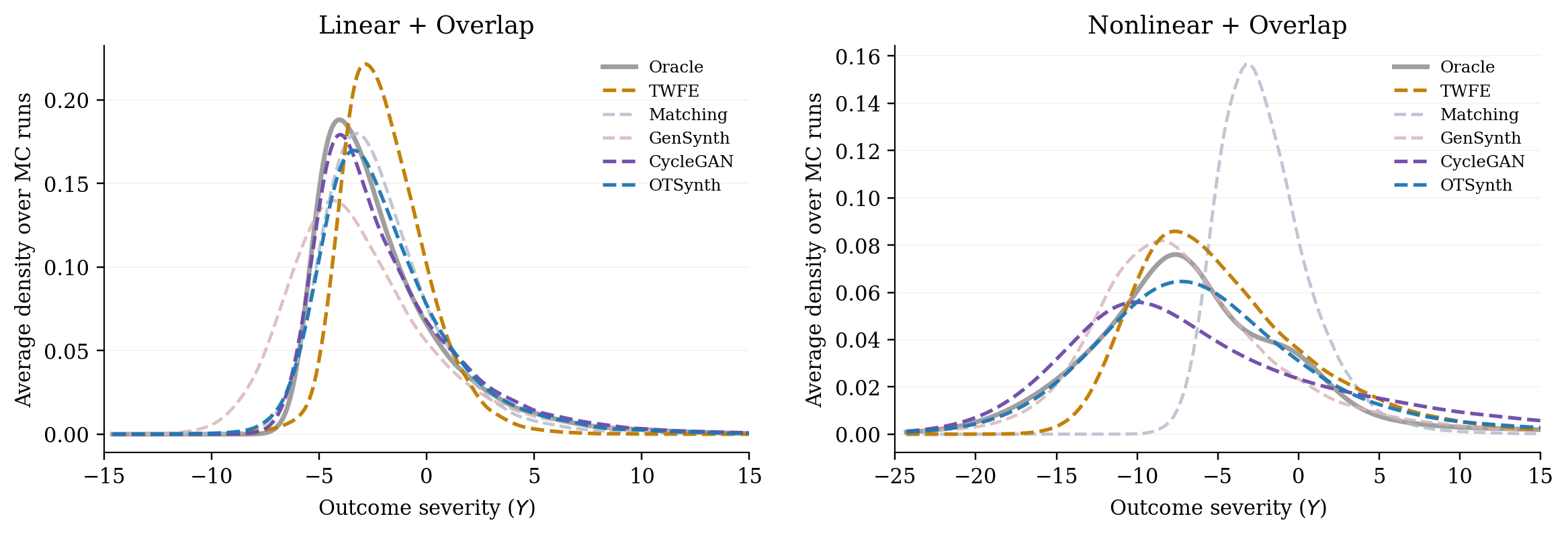}
\caption{Estimated outcome densities under the Limited cross-site overlap design.}
\label{fig:y_density_overlap}
\end{figure}

\newpage

\begin{table}[H]
\centering
\captionsetup{justification=centering}

\caption{Additional diagnostics for the Arm-specific cross-site transformation design.}
\label{tab:dgp_sharedf_additional_1}

\setlength{\tabcolsep}{3.2pt}
\renewcommand{\arraystretch}{1.55}

\begin{adjustbox}{max width=\textwidth}
\begin{tabular}{
ll
*{10}{>{\columncolor{Yshade}}c}
*{2}{>{\columncolor{Xshade}}c}
}
\toprule
&
&
\multicolumn{10}{c}{\cellcolor{Yshade}Outcome \(Y\)}
&
\multicolumn{2}{c}{\cellcolor{Xshade}Features \(X\)}
\\
\cmidrule(lr){3-12}
\cmidrule(lr){13-14}

DGP type
& Model
& \(E_{q_{.01}}\)
& \(E_{q_{.05}}\)
& \(E_{q_{.25}}\)
& \(E_{q_{.50}}\)
& \(E_{q_{.75}}\)
& \(E_{q_{.95}}\)
& \(E_{q_{.99}}\)
& \(D_{\mathrm{KS}}\)
& \(D_{\mathrm{H}}\)
& \(D_{\mathrm{JS}}\)
& \(SW_1(X_Q)\)
& \(SW_1(X_U)\)
\\
\midrule

\multirow{5}{*}{Linear}
& TWFE
& \estold{0.1374}{(0.0133)}
& \estold{0.3643}{(0.0083)}
& \estold{0.5253}{(0.0073)}
& \estold{0.5558}{(0.0096)}
& \estold{0.4764}{(0.0149)}
& \estold{\textbf{0.1575}}{(0.0167)}
& \estold{1.0421}{(0.0649)}
& \estold{0.3204}{(0.0036)}
& \estold{0.3090}{(0.0029)}
& \estold{0.0891}{(0.0016)}
& \estold{\textbf{0.0585}}{(0.0010)}
& \estold{\textbf{0.0590}}{(0.0018)}
\\

& Matching
& \estold{\textbf{0.0545}}{(0.0069)}
& \estold{\textbf{0.1177}}{(0.0065)}
& \estold{\textbf{0.2871}}{(0.0064)}
& \estold{0.3924}{(0.0104)}
& \estold{0.5041}{(0.0162)}
& \estold{0.7608}{(0.0459)}
& \estold{1.1964}{(0.1075)}
& \estold{0.2046}{(0.0037)}
& \estold{\textbf{0.2035}}{(0.0025)}
& \estold{\textbf{0.0381}}{(0.0009)}
& \estold{\textbf{0.0585}}{(0.0010)}
& \estold{\textbf{0.0590}}{(0.0018)}
\\

& GenSynth
& \estold{1.2422}{(0.0264)}
& \estold{0.8190}{(0.0168)}
& \estold{0.2995}{(0.0103)}
& \estold{\textbf{0.0497}}{(0.0046)}
& \estold{\textbf{0.3420}}{(0.0147)}
& \estold{1.2304}{(0.0421)}
& \estold{2.3329}{(0.1065)}
& \estold{\textbf{0.1835}}{(0.0034)}
& \estold{0.3076}{(0.0034)}
& \estold{0.0772}{(0.0016)}
& \estold{0.2299}{(0.0019)}
& \estold{0.3786}{(0.0037)}
\\

& CycleGAN
& \estold{0.2668}{(0.0119)}
& \estold{0.2747}{(0.0082)}
& \estold{0.3368}{(0.0080)}
& \estold{0.4326}{(0.0118)}
& \estold{0.5764}{(0.0197)}
& \estold{0.8865}{(0.0448)}
& \estold{1.2608}{(0.0961)}
& \estold{0.2287}{(0.0042)}
& \estold{0.2400}{(0.0033)}
& \estold{0.0524}{(0.0014)}
& \estold{0.0739}{(0.0016)}
& \estold{0.0916}{(0.0031)}
\\

& \textbf{OTSynth}
& \estold{0.1591}{(0.0133)}
& \estold{0.2528}{(0.0107)}
& \estold{0.3764}{(0.0091)}
& \estold{0.4634}{(0.0128)}
& \estold{0.5400}{(0.0178)}
& \estold{0.5537}{(0.0423)}
& \estold{\textbf{0.7220}}{(0.0681)}
& \estold{0.2466}{(0.0045)}
& \estold{0.2430}{(0.0036)}
& \estold{0.0553}{(0.0016)}
& \estold{0.0712}{(0.0016)}
& \estold{0.0916}{(0.0023)}
\\

\addlinespace[2pt]
\midrule
\addlinespace[2pt]

\multirow{5}{*}{Nonlinear}
& TWFE
& \estold{0.2289}{(0.0170)}
& \estold{0.1304}{(0.0107)}
& \estold{0.0953}{(0.0092)}
& \estold{0.1771}{(0.0108)}
& \estold{0.5313}{(0.0240)}
& \estold{1.0810}{(0.0570)}
& \estold{\textbf{0.7709}}{(0.0888)}
& \estold{0.2176}{(0.0047)}
& \estold{0.2413}{(0.0042)}
& \estold{0.0534}{(0.0017)}
& \estold{\textbf{0.0591}}{(0.0010)}
& \estold{\textbf{0.0610}}{(0.0018)}
\\

& Matching
& \estold{1.3621}{(0.0436)}
& \estold{1.1106}{(0.0363)}
& \estold{0.7536}{(0.0241)}
& \estold{0.5421}{(0.0178)}
& \estold{0.4967}{(0.0180)}
& \estold{0.2939}{(0.0236)}
& \estold{1.4761}{(0.1083)}
& \estold{0.6751}{(0.0026)}
& \estold{0.6039}{(0.0046)}
& \estold{0.2942}{(0.0044)}
& \estold{\textbf{0.0591}}{(0.0010)}
& \estold{\textbf{0.0610}}{(0.0018)}
\\

& GenSynth
& \estold{\textbf{0.1516}}{(0.0162)}
& \estold{\textbf{0.0901}}{(0.0098)}
& \estold{\textbf{0.0677}}{(0.0065)}
& \estold{0.1187}{(0.0079)}
& \estold{\textbf{0.0606}}{(0.0061)}
& \estold{\textbf{0.1173}}{(0.0121)}
& \estold{1.4685}{(0.1059)}
& \estold{\textbf{0.1120}}{(0.0049)}
& \estold{\textbf{0.1501}}{(0.0037)}
& \estold{\textbf{0.0206}}{(0.0010)}
& \estold{0.3040}{(0.0015)}
& \estold{0.4920}{(0.0022)}
\\

& CycleGAN
& \estold{0.2067}{(0.0215)}
& \estold{0.2318}{(0.0152)}
& \estold{0.2245}{(0.0115)}
& \estold{\textbf{0.0627}}{(0.0069)}
& \estold{0.6353}{(0.0325)}
& \estold{2.2008}{(0.0961)}
& \estold{2.6498}{(0.2026)}
& \estold{0.2036}{(0.0039)}
& \estold{0.2953}{(0.0044)}
& \estold{0.0800}{(0.0023)}
& \estold{0.0778}{(0.0017)}
& \estold{0.1327}{(0.0023)}
\\

& \textbf{OTSynth}
& \estold{0.5156}{(0.0330)}
& \estold{0.3274}{(0.0167)}
& \estold{0.0708}{(0.0070)}
& \estold{0.1809}{(0.0109)}
& \estold{0.5621}{(0.0230)}
& \estold{1.4906}{(0.0764)}
& \estold{2.1871}{(0.2113)}
& \estold{0.2200}{(0.0043)}
& \estold{0.2657}{(0.0036)}
& \estold{0.0648}{(0.0017)}
& \estold{0.0709}{(0.0016)}
& \estold{0.0995}{(0.0029)}
\\

\bottomrule
\end{tabular}
\end{adjustbox}
\end{table}

\vspace{-6mm}

\begin{table}[H]
\centering
\setlength{\tabcolsep}{3.8pt}
\renewcommand{\arraystretch}{1.55}

\begin{adjustbox}{max width=\textwidth}
\begin{tabular}{
ll
*{8}{>{\columncolor{Xshade}}c}
*{4}{>{\columncolor{Dshade}}c}
}
\toprule
&
&
\multicolumn{8}{c}{\cellcolor{Xshade}Features \(X\)}
&
\multicolumn{4}{c}{\cellcolor{Dshade}Dependence}
\\
\cmidrule(lr){3-10}
\cmidrule(lr){11-14}

DGP type
& Model
& \(E_{\mu,Q}\)
& \(E_{\mu,U}\)
& \(E_{\Sigma,Q}\)
& \(E_{\Sigma,U}\)
& \shortstack{\(W_1\)\\Agg.\ Sev.}
& \shortstack{\(W_1\)\\Comorb.}
& \shortstack{\(W_1\)\\Acuity}
& \shortstack{\(W_1\)\\Resp.}
& \(E_{\mathrm{corr},Q}\)
& \(E_{\mathrm{corr},U}\)
& \(E_{\mathrm{Sp},Q}\)
& \(E_{\mathrm{Sp},U}\)
\\
\midrule

\multirow{5}{*}{Linear}
& TWFE
& \estold{\textbf{0.0439}}{(0.0016)}
& \estold{\textbf{0.0444}}{(0.0028)}
& \estold{\textbf{0.0527}}{(0.0011)}
& \estold{\textbf{0.0522}}{(0.0017)}
& \estold{\textbf{0.0574}}{(0.0035)}
& \estold{\textbf{0.0541}}{(0.0027)}
& \estold{\textbf{0.0586}}{(0.0036)}
& \estold{\textbf{0.0560}}{(0.0030)}
& \estold{0.1395}{(0.0023)}
& \estold{0.2373}{(0.0045)}
& \estold{0.1064}{(0.0022)}
& \estold{0.1654}{(0.0036)}
\\

& Matching
& \estold{\textbf{0.0439}}{(0.0016)}
& \estold{\textbf{0.0444}}{(0.0028)}
& \estold{\textbf{0.0527}}{(0.0011)}
& \estold{\textbf{0.0522}}{(0.0017)}
& \estold{\textbf{0.0574}}{(0.0035)}
& \estold{\textbf{0.0541}}{(0.0027)}
& \estold{\textbf{0.0586}}{(0.0036)}
& \estold{\textbf{0.0560}}{(0.0030)}
& \estold{0.0641}{(0.0023)}
& \estold{\textbf{0.0604}}{(0.0031)}
& \estold{\textbf{0.0623}}{(0.0020)}
& \estold{\textbf{0.0601}}{(0.0029)}
\\

& GenSynth
& \estold{0.2032}{(0.0025)}
& \estold{0.3687}{(0.0040)}
& \estold{0.1561}{(0.0009)}
& \estold{0.2924}{(0.0020)}
& \estold{0.6650}{(0.0064)}
& \estold{0.4142}{(0.0031)}
& \estold{0.2194}{(0.0038)}
& \estold{0.1537}{(0.0041)}
& \estold{0.1585}{(0.0025)}
& \estold{0.1607}{(0.0027)}
& \estold{0.1643}{(0.0025)}
& \estold{0.1401}{(0.0022)}
\\

& CycleGAN
& \estold{0.0642}{(0.0023)}
& \estold{0.0656}{(0.0043)}
& \estold{0.0701}{(0.0018)}
& \estold{0.1202}{(0.0027)}
& \estold{0.0953}{(0.0054)}
& \estold{0.1065}{(0.0040)}
& \estold{0.1196}{(0.0048)}
& \estold{0.0842}{(0.0031)}
& \estold{\textbf{0.0613}}{(0.0018)}
& \estold{0.0996}{(0.0028)}
& \estold{0.0679}{(0.0021)}
& \estold{0.0941}{(0.0035)}
\\

& \textbf{OTSynth}
& \estold{0.0618}{(0.0023)}
& \estold{0.0625}{(0.0032)}
& \estold{0.0655}{(0.0024)}
& \estold{0.1033}{(0.0039)}
& \estold{0.0861}{(0.0050)}
& \estold{0.0847}{(0.0039)}
& \estold{0.1249}{(0.0043)}
& \estold{0.0945}{(0.0040)}
& \estold{0.0825}{(0.0024)}
& \estold{0.1060}{(0.0040)}
& \estold{0.0699}{(0.0023)}
& \estold{0.0922}{(0.0030)}
\\

\addlinespace[2pt]
\midrule
\addlinespace[2pt]

\multirow{5}{*}{Nonlinear}
& TWFE
& \estold{\textbf{0.0439}}{(0.0016)}
& \estold{\textbf{0.0428}}{(0.0024)}
& \estold{\textbf{0.0578}}{(0.0019)}
& \estold{\textbf{0.0861}}{(0.0032)}
& \estold{\textbf{0.0587}}{(0.0033)}
& \estold{\textbf{0.0501}}{(0.0022)}
& \estold{\textbf{0.0567}}{(0.0030)}
& \estold{\textbf{0.0544}}{(0.0025)}
& \estold{0.1846}{(0.0038)}
& \estold{0.3513}{(0.0054)}
& \estold{0.1410}{(0.0025)}
& \estold{0.2391}{(0.0040)}
\\

& Matching
& \estold{\textbf{0.0439}}{(0.0016)}
& \estold{\textbf{0.0428}}{(0.0024)}
& \estold{\textbf{0.0578}}{(0.0019)}
& \estold{\textbf{0.0861}}{(0.0032)}
& \estold{\textbf{0.0587}}{(0.0033)}
& \estold{\textbf{0.0501}}{(0.0022)}
& \estold{\textbf{0.0567}}{(0.0030)}
& \estold{\textbf{0.0544}}{(0.0025)}
& \estold{0.1029}{(0.0035)}
& \estold{\textbf{0.1067}}{(0.0059)}
& \estold{0.0745}{(0.0020)}
& \estold{\textbf{0.0561}}{(0.0031)}
\\

& GenSynth
& \estold{0.2312}{(0.0021)}
& \estold{0.4275}{(0.0023)}
& \estold{0.2086}{(0.0009)}
& \estold{0.4499}{(0.0013)}
& \estold{0.6083}{(0.0046)}
& \estold{0.4723}{(0.0021)}
& \estold{0.5024}{(0.0035)}
& \estold{0.1562}{(0.0027)}
& \estold{0.1788}{(0.0036)}
& \estold{0.1863}{(0.0038)}
& \estold{0.2152}{(0.0039)}
& \estold{0.2134}{(0.0051)}
\\

& CycleGAN
& \estold{0.0674}{(0.0025)}
& \estold{0.0660}{(0.0036)}
& \estold{0.0785}{(0.0025)}
& \estold{0.1698}{(0.0046)}
& \estold{0.1281}{(0.0041)}
& \estold{0.2031}{(0.0039)}
& \estold{0.1353}{(0.0045)}
& \estold{0.1846}{(0.0038)}
& \estold{\textbf{0.0820}}{(0.0029)}
& \estold{0.2127}{(0.0059)}
& \estold{\textbf{0.0721}}{(0.0020)}
& \estold{0.1036}{(0.0041)}
\\

& \textbf{OTSynth}
& \estold{0.0614}{(0.0022)}
& \estold{0.0598}{(0.0033)}
& \estold{0.0649}{(0.0022)}
& \estold{0.1254}{(0.0040)}
& \estold{0.1101}{(0.0058)}
& \estold{0.1256}{(0.0037)}
& \estold{0.1260}{(0.0051)}
& \estold{0.1393}{(0.0042)}
& \estold{0.0960}{(0.0033)}
& \estold{0.1579}{(0.0066)}
& \estold{0.0734}{(0.0022)}
& \estold{0.1061}{(0.0042)}
\\

\bottomrule
\end{tabular}
\end{adjustbox}

\vspace{2mm}

\begin{minipage}{0.98\textwidth}
\centering
\footnotesize
Entries report means over \(R=50\) Monte Carlo runs, with standard errors in parentheses.
\end{minipage}

\end{table}

\begin{figure}[H]
\centering
\includegraphics[width=0.95\textwidth]{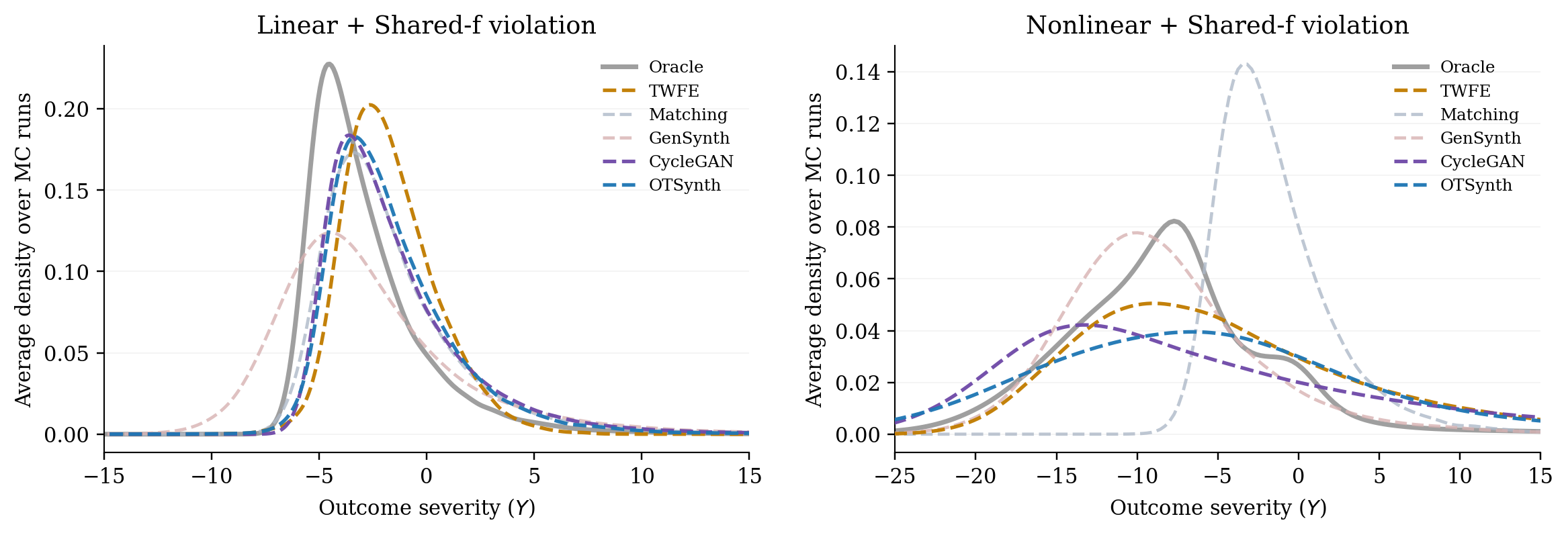}
\caption{Estimated outcome densities under the Arm-specific cross-site transformation design.}
\label{fig:y_density_sharedf}
\end{figure}

\newpage

\begin{table}[H]
\centering
\captionsetup{justification=centering}

\caption{Additional diagnostics for the Heterogeneous care regime design.}
\label{tab:dgp_regime_additional_1}

\setlength{\tabcolsep}{3.2pt}
\renewcommand{\arraystretch}{1.55}

\begin{adjustbox}{max width=\textwidth}
\begin{tabular}{
ll
*{10}{>{\columncolor{Yshade}}c}
*{2}{>{\columncolor{Xshade}}c}
}
\toprule
&
&
\multicolumn{10}{c}{\cellcolor{Yshade}Outcome \(Y\)}
&
\multicolumn{2}{c}{\cellcolor{Xshade}Features \(X\)}
\\
\cmidrule(lr){3-12}
\cmidrule(lr){13-14}

DGP type
& Model
& \(E_{q_{.01}}\)
& \(E_{q_{.05}}\)
& \(E_{q_{.25}}\)
& \(E_{q_{.50}}\)
& \(E_{q_{.75}}\)
& \(E_{q_{.95}}\)
& \(E_{q_{.99}}\)
& \(D_{\mathrm{KS}}\)
& \(D_{\mathrm{H}}\)
& \(D_{\mathrm{JS}}\)
& \(SW_1(X_Q)\)
& \(SW_1(X_U)\)
\\
\midrule

\multirow{5}{*}{Linear}
& TWFE
& \estold{1.0641}{(0.0174)}
& \estold{1.0260}{(0.0144)}
& \estold{0.3249}{(0.0101)}
& \estold{0.1776}{(0.0056)}
& \estold{0.0985}{(0.0071)}
& \estold{0.3426}{(0.0139)}
& \estold{0.8956}{(0.0453)}
& \estold{0.1937}{(0.0034)}
& \estold{0.3793}{(0.0023)}
& \estold{0.1224}{(0.0012)}
& \estold{\textbf{0.0585}}{(0.0010)}
& \estold{\textbf{0.0590}}{(0.0018)}
\\

& Matching
& \estold{\textbf{0.3768}}{(0.0132)}
& \estold{0.4056}{(0.0102)}
& \estold{1.7753}{(0.0122)}
& \estold{1.5525}{(0.0113)}
& \estold{1.1367}{(0.0148)}
& \estold{0.2401}{(0.0254)}
& \estold{0.9991}{(0.0946)}
& \estold{0.6098}{(0.0024)}
& \estold{0.5007}{(0.0022)}
& \estold{0.2328}{(0.0019)}
& \estold{\textbf{0.0585}}{(0.0010)}
& \estold{\textbf{0.0590}}{(0.0018)}
\\

& GenSynth
& \estold{1.1641}{(0.0321)}
& \estold{0.6019}{(0.0202)}
& \estold{1.2996}{(0.0152)}
& \estold{0.6809}{(0.0119)}
& \estold{0.1643}{(0.0120)}
& \estold{2.0364}{(0.0334)}
& \estold{4.1208}{(0.0985)}
& \estold{0.3587}{(0.0031)}
& \estold{0.4135}{(0.0024)}
& \estold{0.1538}{(0.0016)}
& \estold{0.2299}{(0.0019)}
& \estold{0.3786}{(0.0037)}
\\

& CycleGAN
& \estold{0.6561}{(0.0186)}
& \estold{0.7170}{(0.0162)}
& \estold{0.4506}{(0.0115)}
& \estold{0.1722}{(0.0086)}
& \estold{0.1872}{(0.0109)}
& \estold{0.9101}{(0.0332)}
& \estold{1.6700}{(0.0762)}
& \estold{0.2022}{(0.0040)}
& \estold{0.3643}{(0.0030)}
& \estold{0.1199}{(0.0018)}
& \estold{0.0753}{(0.0015)}
& \estold{0.0933}{(0.0032)}
\\

& \textbf{OTSynth}
& \estold{0.4685}{(0.0380)}
& \estold{\textbf{0.2373}}{(0.0194)}
& \estold{\textbf{0.1773}}{(0.0124)}
& \estold{\textbf{0.0325}}{(0.0031)}
& \estold{\textbf{0.0680}}{(0.0058)}
& \estold{\textbf{0.1024}}{(0.0101)}
& \estold{\textbf{0.3272}}{(0.0366)}
& \estold{\textbf{0.0870}}{(0.0020)}
& \estold{\textbf{0.2119}}{(0.0024)}
& \estold{\textbf{0.0410}}{(0.0010)}
& \estold{0.0712}{(0.0016)}
& \estold{0.0940}{(0.0030)}
\\

\addlinespace[2pt]
\midrule
\addlinespace[2pt]

\multirow{5}{*}{Nonlinear}
& TWFE
& \estold{0.6624}{(0.0206)}
& \estold{0.4993}{(0.0143)}
& \estold{\textbf{0.0365}}{(0.0033)}
& \estold{0.1455}{(0.0058)}
& \estold{\textbf{0.0650}}{(0.0073)}
& \estold{0.5679}{(0.0224)}
& \estold{1.7761}{(0.0745)}
& \estold{0.1577}{(0.0035)}
& \estold{0.3040}{(0.0051)}
& \estold{0.0831}{(0.0026)}
& \estold{\textbf{0.0591}}{(0.0010)}
& \estold{\textbf{0.0610}}{(0.0018)}
\\

& Matching
& \estold{0.7009}{(0.0172)}
& \estold{0.3922}{(0.0104)}
& \estold{0.3694}{(0.0085)}
& \estold{0.3620}{(0.0085)}
& \estold{0.5441}{(0.0142)}
& \estold{1.2837}{(0.0317)}
& \estold{2.5587}{(0.0851)}
& \estold{0.5213}{(0.0034)}
& \estold{0.5181}{(0.0072)}
& \estold{0.2359}{(0.0057)}
& \estold{\textbf{0.0591}}{(0.0010)}
& \estold{\textbf{0.0610}}{(0.0018)}
\\

& GenSynth
& \estold{0.4032}{(0.0128)}
& \estold{0.2927}{(0.0101)}
& \estold{0.2280}{(0.0067)}
& \estold{0.0828}{(0.0051)}
& \estold{0.1341}{(0.0086)}
& \estold{0.6580}{(0.0225)}
& \estold{1.7356}{(0.0736)}
& \estold{0.2083}{(0.0040)}
& \estold{0.2829}{(0.0059)}
& \estold{0.0743}{(0.0028)}
& \estold{0.3040}{(0.0015)}
& \estold{0.4920}{(0.0022)}
\\

& CycleGAN
& \estold{0.3129}{(0.0187)}
& \estold{0.2426}{(0.0121)}
& \estold{0.2127}{(0.0071)}
& \estold{\textbf{0.0357}}{(0.0035)}
& \estold{0.1887}{(0.0142)}
& \estold{0.4285}{(0.0428)}
& \estold{0.6491}{(0.0667)}
& \estold{0.1653}{(0.0042)}
& \estold{0.2592}{(0.0060)}
& \estold{0.0641}{(0.0025)}
& \estold{0.0772}{(0.0017)}
& \estold{0.1313}{(0.0023)}
\\

& \textbf{OTSynth}
& \estold{\textbf{0.2685}}{(0.0306)}
& \estold{\textbf{0.0805}}{(0.0101)}
& \estold{0.1277}{(0.0071)}
& \estold{0.0922}{(0.0064)}
& \estold{0.1155}{(0.0111)}
& \estold{\textbf{0.1923}}{(0.0233)}
& \estold{\textbf{0.5594}}{(0.0630)}
& \estold{\textbf{0.1219}}{(0.0032)}
& \estold{\textbf{0.1987}}{(0.0051)}
& \estold{\textbf{0.0379}}{(0.0018)}
& \estold{0.0709}{(0.0016)}
& \estold{0.1016}{(0.0031)}
\\

\bottomrule
\end{tabular}
\end{adjustbox}
\end{table}

\vspace{-6mm}

\begin{table}[H]
\centering
\setlength{\tabcolsep}{3.8pt}
\renewcommand{\arraystretch}{1.55}

\begin{adjustbox}{max width=\textwidth}
\begin{tabular}{
ll
*{8}{>{\columncolor{Xshade}}c}
*{4}{>{\columncolor{Dshade}}c}
}
\toprule
&
&
\multicolumn{8}{c}{\cellcolor{Xshade}Features \(X\)}
&
\multicolumn{4}{c}{\cellcolor{Dshade}Dependence}
\\
\cmidrule(lr){3-10}
\cmidrule(lr){11-14}

DGP type
& Model
& \(E_{\mu,Q}\)
& \(E_{\mu,U}\)
& \(E_{\Sigma,Q}\)
& \(E_{\Sigma,U}\)
& \shortstack{\(W_1\)\\Agg.\ Sev.}
& \shortstack{\(W_1\)\\Comorb.}
& \shortstack{\(W_1\)\\Acuity}
& \shortstack{\(W_1\)\\Resp.}
& \(E_{\mathrm{corr},Q}\)
& \(E_{\mathrm{corr},U}\)
& \(E_{\mathrm{Sp},Q}\)
& \(E_{\mathrm{Sp},U}\)
\\
\midrule

\multirow{5}{*}{Linear}
& TWFE
& \estold{\textbf{0.0439}}{(0.0016)}
& \estold{\textbf{0.0444}}{(0.0028)}
& \estold{\textbf{0.0527}}{(0.0011)}
& \estold{\textbf{0.0522}}{(0.0017)}
& \estold{\textbf{0.0574}}{(0.0035)}
& \estold{\textbf{0.0541}}{(0.0027)}
& \estold{\textbf{0.0586}}{(0.0036)}
& \estold{\textbf{0.0560}}{(0.0030)}
& \estold{0.1260}{(0.0024)}
& \estold{0.1259}{(0.0055)}
& \estold{0.1551}{(0.0021)}
& \estold{0.0933}{(0.0038)}
\\

& Matching
& \estold{\textbf{0.0439}}{(0.0016)}
& \estold{\textbf{0.0444}}{(0.0028)}
& \estold{\textbf{0.0527}}{(0.0011)}
& \estold{\textbf{0.0522}}{(0.0017)}
& \estold{\textbf{0.0574}}{(0.0035)}
& \estold{\textbf{0.0541}}{(0.0027)}
& \estold{\textbf{0.0586}}{(0.0036)}
& \estold{\textbf{0.0560}}{(0.0030)}
& \estold{0.2642}{(0.0035)}
& \estold{0.2508}{(0.0052)}
& \estold{0.2357}{(0.0026)}
& \estold{0.2248}{(0.0049)}
\\

& GenSynth
& \estold{0.2032}{(0.0025)}
& \estold{0.3687}{(0.0040)}
& \estold{0.1561}{(0.0009)}
& \estold{0.2924}{(0.0020)}
& \estold{0.6650}{(0.0064)}
& \estold{0.4142}{(0.0031)}
& \estold{0.2194}{(0.0038)}
& \estold{0.1537}{(0.0041)}
& \estold{0.2895}{(0.0025)}
& \estold{0.3055}{(0.0039)}
& \estold{0.2552}{(0.0026)}
& \estold{0.2238}{(0.0043)}
\\

& CycleGAN
& \estold{0.0654}{(0.0022)}
& \estold{0.0693}{(0.0046)}
& \estold{0.0720}{(0.0018)}
& \estold{0.1206}{(0.0032)}
& \estold{0.1040}{(0.0061)}
& \estold{0.1141}{(0.0038)}
& \estold{0.1180}{(0.0050)}
& \estold{0.0849}{(0.0035)}
& \estold{0.1073}{(0.0030)}
& \estold{0.1142}{(0.0044)}
& \estold{0.0907}{(0.0025)}
& \estold{0.1282}{(0.0054)}
\\

& \textbf{OTSynth}
& \estold{0.0618}{(0.0023)}
& \estold{0.0652}{(0.0039)}
& \estold{0.0655}{(0.0024)}
& \estold{0.1027}{(0.0036)}
& \estold{0.0851}{(0.0059)}
& \estold{0.0905}{(0.0045)}
& \estold{0.1285}{(0.0057)}
& \estold{0.1008}{(0.0040)}
& \estold{\textbf{0.0736}}{(0.0030)}
& \estold{\textbf{0.0651}}{(0.0037)}
& \estold{\textbf{0.0658}}{(0.0025)}
& \estold{\textbf{0.0817}}{(0.0045)}
\\

\addlinespace[2pt]
\midrule
\addlinespace[2pt]

\multirow{5}{*}{Nonlinear}
& TWFE
& \estold{\textbf{0.0439}}{(0.0016)}
& \estold{\textbf{0.0428}}{(0.0024)}
& \estold{\textbf{0.0578}}{(0.0019)}
& \estold{\textbf{0.0861}}{(0.0032)}
& \estold{\textbf{0.0587}}{(0.0033)}
& \estold{\textbf{0.0501}}{(0.0022)}
& \estold{\textbf{0.0567}}{(0.0030)}
& \estold{\textbf{0.0544}}{(0.0025)}
& \estold{0.1698}{(0.0033)}
& \estold{0.2246}{(0.0058)}
& \estold{0.1468}{(0.0024)}
& \estold{0.1388}{(0.0064)}
\\

& Matching
& \estold{\textbf{0.0439}}{(0.0016)}
& \estold{\textbf{0.0428}}{(0.0024)}
& \estold{\textbf{0.0578}}{(0.0019)}
& \estold{\textbf{0.0861}}{(0.0032)}
& \estold{\textbf{0.0587}}{(0.0033)}
& \estold{\textbf{0.0501}}{(0.0022)}
& \estold{\textbf{0.0567}}{(0.0030)}
& \estold{\textbf{0.0544}}{(0.0025)}
& \estold{0.2203}{(0.0037)}
& \estold{0.2294}{(0.0043)}
& \estold{0.2648}{(0.0022)}
& \estold{0.2924}{(0.0040)}
\\

& GenSynth
& \estold{0.2312}{(0.0021)}
& \estold{0.4275}{(0.0023)}
& \estold{0.2086}{(0.0009)}
& \estold{0.4499}{(0.0013)}
& \estold{0.6083}{(0.0046)}
& \estold{0.4723}{(0.0021)}
& \estold{0.5024}{(0.0035)}
& \estold{0.1562}{(0.0027)}
& \estold{0.2730}{(0.0034)}
& \estold{0.3035}{(0.0042)}
& \estold{0.2921}{(0.0025)}
& \estold{0.3129}{(0.0047)}
\\

& CycleGAN
& \estold{0.0670}{(0.0025)}
& \estold{0.0654}{(0.0036)}
& \estold{0.0766}{(0.0023)}
& \estold{0.1638}{(0.0045)}
& \estold{0.1288}{(0.0039)}
& \estold{0.2097}{(0.0040)}
& \estold{0.1369}{(0.0046)}
& \estold{0.1894}{(0.0041)}
& \estold{0.1003}{(0.0030)}
& \estold{0.1810}{(0.0056)}
& \estold{0.1138}{(0.0023)}
& \estold{0.0968}{(0.0044)}
\\

& \textbf{OTSynth}
& \estold{0.0614}{(0.0022)}
& \estold{0.0615}{(0.0033)}
& \estold{0.0649}{(0.0022)}
& \estold{0.1376}{(0.0070)}
& \estold{0.1108}{(0.0050)}
& \estold{0.1298}{(0.0037)}
& \estold{0.1307}{(0.0062)}
& \estold{0.1436}{(0.0041)}
& \estold{\textbf{0.0789}}{(0.0029)}
& \estold{\textbf{0.0874}}{(0.0050)}
& \estold{\textbf{0.0732}}{(0.0024)}
& \estold{\textbf{0.0672}}{(0.0037)}
\\

\bottomrule
\end{tabular}
\end{adjustbox}

\vspace{2mm}

\begin{minipage}{0.98\textwidth}
\centering
\footnotesize
Entries report means over \(R=50\) Monte Carlo runs, with standard errors in parentheses.
\end{minipage}

\end{table}

\begin{figure}[H]
\centering
\includegraphics[width=0.95\textwidth]{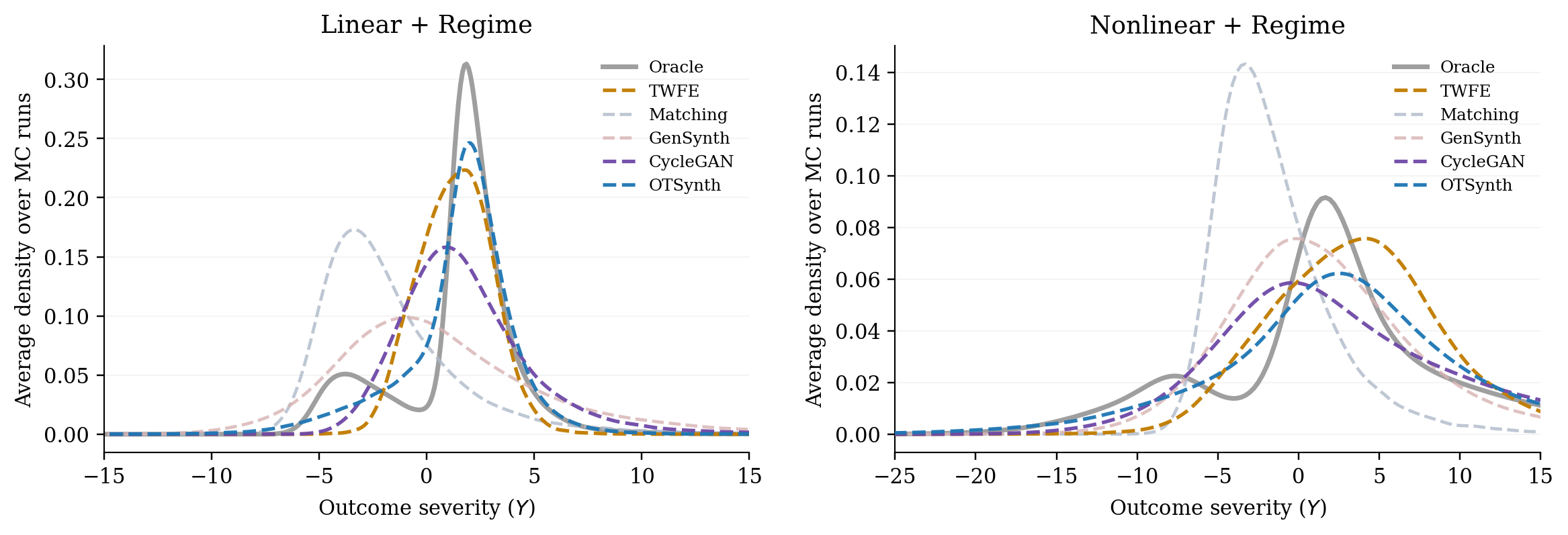}
\caption{Estimated outcome densities under the Heterogeneous care regime design.}
\label{fig:y_density_regime}
\end{figure}

\bibliographystyle{chicago} 
\bibliography{refs}

%% file: final_preamble.tex
\usepackage{amsmath}
\usepackage{graphicx}
\usepackage{enumerate}
\usepackage{natbib}
\usepackage{url} 

\usepackage[draft]{pdfcomment}

\newcommand{\blind}{1}

\usepackage{array, amsmath, amssymb, amsthm, latexsym, enumitem, multirow, caption, graphicx, color, hyperref, babel, enumerate, float, booktabs,  mathtools, amscd, bbm, url, setspace, inputenc, authblk, subcaption, natbib,tikz, bbm, dsfont, algorithm, algpseudocode, comment, amsfonts, amsfonts,amsmath,amsthm,amssymb,bbm,mathrsfs}

\newtheorem{innercustomgeneric}{\customgenericname}
\providecommand{\customgenericname}{}
\newcommand{\newcustomtheorem}[2]{%
\newenvironment{#1}[1]
{%
\renewcommand\customgenericname{#2}%
\renewcommand\theinnercustomgeneric{##1}%
\innercustomgeneric
}
{\endinnercustomgeneric}
}

\newcustomtheorem{customlemma}{lemma}
\theoremstyle{definition}

\newtheorem{theorem}{Theorem}[section]

\newtheorem{lemma}{Lemma}
[section]

\newtheorem{assumption}{Assumption}[section]

\newcommand{\model}[1]{\emph{{#1}}}

\renewcommand{\baselinestretch}{1.46}
\usepackage{soul}
\usepackage{colortbl,xcolor}

\colorlet{lightorange}{orange!50}
\def\Z{\mathcal{Z}}

\def\1{\mathbbm{1}}

\def\1{\mathbbm{1}}
\def\FGW{\operatorname{FGW}}
\def\RFGW{{\operatorname{RFGW}}}
\def\mh{\widehat{\mu}}

\def\f{\phi}

\usepackage{graphicx}    
\usepackage[export]{adjustbox} 
\usepackage{float}       
\usepackage{xcolor}      
\usepackage{caption}     